\documentclass[aps,prd,showpacs,amsfonts,notitlepage,amssymb,amsmath,nofootinbib,superscriptaddress]{revtex4-1}
\usepackage[utf8]{inputenc}
\usepackage[english]{babel}
\usepackage[titletoc,toc,title]{appendix}
\usepackage{amsmath}
\usepackage{amsfonts}
\usepackage{amssymb}
\usepackage[colorlinks=true,linkcolor=blue,citecolor=blue,urlcolor=blue]{hyperref}
\usepackage{graphicx}
\usepackage{enumerate}
\usepackage{csquotes}
\usepackage[caption=false]{subfig}
\usepackage{dsfont}
\usepackage{color}
\usepackage{bbold}
\usepackage{mathtools}
\usepackage{tensor}
\usepackage{gensymb}

\begin{document}

\title{Four-wave mixing and enhanced analogue Hawking effect \\ 
in a nonlinear optical waveguide}
\date{\today}

\author{Scott Robertson}
\affiliation{Laboratoire de l'Acc\'{e}l\'{e}rateur Lin\'{e}aire (UMR 8607), IN2P3/CNRS, Univ. Paris-Sud, Universit\'e Paris-Saclay, 91405 Orsay, France}
\author{Charles Ciret}
\affiliation{Laboratoire de Photonique d'Angers EA 4464, Universit\'e d'Angers, 2 Bd. Lavoisier, 49000 Angers, France}
\author{Serge Massar}
\affiliation{Laboratoire d'Information Quantique CP 224, Universit\'e Libre de Bruxelles (ULB), Av. F. D. Roosevelt 50, B-1050 Bruxelles, Belgium}
\author{Simon-Pierre Gorza}
\affiliation{OPERA-Photonique CP 194/5, Universit\'e Libre de Bruxelles (ULB), Av. F. D. Roosevelt 50, B-1050 Bruxelles, Belgium}
\author{Renaud Parentani}
\affiliation{Laboratoire de Physique Th\'eorique (UMR 8627), CNRS, Univ. Paris-Sud, Universit\'e Paris-Saclay, 91405 Orsay, France}

\begin{abstract}
We study 
in detail
the scattering of light on a soliton propagating in a waveguide
which has been proposed as an experimental system in which one could observe the analogue Hawking effect.
When not applying the rotating wave approximation, we show that the linearized wave equation governing perturbations has the same structure as that governing phonon propagation in an atomic Bose condensate. 
By taking into account the full dispersion relation, we then show that the scattering coefficients encoding the 
production of photon pairs
are amplified by a resonance effect related to the modulation instability occurring in the presence of a continuous wave. 
When using a realistic example of a silicon nitride waveguide on a silica substrate, we find that 
a soliton of duration $10\,{\rm fs}$ would spontaneously emit about one photon pair for every cm it travels,
which makes the effect readily observable. 
This result is confirmed by numerically solving the equation encoding the Kerr nonlinearity and governing the evolution of the full field (soliton plus perturbations).
We discuss the link with previous works devoted to the analogue Hawking effect where the pair creation rates were about six orders of magnitude smaller.
\end{abstract}

\maketitle

\section{Introduction
\label{sec:introduction}}

In 1981, William Unruh pointed out that an 
analogue version of the Hawking effect could be observed in a stationary flow of a moving medium when the flow velocity crosses the speed of quasiparticle excitations~\cite{Unruh:1980cg}.
Since then, several proposals have been made in various media, including 
atomic Bose-Einstein condensates (BEC)~\cite{Garay:1999sk} and 
water in a flume~\cite{Schuetzhold-Unruh-2002}; 
see~\cite{Barcelo-2018,Barcelo:2005fc} 
for other examples and for the link with black hole physics. 
In 2008, the group led by Ulf Leonhardt showed the existence of a fiber-optical 
version 
characterizing the scattering of light on a 
soliton~\cite{Philbin:2007ji,Philbin:2007jj}, see~\cite{Belgiorno:2010wn,Ciret-et-al-2016-I,Ciret-Gorza-2016,Drori:2018ivu} 
for subsequent 
experimental works based on this proposal. 
In these settings, dispersive effects play a key role. 
Indeed, the stability of the soliton requires 
the dispersion relation 
of light propagating in 
the waveguide to display 
an anomalous behavior in the spectral region occupied by the soliton~\cite{Agrawal}. 
Furthermore, the smallness of $\delta n$, the 
modification of the refractive index induced by the soliton 
(typically $\sim 10^{-4}$), implies that the scattering coefficients are 
correspondingly small~\cite{Robertson-thesis-2011} and that one effectively works 
in a highly dispersive regime where the emitted spectrum is not governed by the 
``surface gravity'' 
of an 
analogue horizon~\cite{Finazzi:2012iu}. 
In contrast to 
flowing BEC 
where the calculation of scattering coefficients 
is rather straightforward as 
the spatial gradient near the sonic horizon is dominant~\cite{Leonhardt-Kiss-Ohberg-2003-HE,Macher:2009nz}, 
the smallness of $\delta n$ in the present settings 
complicates the calculation since it is not {\it a priori} clear which approximations are legitimate. 

In this paper, we shall clarify this point 
by computing the scattering coefficients 
by two different means. 
As in~\cite{Leonhardt-Kiss-Ohberg-2003-HE,Macher:2009nz}, 
we shall first work in the background field approximation by solving the linear wave equation describing mode propagation on top of the background, here an optical soliton propagating in a nonlinear waveguide (WG). We shall 
then numerically solve the nonlinear Schr\"{o}dinger equation (NLSE) 
governing the propagation of all the optical fields in the WG, i.e. both the soliton and small perturbations, 
along lines close to those adopted when 
numerically 
studying the analogue Hawking effect in a flowing BEC~\cite{Carusotto:2008ep}. 

In the first approach, the wave equation governing the
mode mixing on the soliton is carefully derived by a linearization of the NLSE. In this we follow and generalize
the treatment of four-wave mixing (FWM) presented in~\cite{Yulin-Skryabin-Russell-2004}. 
As in the standard Bogoliubov treatment of linear perturbations in atomic BEC~\cite{Macher:2009nz} or in polariton systems~\cite{PhysRevA.78.063804,Gerace:2012an}, the wave equation contains a term which mixes configurations of opposite detuned frequencies, and which would be dropped if performing the rotating wave approximation (RWA). 
With respect to atomic BEC,
the novel element here arises from the non-monotonic character of the dispersion relation, which is here expressed as a relation between the 
detuned wave number (measured with respect to the carrier wave number of the soliton) and 
the detuned frequency. 
Crucially, when keeping the fourth-order term in the Taylor expansion of the dispersion relation, 
the positive- and negative-norm branches cross each other, a condition known as {\it phase matching} in optics, see e.g.~\cite{Wadsworth-2004,Rarity-2005}. 
In the presence of 
a continuous wave (CW) laser beam, this would induce a modulation 
instability in a narrow frequency window near 
the crossing, as was shown in Refs.~\cite{Abdullaev-et-al-1994,Pitois-Millot-2003}. 
When replacing the CW by a soliton, we here show that the crossing leads 
to 
a significant 
increase of 
the pair production.
As a result, we obtain a broad spectrum centered around the frequency of the crossing, and not at the frequency at which one might have expected the analogue Hawking signal to reach its maximum value, namely with one photon in the frequency domain where there is a strong conversion from probe to idler, or (in analogue terms) where there is a group velocity horizon.~\footnote{When comparing the spectrum to those found in previous studies of the analogue Hawking effect in waveguides~\cite{Philbin:2007ji,Robertson-thesis-2011,Finazzi:2012bn,Bermudez-Leonhardt-2016,Jacquet-Koenig-2018,Drori:2018ivu}, there is an overall enhancement factor of $\sim 10^{6}$.  This large factor is due to the combination of two effects.  The first is the crossing {\it per se}, as just described.  The second requires 
a distiction to be made %
between ``soft'' and ``hard'' photon production processes.  By ``soft'' we mean the processes that involve the standard FWM known in the nonlinear optics literature and which conserve the total number of photons (soliton plus perturbations), see our Eq.~(\ref{eq:wave_eqn_full}).  Instead, 
``hard'' processes stem from the second-order character of the original Maxwell equations, see e.g. Ref.~\cite{Amiranashvili-2016}. 
It is worth noticing that 
the above-mentioned 
works 
have focused on ``hard'' processes and ignored ``soft'' ones.  It remains to identify the appropriate treatment in which both ``soft'' and ``hard'' processes are taken into account, thereby providing a complete description of the scatttering on the soliton.  We are currently working on this. \label{fn:soft-hard}}

To confirm this result, 
which is obtained from 
the linearized wave equation, we then perform numerical simulations describing the evolution of the full field 
(i.e. soliton $+$ perturbations), governed by the NLSE. We 
first study 
stimulated processes triggered by sending a probe wave
near the phase matching condition. We 
recover the enhanced photon pair creation rate 
and establish that the total number of photons is conserved. As in a BEC, 
the produced quanta originate from particles extracted from the ``condensate'' (i.e. photons from 
the soliton). 
By replacing the probe wave by vacuum fluctuations, we then present 
(in a preliminary analysis) 
the whole 
spectrum and the correlations of spontaneously produced pairs. 
This is 
technically achieved by implementing 
the doublet formalism of Ref.~\cite{Leonhardt-Kiss-Ohberg-2003} at the level of the NLSE. 

The paper is organized as follows.
In Section~\ref{sec:settings}, we first present the nonlinear wave equation and 
its soliton solution. 
We then derive its linearized version describing waves scattered on the soliton, and 
briefly present 
the realistic example of a silicon nitride waveguide used in the numerical simulations. 
In Section~\ref{sec:scattering}, we present the results displaying the above-mentioned enhanced anomalous 
coefficients by numerically 
solving for the
eigenmodes 
of 
the linearized 
equation. 
We 
summarize and 
conclude in Section~\ref{sec:conclusion}.
In Appendix~\ref{app:homogeneous}, we consider 
the propagation of quasiparticle excitations on top of a 
continuous laser beam at the soliton frequency. 
Appendix~\ref{sec:scattering-NL} is concerned with 
numerical integration of the NLSE.

\section{Settings
\label{sec:settings}}

In this section we provide the theoretical background behind the two wave equations 
used in the following sections.
We begin with the nonlinear Schr\"{o}dinger equation (NLSE) describing the evolution of the total field (background + fluctuations). The soliton propagation, which provides the background field, is a 
solution of 
this equation when omitting higher order terms in the dispersion relation. 
The NLSE is then linearized in order to describe fluctuations propagating on top of the soliton. 
Finally, the dispersion relation of the linearized modes and the scattering coefficients relating asymptotic 
modes are discussed. 

\subsection{The nonlinear wave equation} 

Let $z$ and $t$ be the standard space and time coordinates in the lab frame, in which the waveguide is at rest. 
Given our purposes, it is appropriate to write 
the (complex) electric field $E(t,z)$ 
as a slowly-varying envelope $A(t,z)$ multiplying a carrier continuous 
wave: 
\begin{equation}
E(t,z) = A(t,z) \, \mathrm{exp}\left( i \beta_{0} z - i \omega_{0} t\right) \, ,
\label{eq:amp-and-carrier}
\end{equation} 
where $\beta_{0} = \beta\left(\omega_{0}\right)$ is the wave number 
solution of the dispersion relation of linear waves in the waveguide.~\footnote{Conventionally in guided 
optics the dispersion relation between the wave number $k$ and the angular frequency $\omega$ is written $k = \beta(\omega) > 0$. The absence of a square conveys the fact that only forward-propagating waves are considered.  Note also that we have selected a single mode (of fixed polarization and with no nodes in the transverse direction) within the waveguide, and we do not consider processes which couple this mode to others.
\label{fn:beta}} 
In this work, we shall use the following effective 
equation for $A$ (see e.g. Ref.~\cite{Agrawal} for its derivation): 
\begin{equation}
-i\partial_{z} A = \left(\beta\left(\omega_{0} + i\partial_{t}\right)-\beta_{0}\right) A + \gamma \left| A \right|^{2} A \,.
\label{eq:starting_eqn}
\end{equation}
Here, $\gamma > 0$ is the nonlinear coefficient, while the operator $\beta(\omega_{0}+i\partial_{t})-\beta_{0}$ gives the detuned wave number $\beta-\beta_{0}$ as a function of the detuned frequency $i \partial_t = \Delta\omega = \omega-\omega_{0}$ of linear waves. 
In Eq.~(\ref{eq:starting_eqn}) several terms 
have been neglected, such as those encoding linear and nonlinear loss, 
retardation, etc.; see 
Refs.~\cite{Abdullaev-et-al-1994,Dudley-Genty-Coen-2006,Ciret-et-al-2016-I} for works in which some of these terms are 
discussed. 

To prepare the analysis where the carrier wave is replaced by a soliton solution, it 
is appropriate to introduce 
a new coordinate system in which homogeneity (i.e. invariance under translations in $z$) will be preserved as an exact symmetry. 
To this end, we write the operator on the right-hand side 
as a Taylor series: 
\begin{equation}
\beta\left(\omega_{0} + i\partial_{t}\right) - \beta_{0} = \sum_{n=1}^{\infty} \frac{\beta_{n}}{n!} \, \left(i\partial_{t}\right)^{n} \,, 
\label{Taylor-beta}
\end{equation}
where $\beta_{1} = \beta^{\prime}(\omega_{0})$, $\beta_{2} = \beta^{\prime\prime}(\omega_{0})$, etc., and then bring 
the term in $\beta_{1}$ over to the left-hand side. 
This gives 
\begin{equation}
-i\left(\partial_{z} + \beta_{1} \partial_{t}\right) A = \left(-\frac{1}{2}\beta_{2} \partial_{t}^{2} + ...\right) A + \gamma \left| A \right|^{2} A \,.
\label{zt-eq} 
\end{equation}
Note that $\beta_{1} = dk/d\omega$ 
is the inverse group velocity of the carrier wave.  
One then introduces a new time coordinate 
$\tau \doteq 
t - \beta_{1} z$, 
so that $\partial_{z} \vert_{\tau} = \partial_{z} \vert_{t} + \beta_{1} \partial_{t}$. 
Eq.~(\ref{zt-eq}) thus reads 
\begin{equation}
-i \partial_{z} A = B\left(i\partial_{\tau}\right) A + \gamma \left| A \right|^{2} A \,, 
\label{eq:wave_eqn_full}
\end{equation}
where $B\left(\Delta\omega\right)$ is the detuned wave number 
of linear waves 
in the $(z,\tau)$ coordinate system 
as a function of the detuned frequency: 
\begin{eqnarray}
B\left(\Delta\omega\right) & \doteq & \beta\left(\omega_{0}+\Delta\omega\right) - \beta_{0} - \beta_{1} \, \Delta\omega \, . 
\label{eq:B_defn}
\end{eqnarray}
As it is more appropriate in this work, we shall 
use the 
function $B\left(\Delta\omega\right)$ itself rather than
its Taylor development.
In this respect, 
however, 
we should point out 
that the term in Eq.~(\ref{Taylor-beta}) 
proportional to $\beta_{4}$ 
(which is 
often neglected) plays 
a key role in the following analysis; see~\cite{Abdullaev-et-al-1994,Pitois-Millot-2003} 
for works where the narrow resonance induced by 
$\beta_{4}$ when sending a CW (rather than a soliton) 
is discussed. 

In the sequel, several 
conserved quantities will play important roles. 
When dealing with the nonlinear equation (\ref{eq:wave_eqn_full}),
for any 
complex solution $A$ 
the norm 
\begin{equation}
N_A = \langle A | A \rangle \doteq \int {\rm d}\tau \left| A(\tau,z)\right|^{2} \, , 
\label{A1-A2}
 \end{equation}
is $z$-independent 
and positive definite.   
In quantum settings, when properly normalized, 
this conserved quantity gives the 
total 
number of (forward-propagating) 
photons injected in the waveguide. 
This conservation law follows from 
Eq.~(\ref{eq:wave_eqn_full}) 
and prevents the inclusion of the ``hard'' processes mentioned in footnote~\ref{fn:soft-hard}. 
It should also be mentioned that there is another 
conserved quantity, 
namely 
\begin{equation}
{K}_A \doteq \int {\rm d}\tau \left\{ A^*(\tau,z) B\left(i\partial_{\tau}\right)A(\tau,z) + \frac{\gamma}{2} \left|A(\tau,z)\right|^{4}\right\}\, .
\label{K_A}
\end{equation}
This real quantity 
gives the total momentum carried by $A$  
in the 
($z,\tau$) coordinate system. 
In the present settings, it acts as a Hamiltonian, 
the right-hand side of Eq.~(\ref{eq:wave_eqn_full}) being 
equal to 
$\delta K_{A} / \delta A^{\star}(\tau,z)$ (see Eq.~(5.7) in~\cite{Pitaevskii-Stringari-BEC} for the corresponding equation in an atomic BEC).
Therefore, the coordinate $z$ parameterizes the evolution of the field configurations in the waveguide, thus acting as a temporal coordinate.

\subsection{Silicon nitride waveguides} 

In this work we consider more specifically the propagation in a silicon nitride (SiN) waveguide.  Following the continuous improvement of their fabrication processes, SiN WGs are nowadays one of the best nonlinear integrated platform avaiable thanks to a high nonlinear index, very low linear ($<$1dB/m) and nonlinear propagation losses and a broad transparency window from the visible to the mid-IR ~\cite{Halir-2012,Epping-2015, Ji-2017}. The combine material and geometric dispersion allows for dispersion engineering that enables, together with low Raman scattering, the propagation of ultra-short fundamental solitons that do not suffer from soliton self frequency shift as in silica-based optical fibres \cite{Klenner-2016}. 
In Sec.~\ref{sec:scattering-NL}, we validate the neglect of linear losses in some dedicated numerical simulations based on the 
NLSE. Nonlinear losses will not be considered in this work.

\begin{figure}
\includegraphics[width=0.45\columnwidth]{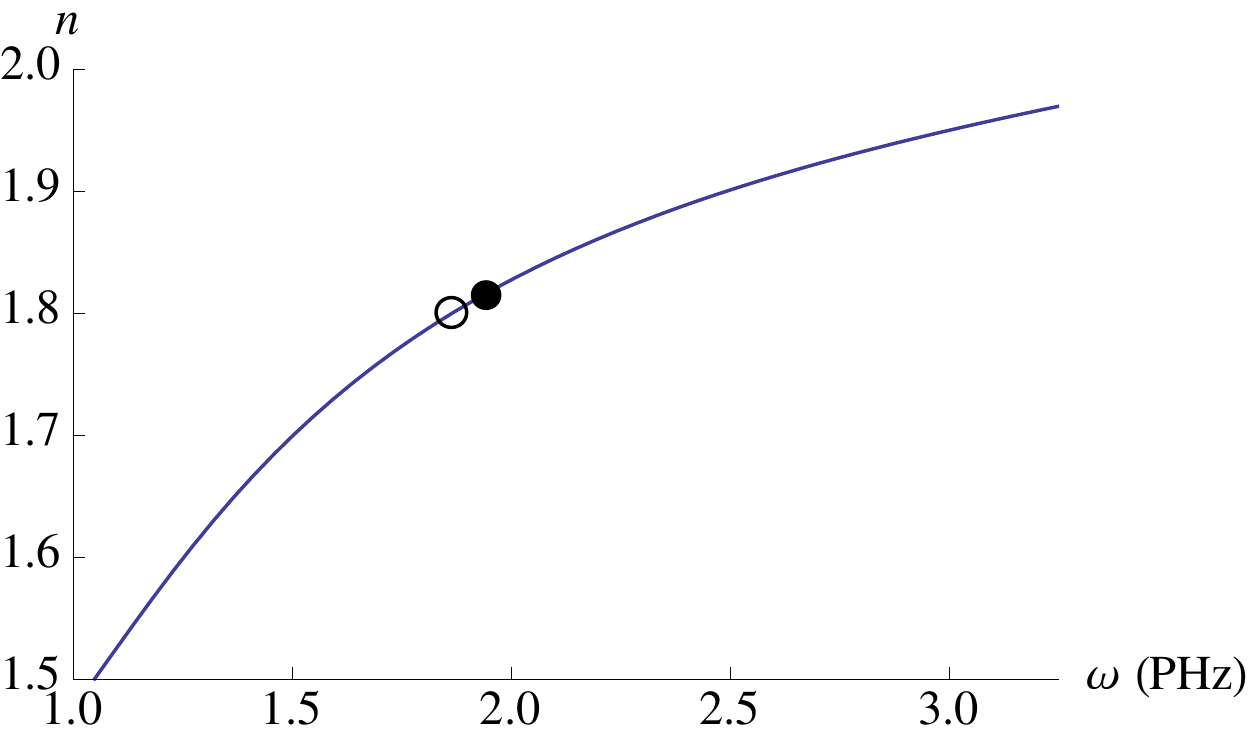} \, \includegraphics[width=0.45\columnwidth]{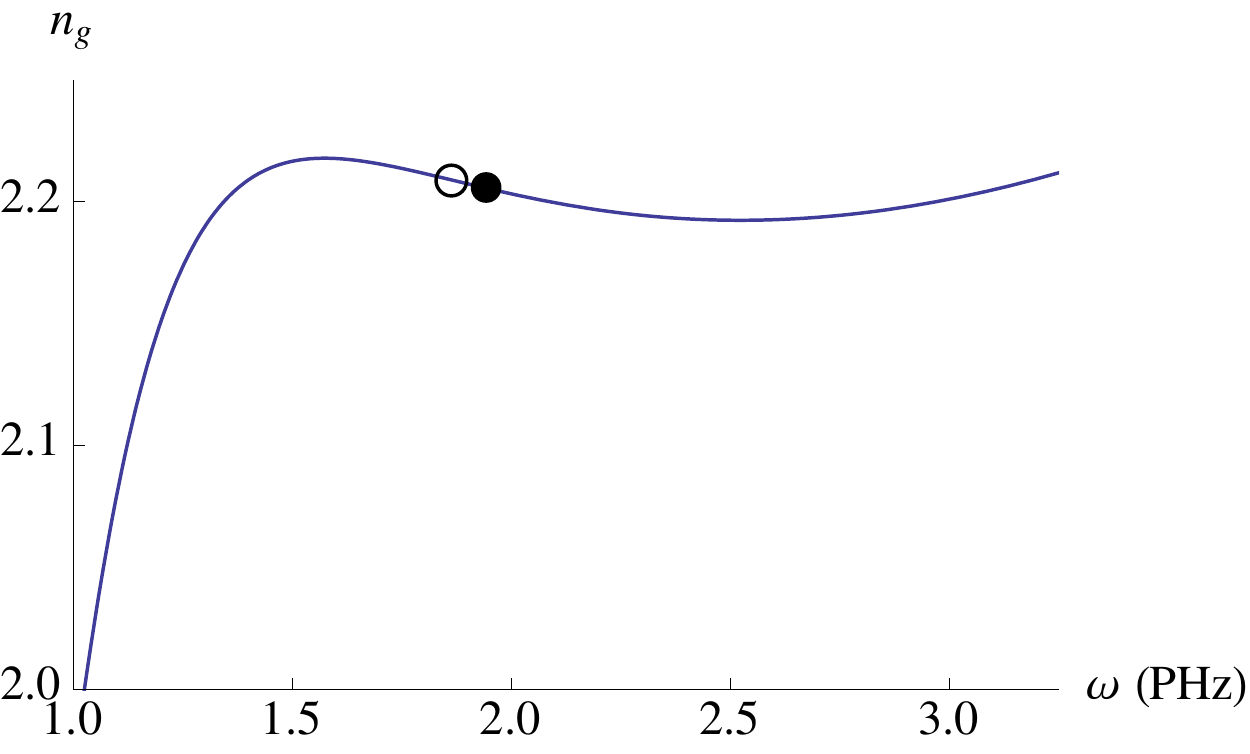}
\caption{Computed dispersion relation (of one particular mode, see footnote~\ref{fn:beta}) in a 500 nm $\times$ 1200 nm silicon nitride waveguide on a silica substrate, 
see text for more details. 
On the left is plotted $n(\omega) = c\,\beta(\omega)/\omega$, 
the index as a function of the frequency; on the right is plotted the group index $n_{g}(\omega) = c\,\beta^{\prime}(\omega)$, again as a function of frequency. 
The black dot shows the position of the pulse engendering the soliton (at $\lambda = 970\,{\rm nm}$, or $\omega_{0} = 1.94\,{\rm PHz}$), while the open circle shows where 
the third derivative 
$\beta_{3}(\omega) = 0$ (at $\omega = 1.86\,{\rm PHz}$). 
We have adopted this frequency range because, as we shall later see, 
there will only be significant pair production for $\Delta \omega = \omega-\omega_{0} \in \left[ -1 \,,\, 1 \right]$ PHz. 
\label{fig:beta}}
\end{figure}

We work with a realistic example of a 
rectangular silicon nitride waveguide (of height $0.5\,\mu{\rm m}$ and width $1.2\,\mu{\rm m}$) on a silica substrate. 
We took into account the profile of monochromatic probe waves in the perpendicular directions to numerically
derive (using the software {\it Lumerical}) the effective one-dimensional dispersion relation in the longitudinal direction. 
This 
dispersion relation 
is shown  
in Figure~\ref{fig:beta}. 
As 
$\beta(\omega)$ is approximately linear, 
we present the index $n(\omega)$ defined by $n(\omega) = c \beta(\omega) /\omega$, 
which varies more significantly.  We also present the group index $n_{g}(\omega)$, which gives the group velocity through the relation $n_{g}(\omega)= c/v_{g}(\omega)$. 
Of key importance for the sequel is that the dispersion is {\it anomalous} in the vicinity of $\omega = 2\,{\rm PHz}$, by which we mean the group velocity $v_g$ 
increases with $\omega$. 

\subsection{Solitons as 
approximate background solutions}

With the exception of plane waves (which are discussed in Appendix~\ref{app:homogeneous}), we 
are not aware 
of any stationary solution of 
Eq.~(\ref{eq:wave_eqn_full}) when the dispersion relation is given by that shown in Fig.~\ref{fig:beta}. 
However, if we restrict the Taylor series of Eq.~(\ref{Taylor-beta}) to second order, 
i.e. if we take $B(\Delta \omega)= B_{2}(\Delta\omega) \doteq \beta_2 (\Delta \omega)^2/2$, then there 
are well-known solutions which describe a one-parameter family of solitons~\cite{Agrawal}. 
That is, when considering the nonlinear equation
\begin{equation}
-i\partial_{z}A_{0} = -\frac{1}{2}\beta_{2} \, \partial_{\tau}^{2} A_{0} + \gamma \left| A_{0} \right|^{2} A_{0} \,,
\label{eq:non-lin-simplified}
\end{equation}
it is straightforward to show that this is solved by
\begin{equation}
A_{0}(z,\tau) = \sqrt{P_{0}} \, \mathrm{sech}\left(\frac{\tau}{\tau_{0}}\right) e^{i 
\delta \beta_0 
z} \,,
\label{eq:soliton}
\end{equation}
where
\begin{equation}
\delta\beta_{0} = \frac{1}{2} \gamma P_{0} = \frac{-\beta_{2}}{2\tau_{0}^{2}} > 0 \,.
\label{eq:solitonNLshift}
\end{equation}
These can be parameterized by the soliton duration $\tau_{0}$; indeed, the linearized wave equation will be seen to depend only on $\tau_{0}$ and $\delta\beta_{0}$, so that we shall have no need to separately consider $P_{0}$ or the nonlinear coefficient $\gamma$. 
Such solitons will be fair 
approximations to exact solutions (so long as their spectral width is not too large, or equivalently, their duration $\tau_{0}$ is not too small), 
and we shall treat a particular soliton as the background when considering
the propagation of fluctuations in the next section.
Hence it will play the same role 
as, say, the inhomogeneous flow over 
an obstacle in a water flume~\cite{Lawrence,Schuetzhold-Unruh-2002,Weinfurtner-2011,Unruh-2013,Coutant:2012mf,Michel:2014zsa,Euve:2015vml} when considering the scattering of surface waves.

It is here 
worth mentioning that the inexactness of the soliton as a solution 
of Eq.~(\ref{eq:wave_eqn_full}) will mostly manifest itself by a significant 
emission of Cherenkov radiation (CR)~\cite{Akhmediev-Karlsson-1995}, i.e. waves with the same detuned 
wave number as the soliton, 
in close analogy with the 
zero-frequency 
macroscopic undulation found in 
stationary 
flows over 
an obstacle~\cite{Lawrence,Coutant:2012mf,Euve:2015vml}. 
The CR will not enter the calculations of 
the scattering coefficients describing 
linear 
perturbations propagating on top of the soliton, because the soliton will be treated as an exact background solution when linearizing the wave equation.
However, it will be explicitly seen when numerically solving the NLSE (see Appendix~\ref{sec:scattering-NL}).  

As is well known, 
and as indicated by Eq.~(\ref{eq:solitonNLshift}), 
$\beta_{2}<0$ is a necessary condition for the existence of stable 
soliton solutions, i.e. the dispersion at the soliton frequency must be anomalous.  
Since $\beta_{1} = \beta^{\prime}(\omega)$ is the inverse group velocity, this means that the soliton can only exist in a region where the group velocity is increasing with frequency (as shown 
in the right panel of Figure~\ref{fig:beta}).  
Importantly, the finite size of the anomalous region implies a particular structure of the dispersion relation, which in turn necessarily leads to the narrow resonance mentioned in the introduction.

\subsection{
The linear wave equation 
\label{sub:linear_wave_eqn}}

As in 
other treatments in media, we now linearize Eq.~(\ref{eq:wave_eqn_full}) around a 
background solution so as to get the equation governing the evolution 
of probe waves. 
In this section, to separate the emission of CR from the scattering of linear perturbations,
we assume 
we have a homogeneous 
background field 
$e^{i \delta\!\beta_{0} z} A_{0}(\tau)$ that exactly solves Eq.~(\ref{eq:wave_eqn_full}). 
Perturbations on top of such a 
solution 
are conveniently described by 
the decomposition
\begin{equation}
A(z,\tau) = e^{i \delta\beta_{0} z} \left( A_{0}(\tau) + \delta A(z,\tau) \right) \,,
\label{eq:A_decomp}
\end{equation}
where the $z$-oscillation of the background governed by $\delta\beta_{0}$ has been factored out.
As a result,  
the dispersion relation of asymptotic modes propagating on top of the soliton will explicitly depend on $\delta\beta_0$ (see the next subsection).
The homogeneity 
of the background $e^{i \delta\!\beta_{0} z} A_{0}(\tau)$ 
ensures that the detuned 
wave number of linear probe waves will be conserved.
Its value, which 
we shall call $D$, 
therefore plays the role of the conserved (Killing) frequency in 
the Hawking effect, 
both in the relativistic calculation~\cite{Hawking1975,Primer}
and 
in its 
analogue/dispersive version~\cite{Brout:1995wp,Schuetzhold-Unruh-2002,
Leonhardt-Kiss-Ohberg-2003-HE,Robertson-thesis-2011}. 
 
If the magnitude of $\delta A$ is small enough, it will obey the linearized form of Eq. (\ref{eq:wave_eqn_full}):
\begin{equation}
-i\partial_{z}\left(\delta A\right) = B(i\partial_{\tau}) \, \delta A - \delta\beta_{0} \, \delta A + 2 \gamma \, \vert A_{0} \vert^{2} \, \delta A + \gamma \, A_{0}^{2} \, \delta A^{\star} \,.
\label{eq:wave_eqn_lin}
\end{equation}
Performing now the RWA implies the dropping of 
the last 
term in $\delta A^{\star}$. 
This is equivalent to assuming no ``phase matching''~\cite{Agrawal}, i.e. that the oscillating part of $A_{0}^{2} \, \delta A^{\star}$ is off-shell, lying far from the dispersion relation. 
Consideration of 
existent works on the analogue Hawking effect in waveguides~\cite{Philbin:2007ji,Finazzi:2012bn,Bermudez-Leonhardt-2016,Jacquet-Koenig-2018} indicates that 
this approximation was effectively 
applied, thereby missing a resonance 
that shall play an important role in the scattering of linear waves. 

Keeping the term in $\delta A^{\star}$, 
the solutions of the wave equation are naturally doublets of the form $w \equiv \left[ w_{+},\,w_{-}^{\star} \right]$.  
Details are given in Ref.~\cite{Leonhardt-Kiss-Ohberg-2003} in the context of 
BECs. 
(Note in particular that $w_{+}$ and $w_{-}$ play roles similar 
to those of 
$u_k$ and $v_k$ in the Bogoliubov-de Gennes formalism.)
As in~\cite{Leonhardt-Kiss-Ohberg-2003}, the doublet obeys the same equation as that obeyed by $\left[ \delta A , \, \delta A^{\star} \right]$, i.e.
\begin{equation}
-i\partial_{z} w = {O} w \,,
\label{eq:wave_eqn_2b2}
\end{equation}
where the differential operator $O$ is
\begin{equation}
{O} = \left[ \begin{array}{cc} B(i\partial_{\tau}) - \delta\beta_{0} + 
2 \gamma \vert A_{0}(\tau) 
\vert^{2} & \gamma A_{0}^{2}(\tau)  \\ -\gamma \left(A_{0}^{\star}(\tau) \right)^{2} & -B(-i\partial_{\tau}) + \delta\beta_{0} - 2\gamma \vert A_{0}(\tau)  \vert^{2} \end{array} \right] \,.
\label{eq:diff_operator}
\end{equation}
Note that, for this form of ${O}$ to make sense, $w$ should be understood as a column vector in Eq.~(\ref{eq:wave_eqn_2b2}). 
We define the conjugate $\bar{w}$ of $w$ as follows: if $w = \left[ w_{+},\,w_{-}^{\star} \right]$, then $\bar{w} \equiv \left[ w_{-},\,w_{+}^{\star} \right]$.  It is easily shown that, if $w$ is a solution of Eq. (\ref{eq:wave_eqn_2b2}), then so is $\bar{w}$ (i.e. the equation is invariant under the above conjugation).
It is also worth noting that, had we not factored out $e^{i \delta\beta_{0} z}$ from $\delta A$ in Eq.~(\ref{eq:A_decomp}), the off-diagonal elements of $O$ would be proportional to $e^{\pm 2i \delta\beta_{0} z}$.  In this way, the 
$z$-independence of $O$ as defined above justifies the decomposition in Eq.~(\ref{eq:A_decomp}).

Our aim now is to perform second quantization, so that annihilation and creation operators are associated with modes of positive and negative norm, respectively.  
To this end, we turn to the scalar product which is used to define the norm of the modes. 
Equation (\ref{eq:wave_eqn_2b2}) possesses indeed a conserved 
scalar product: given two solutions $w_{1} = \left[ w_{1,+} \,,\, w_{1,-}^{\star} \right]$ and $w_{2} = \left[ w_{2,+} \,,\, w_{2,-}^{\star} \right]$, it is given by 
\begin{equation}
\left( w_{1},\, w_{2} \right) \equiv \int_{-\infty}^{+\infty} d\tau \left( w_{1,+}^{\star}(\tau) w_{2,+}(\tau) - w_{1,-}(\tau) w_{2,-}^{\star}(\tau) \right) \,.
\label{eq:scalar_product}
\end{equation}
Using Eqs.~(\ref{eq:wave_eqn_2b2}) and~(\ref{eq:diff_operator}), one 
verifies that 
\begin{equation}
\partial_{z} \left(w_{1},\,w_{2}\right) = 0\,.
\label{eq:norm_conservation}
\end{equation}
The norm of a doublet $(w,\,w)$ is thus conserved~\footnote{It is to be noted that, when considering a more general $z$-dependent background solution, 
Eq.~(\ref{eq:norm_conservation}) is still satisfied, as is the Klein-Gordon norm in a time-dependent spacetime geometry.}, 
with the two components $w_{+}$ and $w_{-}^{\star}$ giving respectively the positive- and negative-norm content of $w$.~\footnote{If 
$w$ is of the form $\left[ \delta A \,,\, \delta A^{\star} \right]$, then it necessarily has zero norm.  This justifies considering the full space of complex solutions, so as to have well-defined positive- and negative-norm modes that are necessarily involved in the second quantization of the field. One should also note that the above norm should be distinguished from that of the nonlinear equation given in Eq.~(\ref{A1-A2}). Nevertheless, as we shall show below, these conserved norms are intimately related when considering asymptotic configurations. 
}
The non-positive character of the above scalar product characterizes excitations of bosonic fields, and is also found when considering the solutions of the Bogoliubov-de Gennes equation 
in BEC 
physics or those of the Klein-Gordon equation in relativistic Quantum Field Theory. 
The mixing of modes of opposite norm is called anomalous scattering, and corresponds to the creation of pairs in quantized settings.  When the initial state is vacuum, one then gets spontaneous pair creation, as on-shell particles (photons) are found in the output channel.

\subsection{The two dispersion relations governing ``soft'' processes
\label{sub:disp_rel}} 

Since the  operator $O$ in Eq.~(\ref{eq:wave_eqn_2b2}) in independent of $z$, 
one can work at fixed detuned wave number $D$ and 
consider the 
{\it globally defined} 
doublets $e^{i D z} w_{D,\,j}(\tau)$, 
where $j$ labels the various linearly independent solutions.
To compute the scattering coefficients encoded in these 
doublets, 
it is necessary to identify the complete set of {\it asymptotic} solutions defined far away from the soliton, i.e. for $\tau \rightarrow \pm \infty$.  

In these asymptotic regions, 
$A_{0}(\tau)$ vanishes and Eq. (\ref{eq:wave_eqn_2b2}) 
becomes independent of $\tau$.
As a result, the two components 
of any doublet $w =  (w_{+}, w_{-}^*)$
decouple, as is the case in atomic Bose condensates when interactions can be neglected after having opened the trap~\cite{Tozzo-Dalfovo-2004,Robertson:2016evj}. 
Hence the asymptotic values of the 
detuning frequency $\Delta\omega$ 
are also conserved.
Far away from the soliton,
both components of doublets
will thus be superpositions of plane waves of the form $e^{i Dz} \, e^{-i\,\Delta\omega_j(D)\,\tau}$ 
where 
the various $\Delta\omega_j(D)$ will be algebraically related 
to $D$.
Because positive- and negative-norm 
modes mix with each other when scattered, we must consider 
both sets of asymptotic  
modes, and thus two sets of roots: $\Delta\omega_{j,+}^{\rm as}(D)$ for the former and
$\Delta\omega_{j,-}^{\rm as}(D)$ for the latter.

These roots 
are immediately obtained by considering separately the 
diagonal terms of the matrix $O$ of Eq.~(\ref{eq:wave_eqn_2b2}) with $A_0 = 0$. 
The upper (positive-norm) components 
$w_{D,\,j,\,+}(\tau)$ are characterized by 
detuned frequencies 
$\Delta\omega_{j,+}^{\rm as}(D)$ which satisfy
\begin{equation}
D^{\rm as}_+(\Delta \omega) \doteq B(\Delta\omega) - \delta\beta_{0} \, .
\label{eq:D_defn}
\end{equation}
Instead, 
lower (negative-norm) components 
$w^{\star}_{-D,\,j,\,-}(\tau) \propto {\rm exp}(-i\,\Delta\omega_{j,-}^{\rm as}(D)\,\tau)$ have their detuned frequencies 
$\Delta\omega_{j,-}^{\rm as}(D)$ 
obeying 
\begin{equation}
D_-^{\rm as}(\Delta \omega) \doteq -B(-\Delta\omega)+\delta\beta_{0} \,.
\label{eq:D_negnorm}
\end{equation}
One easily verifies 
that $D_{-}^{\rm as}(\Delta\omega) = -D_{+}^{\rm as}(-\Delta\omega)$, which expresses the fact that negative-norm modes are conjugates of positive-norm modes. Hence the two dispersion relations are related by a $\pi$ rotation around $\left(\Delta\omega=0\,,\,D=0\right)$, or equivalently the point $(\omega=\omega_{0}\,,\,k=\beta_{0}+\delta\beta_{0})$ describing the soliton. 
Note that this is {\it not} the same as 
in Refs.~\cite{Philbin:2007ji,Robertson-thesis-2011,Finazzi:2012bn,Bermudez-Leonhardt-2016,Jacquet-Koenig-2018,Drori:2018ivu}, 
where positive- and negative-norm modes are related by a $\pi$ rotation around the ``absolute'' origin $(\omega=0\,,\,k=0)$. 
This key difference distinguishes what we referred to in footnote~\ref{fn:soft-hard} as ``soft'' and ``hard'' photon production processes: ``soft'' processes (as studied here) involve the creation of collective excitations on top of the soliton (as for phonons in atomic BEC), while ``hard'' processes involve the spontaneous creation of additional photons.  While both the total momentum $K_{A}$ of Eq.~(\ref{K_A}) 
and 
the total number of photons $N_{A}$ %
are conserved by the ``soft'' processes we consider, it remains to be seen what are the conserved quantities when including ``hard'' processes. 
However, we expect that any $z$-dependence of $K_{A}$ and $N_{A}$ due to the latter will be very small because of their 
strongly non-adiabatic character, which %
greatly suppresses the corresponding scattering amplitudes. (The interested reader is invited to consult Chapters~9 and~10 of Ref.~\cite{Robertson-thesis-2011} for typical values of the pair production rates due to ``hard'' processes,
which seem to be in agreement with the recent observations reported in~\cite{Drori:2018ivu}.) %

\begin{figure}
\includegraphics[width=0.45\columnwidth]{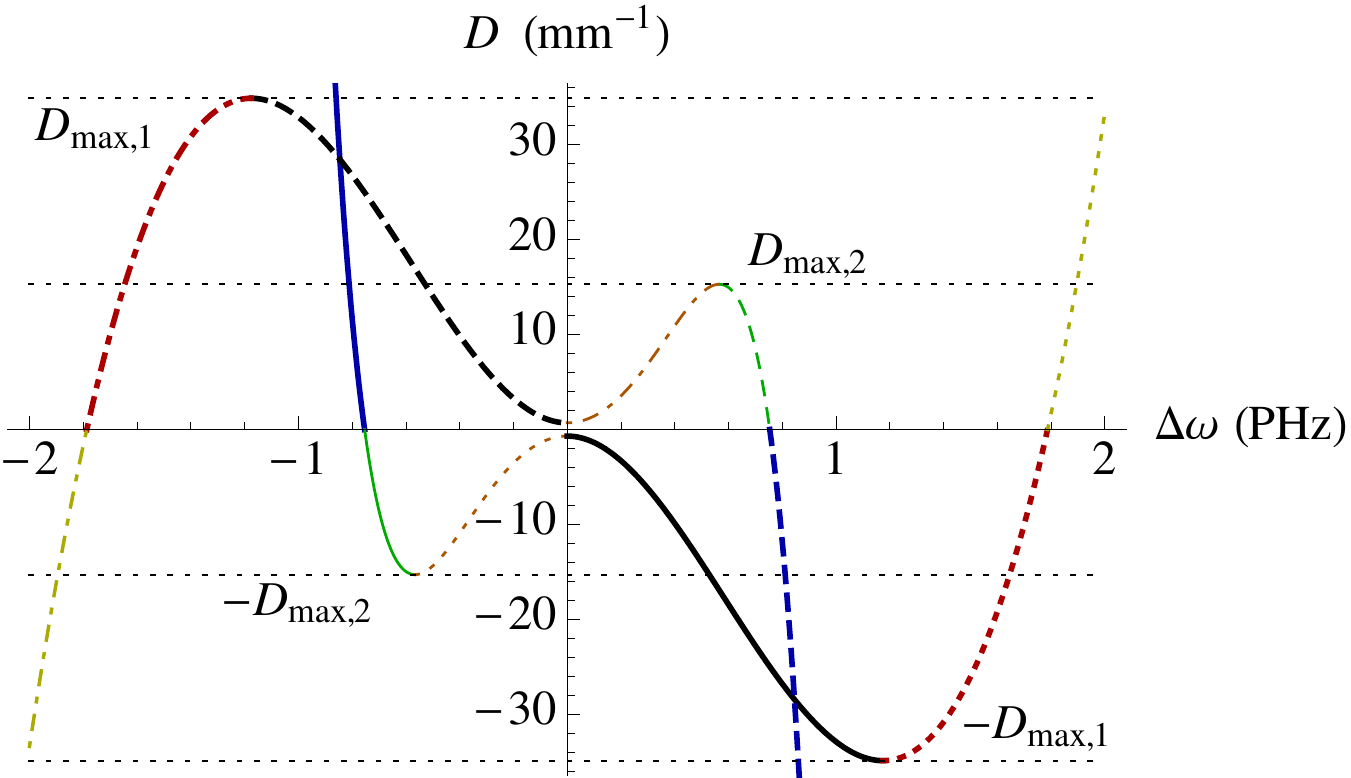} 
\caption{Dispersion relation in the silicon nitride waveguide, 
using the detuned 
variables $D$ and $\Delta\omega$ 
giving, respectively, the difference in wave number and frequency of the modes with respect to those of the soliton (equal to $\beta_{0} + \delta\beta_{0}$ and $\omega_{0}$), 
whose carrier (at 1.94 PHz) 
is indicated by the black dot in Fig.~\ref{fig:beta}. 
The solid and dotted curves correspond to the positive-norm modes of Eq.~(\ref{eq:D_defn}), while the dashed and dot-dashed curves show the negative-norm modes of Eq.~(\ref{eq:D_negnorm}). 
The line styles and the various colors 
are used to ease the distinguishing of 
the different branches, 
with solid and dashed (dotted and dot-dashed) curves showing left-moving (right-moving) modes.
Thicker line styles have been used to highlight the three modes (and their conjugates) most relevant in the present study. %
The extremal values of the curves are indicated by horizontal dotted 
lines. 
Here their values are 
$ D_{\rm max, 1} = 34.9 \, {\rm mm}^{-1}$ and $D_{\rm max, 2} = 15.3 \, {\rm mm}^{-1}$, 
while $\delta\beta_{0} = 0.7\,{\rm mm}^{-1}$ (which corresponds to a soliton duration $\tau_{0} = 10\,{\rm fs}$). 
Of key importance is the fact 
that the negative-norm branch (shown for $D<0$ by the thick 
dashed 
blue line) crosses the positive-norm solid (thick black) line for $D_{\rm cross} = -28.4 \, \rm{mm}^{-1}$, as this {\it phase matching}
is responsible for the enhancement of the anomalous scattering coefficient. 
\label{fig:dispersion_SiN}}
\end{figure}

Using again the example waveguide whose dispersion relation is shown in Fig.~\ref{fig:beta}, 
and choosing for the soliton the carrier frequency indicated by the black dot in that figure with 
$\delta\beta_{0} = 0.7\,{\rm mm}^{-1}$, 
$D_{+}^{\rm as}(\Delta\omega)$ ($D_{-}^{\rm as}(\Delta\omega)$) is shown in Fig.~\ref{fig:dispersion_SiN} by continuous and dotted (dashed and dot-dashed) curves. 
Importantly, the 
structure of 
$D_{+}^{\rm as}(\Delta\omega)$ possesses generic features. As we shall explain,
these stem from the fact the soliton lives in a finite 
frequency window where the waveguide exhibits anomalous dispersion.
To start the analysis, 
we 
first notice that at $\Delta\omega = 0$, the positive-norm branch starts at $D_+(0)=-\delta\beta_{0}$, and because the soliton lives in a region of anomalous dispersion, this branch necessarily dips down to more negative values for small $\left| \Delta \omega \right|$.
On leaving the window of anomalous dispersion, the curvature of the positive-norm branch then flips sign, 
rising to positive values of $D$ on {\it both} 
sides.  As a result, it has a negative minimum for $\Delta \omega < 0$ and another for $\Delta \omega > 0$. 
Crucially, 
since the positive-norm branch is below the negative-norm branch at $\Delta\omega = 0$ and above it for large $\Delta\omega$, it is necessarily the case that these two branches cross. 
In Fig.~\ref{fig:dispersion_SiN}, the crossing occurs at $D_{\rm cross} 
 = -28.4\,{\rm mm}^{-1}$ and $\Delta \omega_{\rm cross} 
= 0.846\,{\rm PHz}$. (Of course there is another crossing for opposite values of $D$ and $\Delta \omega$, but this describes the same phase 
matching condition.) 
This crossing 
induces a modulation instability when sending a CW in the WG, as shown in App.~\ref{app:homogeneous}, and 
an enhancement of the anomalous scattering coefficient when sending a soliton, as we shall see below.

\subsection{The various wave number bands}

By examining 
Fig.~\ref{fig:dispersion_SiN}, 
one also 
sees that for any 
$\Delta\omega$ 
there are 
two values for $D$, which correspond to modes of opposite norm. 
Instead, at fixed $D$ 
the number of roots $\Delta\omega_j$ 
varies between two and six, depending on the value of $D$. 
The full range of $D$ is thus split into a series of bands, in each of which the number of roots 
is fixed. 

The domain of interest of these branches can be restricted to $D < 0$, as the half plane with $D > 0$ contains exactly the same modes (but with the signs of their norms flipped). 
The presence 
of the two minima and the particular behavior near $D=0$ due to the wave number shift $\delta \beta_{0}$ 
mean that 
there are four domains
in total: 
\begin{itemize}
\item For $D < -D_{\rm max, 1}$, only the 
negative-norm branches exist. They are shown in blue dashed %
(for $\Delta\omega > 0$) and yellow dot-dashed %
(for $\Delta\omega < 0$) 
in Fig.~\ref{fig:dispersion_SiN}, 
and the scattering matrix is 
an element of the unitary group $U(2)$.  Since the two 
branches are 
well separated in frequency, there will not 
be any significant mixing between them. 
\item For $-D_{\rm max, 1} < D < -D_{\rm max, 2}$, there are four solutions of the dispersion relation: the two negative-norm 
plus two positive-norm branches shown in solid black and dotted red. %
The scattering matrix is now an element of the pseudo-unitary group $U(2,2)$.
The merging of the two positive-norm 
branches for $D$ near $-D_{\rm max, 1}$ means we can expect significant mixing between these modes.  We can also expect significant mode mixing between the dashed blue and solid black %
branches near the value of $D_{\rm cross}$. 
Crucially, 
since these modes have opposite norm, 
we have here an enhanced 
anomalous scattering.
\item For $-D_{\rm max, 2} < D < -\delta\beta_{0}$, all six of the visible branches of the dispersion relation are present. 
The scattering matrix is an element of $U(4,2)$ since the two new modes have 
positive norm. 
Note that they 
merge when $D \to -D_{\rm max, 2}$. 
We can thus expect significant mixing between 
them 
for $D$ slightly above 
$-D_{\rm max, 2}$.
We can 
also expect some significant mixing for $D \to -\delta\beta_{0}$ where the positive-norm branches
shown in light dotted bronze and thick solid black merge.  %
\item Finally, for $-\delta\beta_{0} < D < 0$, 
there are four solutions of the dispersion relation: 
the negative-norm dot-dashed yellow and dashed blue branches, and the positive-norm solid green and dotted red branches. 
The scattering matrix in this range will thus belong to the group $U(2,2)$.  This range is not particularly interesting for the scattering of linearized perturbations, 
because it is 
narrow in $D$ and because the scattering coefficients are 
small since the various roots 
are quite well separated in frequency. 
It is, however, the range in which Cherenkov radiation is produced; see Sec.~\ref{sec:scattering-NL}. 
\end{itemize}

\subsection{Scattering coefficients} 

We now turn to 
the scattering coefficients on the soliton. 
These describe the asymptotic mode content of 
the global 
solutions of Eq.~(\ref{eq:wave_eqn_2b2}), taking 
as background solution the soliton of Eq.~(\ref{eq:soliton}). 
As already explained, the soliton is not an exact background solution, but it is very close to one. The exact solution will not be exactly stationary, in particular because of the emission of Cherenkov radiation. However, the deviations are very small. Taking the soliton as (approximate) background allows us to use the formalism of Section~\ref{sub:linear_wave_eqn}, and obtain the scattering coefficients. 

As 
mentioned below Eq.~(\ref{eq:diff_operator}), 
if a doublet $e^{i D z} w_{D,\,j}(\tau)$ is a globally defined 
solution of Eq.~(\ref{eq:wave_eqn_2b2}), 
then its 
conjugate doublet $e^{-i D z} \bar{w}_{D,\,j}(\tau)$ is also a solution, 
although its norm has changed sign. 
Negative (positive) norm doublets will thus always be written with (without) a bar.
This notation is in agreement with that used to describe a real scalar field in 
relativistic second quantized settings, namely negative, i.e. complex conjugated, (positive) norm modes
are associated with creation (destruction) operators, and not treated as independent solutions~\cite{Wald-book}.

The {\it in-} and {\it out-} doublets 
are respectively defined as having a single incoming mode (i.e. with a group velocity directed towards the soliton) 
in the asymptotic ``past'' (i.e $z \to -\infty$) 
and a single outgoing mode in the 
``future''. 
Because $A_0$ vanishes on both sides, the same set of roots characterizes both in- and out-modes. 
Therefore, when working at fixed $D$, the asymptotic postive- and negative-norm doublets read~\footnote{It should be noticed that the modes $w_{D,\,j}$ do not contain the standard relativistic normalization factor of $1/\sqrt{\omega}$.  The origin of this is to be found in the quasi-monochromatic approximation for the slowly-varying envelope $A$ used 
to derive Eq.~(\ref{eq:starting_eqn}); 
see Ref.~\cite{Brainis-2009} for a detailed discussion of this point. 
}
\begin{eqnarray}
w^{\rm in/out}_{D,\,j}(\tau) & \rightarrow & \frac{1}{\sqrt{2\pi}} \left| \frac{dD_{+}^{\rm as}}{d\Delta\omega}\left(\Delta\omega^{\rm as}_{j,+}(D)\right) \right|^{-1/2} e^{-i\,\Delta\omega^{\rm as}_{j,+}(D)\,\tau} \left[ \begin{array}{c} 1 \\ 0 \end{array} \right] \,, \nonumber \\
\bar{w}^{\rm in/out}_{D,\,j}(\tau) & \rightarrow & \frac{1}{\sqrt{2\pi}} \left| \frac{dD_{-}^{\rm as}}{d\Delta\omega}\left(\Delta\omega^{\rm as}_{j,-}(D)\right) \right|^{-1/2} e^{-i\,\Delta\omega^{\rm as}_{j,-}(D)\,\tau} \left[ \begin{array}{c} 0 \\ 1 \end{array} \right] \,,
\label{eq:unit-norm-mode}
\end{eqnarray}
so that the set of in- or out-modes forms a complete and orthonormal basis with respect to the scalar product~(\ref{eq:scalar_product}), i.e. 
\begin{eqnarray}
\left( w^{\rm in/out}_{D,\,i} \,,\, w^{\rm in/out}_{D^{\prime},\,j} \right) & = & \delta_{i j} \, \delta\left(D-D^{\prime}\right) \,, \nonumber \\
\left( \bar{w}^{\rm in/out}_{D,\,i} \,,\, \bar{w}^{\rm in/out}_{D^{\prime},\,j} \right) & = & - \delta_{i j} \, \delta\left(D-D^{\prime}\right) \,, \nonumber \\
\left( w^{\rm in/out}_{D,\,i} \,,\, \bar{w}^{\rm in/out}_{D^{\prime},\,j} \right) & = & 0 \,.
\label{eq:normalization}
\end{eqnarray}
The normalization imposed here explains the presence of the square root in Eq.~(\ref{eq:unit-norm-mode}) (as also in other media, see~\cite{Macher-Parentani-2009,Robertson-thesis-2011}). 
These bases are complete, so that the 
quantum field can be written as
\begin{eqnarray}
\hat{w}(z,\tau) & = & \int_{-\infty}^{+\infty} \mathrm{d}D \left( \sum_{j \in J_{+}(D)} \hat{a}_{D,j}^{\rm in} \, w_{D,j}^{\rm in}(\tau) + \sum_{j \in J_{-}(D)} \left(\hat{a}_{-D,j}^{\rm in}\right)^{\dagger} \bar{w}_{-D,j}^{\rm in}(\tau) \right) \, e^{i D z} \nonumber \\
& = & \int_{-\infty}^{+\infty} \mathrm{d}D \left( \sum_{j \in J_{+}(D)} \hat{a}_{D,j}^{\rm out} \, w_{D,j}^{\rm out}(\tau) + \sum_{j \in J_{-}(D)} \left(\hat{a}_{-D,j}^{\rm out}\right)^{\dagger} \bar{w}_{-D,j}^{\rm out}(\tau) \right) \, e^{i D z} \,,
\end{eqnarray}
where $J_{+}(D)$ and $J_{-}(D)$ respectively represent the sets of available positive- and negative-norm solutions at the given value of $D$.
The requirement that $w$ belong to the conjugation-invariant subspace is automatically satisfied since the creation operator $\hat{a}^{\dagger}$ is the hermitian conjugate of the annihilation operator $\hat{a}$ (as is the case for phononic field operators in atomic BEC). 
Namely, we adopt the standard bosonic commutation relation $\left[ \hat{a}_{D,j} \,,\, \hat{a}_{D^{\prime},j^{\prime}}^{\dagger} \right] = \delta_{j, j^{\prime}} \, \delta\left(D-D^{\prime}\right)$ for both in- and out-modes.

Each in-mode, after scattering, becomes a linear superposition of out-modes.  We thus have the transformation:
\begin{eqnarray}
w^{\rm in}_{D,\,i} &=& \sum_{j \in J_{+}(D)} \alpha_{D,\, i j} \, w^{\rm out}_{D,\,j} + \sum_{j \in J_{-}(D)} \beta^{\star}_{D, \, i j} \, \bar{w}^{\rm out}_{-D,\,j} \,, \nonumber \\
\bar{w}^{\rm in}_{-D,\,i} &=& \sum_{j \in J_{+}(D)} \beta_{-D, \, i j} \, w^{\rm out}_{D,\,j} + \sum_{j \in J_{-}(D)} \alpha_{-D,\, i j}^{\star} \, \bar{w}^{\rm out}_{-D,\,j} \,,
\label{eq:in-out_modes}
\end{eqnarray}
for positive- and negative-norm incident modes, respectively.
The coefficients $\alpha_{D, \, i j}$, $\alpha_{-D,\, i j}^{\star}$, $\beta^{\star}_{D,\, i j}$,  and $\beta_{-D, \, i j} $  collectively form the scattering matrix restricted to a particular value of $D$.
The normalization (\ref{eq:normalization}) implies the unitarity relation
\begin{equation}
\sum_{j \in J_{+}(D)} \left|\alpha_{D,\, i j}\right|^{2} - \sum_{j \in J_{-}(D)} \left|\beta_{D,\, i j}\right|^{2} = 1 
\label{eq:unitarity}
\end{equation}
between the scattering coefficients of positive-norm doublets; an exactly analogous 
relation applies to $\alpha_{-D,\, ij}^{\star}$ and $\beta_{-D,\, i j}$ governing the scattering 
of negative-norm doublets.
The $\alpha$ coefficients, which multiply out-modes of the same norm as the incident mode, are 
the standard amplitudes describing elastic scattering. 
On the other hand, the $\beta$ coefficients multiply out-modes of opposite norm to that of the incident mode and 
count negatively towards the unitarity relation.  

Physically, $\left|\beta\right|^{2}$ gives the mean number (per unit wave number per unit length) of spontaneously created pairs 
of photons with {\it opposite} values of $D$.
Returning to the lab frame, and considering the enhanced mode mixing between the solid black and dashed blue %
branches (with $D<0$ near $D_{\rm cross} = -28.4\,{\rm mm}^{-1}$), this means that the two wave numbers will be given by (see Eqs.~(\ref{eq:B_defn}) and~(\ref{eq:D_defn}))
\begin{eqnarray}
k_{\rm black}(D) & = & \beta_{0} + \delta\beta_{0} + \beta_{1} \Delta\omega^{\rm as}_{+,\rm black}(D) + D \,, \nonumber \\
k_{\rm blue}(-D) & = & \beta_{0} + \delta\beta_{0} + \beta_{1} \Delta\omega^{\rm as}_{+,\rm blue}(-D) - D \,,
\label{bb}
\end{eqnarray}
where for both solutions we have used the positive-norm branch solution $\Delta\omega^{\rm as}_{+}(D)$.  
When considering $D=D_{\rm cross}$, we find the simple relation $k_{\rm black} + k_{\rm blue} = 2 \left( \beta_{0} + \delta\beta_{0} \right)$, which is in agreement with the phase matching condition of Refs.~\cite{Wadsworth-2004,Rarity-2005} and which tells us 
that each pair of 
quanta 
is extracted from a pair of ``condensed'' photons of the soliton. 
As far as we know, 
the kinematical relationships of Eqs.~(\ref{bb}) 
have not been 
found in previous 
works on 
the analogue Hawking effect in nonlinear optics, since previous works focus on ``hard'' processes (following the distinction presented in footnote~\ref{fn:soft-hard}).


\section{Elastic and 
anomalous scattering coefficients 
\label{sec:scattering}}

In this section, 
we 
numerically calculate the scattering coefficients given the example dispersion relation and soliton carrier wave of Figs.~\ref{fig:beta} and~\ref{fig:dispersion_SiN}.  We 
focus on two particular 
processes which emphasize the roles played by the $\alpha$ and $\beta$ 
types of 
coefficients defined by Eq.~(\ref{eq:in-out_modes}). 
The plots have been obtained by numerically solving Eq.~(\ref{eq:wave_eqn_2b2}) at fixed $D$. 
The techniques used are the same as those described in~\cite{Robertson-Leonhardt-2014}, generalized to the case of a doublet.  
As in that paper, the wave equation is re-expressed as an integral equation in Fourier space, so that the dispersion term $B(i\partial_{\tau})$ becomes a multiplicative term and the Fourier transform of the squared soliton profile $\left|A_{0}(\tau)\right|^{2}$ becomes the kernel of an integral operator.  Upon discretization, this integral equation becomes a matrix equation, which can be solved numerically after suitable regularization is performed. 
The validity of the results has been verified by checking their agreement with the unitarity relation~(\ref{eq:unitarity}). 
The interested reader is invited to consult~\cite{Robertson-Leonhardt-2014} 
for a detailed description of the techniques employed.

\subsection{Elastic scattering: From total transmission to 
total reflection
\label{sub:Elastic_scattering}} 

First, let us consider what happens for $D$ near $-D_{\rm max, 1}$, where the solid black and dotted red branches merge (see Fig.~\ref{fig:dispersion_SiN}).
When the probe interacts with the soliton, there is an effective deformation of the ``local'' dispersion relation due to the presence of the pulse.
In effect, the dispersion relation is tilted such that the extremum at $-D_{\rm max, 1}$ is reduced in magnitude (see the left panel of Fig.~\ref{fig:solPeak_D} in Appendix~\ref{app:homogeneous}). 
At the peak of the soliton, for $A_0^{\rm max} = \sqrt{P_0}$, this tilting reaches its maximum extent, and we refer to the deformed extremum at the soliton peak as $-D_{\rm min, 1}$.~\footnote{The value of this extremum is 
found by computing the 
eigenvalue of the matrix ${O}$ of Eq.~(\ref{eq:diff_operator}) when using the 
maximal value of $A_{0}$.  A similar concept is found when considering non-monotonic subcritical flows over an obstacle in a flume.  In that case, 
the flow velocity reaches 
a maximal value which determines an extremum 
$\omega_{\rm min}$ of the ``local'' dispersion relation on top of the obstacle~\cite{Michel:2014zsa,Robertson:2016ocv}. 
Frequencies below $\omega_{\rm min}$ are essentially transmitted across the obstacle, while those above $\omega_{\rm min}$ are essentially reflected; 
see~\cite{Euve:2014aga,Euve:2015vml} for experimental verifications in these settings.
The similarity between the present settings and subcritical flows will also be found when considering the behaviour of the norm of some scattering coefficients, see footnote~\ref{fn:sim2}.} 
Then, for $-D_{\rm max, 1} < D < -D_{\rm min, 1}$, there exist no real solutions of the ``local'' dispersion relation at the center of the pulse
which are continuously connected to the probe frequency, 
and so the probe is (partially) blocked.  It reflects from the pulse at a different frequency, whose asymptotic value is 
determined by 
the conservation of $D$: 
the solid black branch is shifted onto the dotted red. %
This is exactly the situation that was studied in Refs.~\cite{Philbin:2007ji,Ciret-et-al-2016-I},
where the frequency-shifting from the probe to the ``idler'' was observed,
thereby revealing the presence of a ``group velocity horizon'' (see footnote~\ref{fn:sim2} for more details, and also Ref.~\cite{Choudhary-Koenig-2012} for a discussion of the elastic scattering coefficients).

This situation is illustrated in the left panel 
of Fig.~\ref{fig:Scattering}.  
The unit positive-norm incident mode (see Eq.~(\ref{eq:unit-norm-mode})) lives on the dotted red branch, %
is characterized by $D$, and acts as a probe wave.  When the magnitude of $D$ is far below $D_{\rm min, 1}$ (indicated by the dashed vertical line in Fig.~\ref{fig:Scattering}), it is essentially transmitted across the pulse.  The transmission coefficient is equal to $1$, with all other scattering coefficients being negligible.  However, there is a clear reversal around $D_{\rm min, 1}$, with the transmission coefficient dropping to zero for $D$ significantly larger than $D_{\rm min, 1}$, and the scattering coefficient describing red-to-black mixing (which is a reflection coefficient) climbing to $1$.  In this regime, the incident wave is completely reflected onto 
the solid black branch.  Note that 
the scattering is essentially 
a two-mode mixing process between modes of positive norm. 
The 
unitarity relation takes thus the form
\begin{equation}
\left|R_{D}\right|^{2} + \left|T_{D}\right|^{2} \approx 1 
\end{equation}
for all values of $D$ in the above-discussed intervals.   The numerical analysis shows that the deviation 
from $1$ 
is 
bounded by $5\times 10^{-4}$.  This 
is mostly due to there being a small amount of anomalous scattering not visible on the linear scale used here, 
but 
is 
clearly visible in 
the upper left panel of Fig.~\ref{fig:MoreScattering} below, where the same results are shown on a logarithmic scale.

\subsection{Enhanced 
anomalous scattering due to phase matching} 

\begin{figure}
\includegraphics[width=0.45\columnwidth]{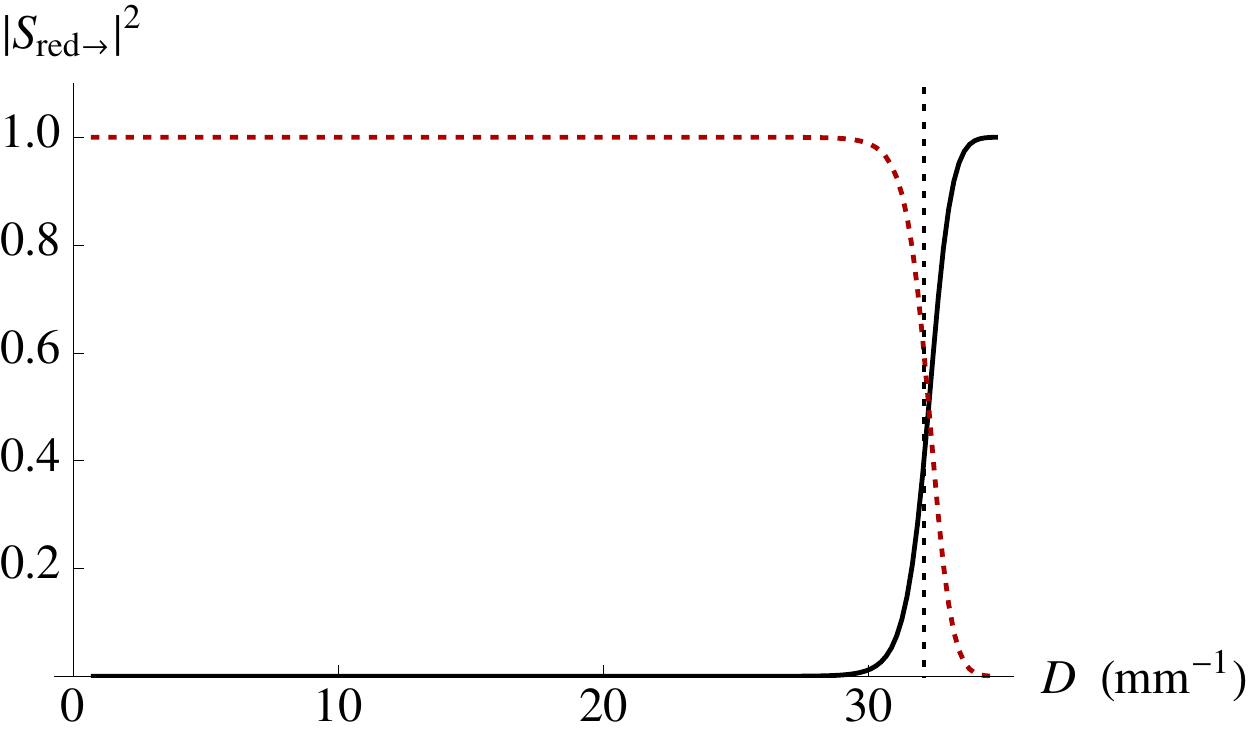} \, \includegraphics[width=0.45\columnwidth]{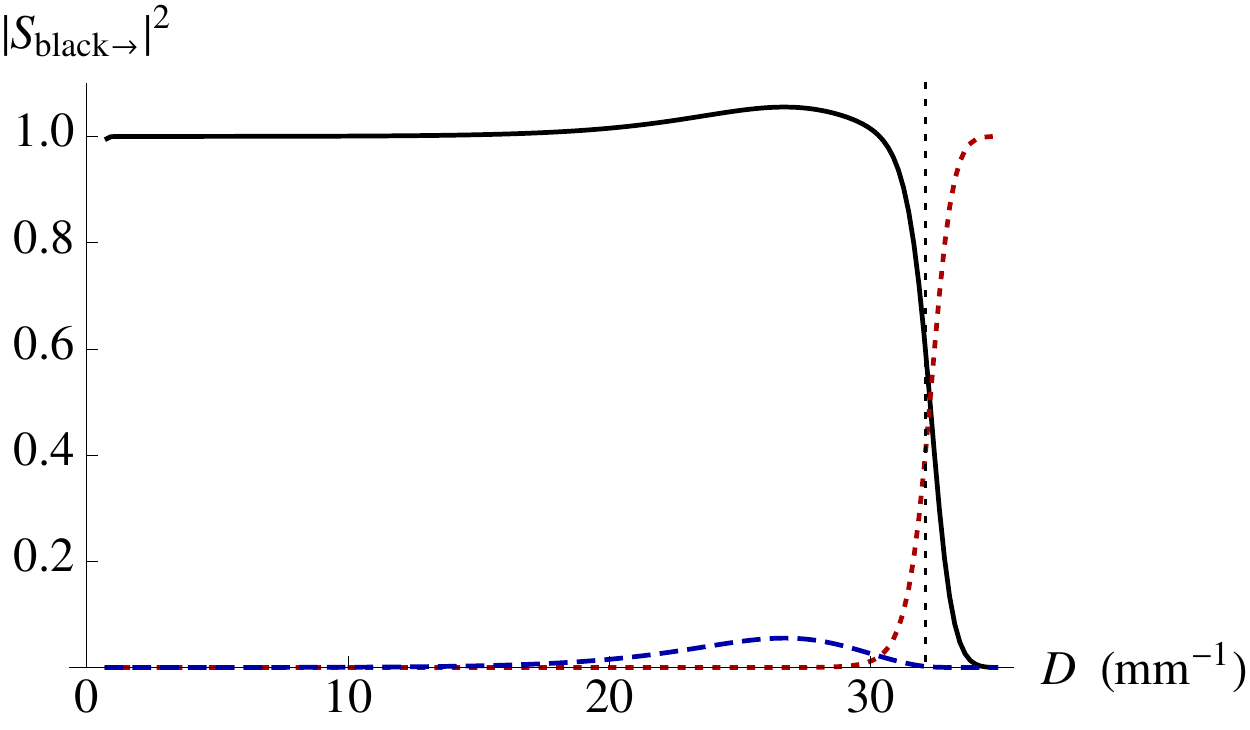}
\caption{Squared magnitudes of the scattering coefficients on a soliton of duration $\tau_{0} = 10\,{\rm fs}$, 
with carrier frequency indicated by the black dot in the waveguide dispersion relation of Fig.~\ref{fig:beta}. 
The vertical dotted line indicates the value of $D_{{\rm min},1}$ described in Sec.~\ref{sub:Elastic_scattering} (see also the left panel of Fig.~\ref{fig:solPeak_D} in Appendix~\ref{app:homogeneous}). 
On the left plot, the incident mode lives on the dotted red branch with positive group velocity.  In that case, one faces an elastic linear mode mixing involving the probe and the idler (here described by a mode living on the solid black branch).  We notice a sharp transition from pure transmission to total reflection occurring very near $D_{{\rm min},1}$.  On the right plot, the incoming mode lives on the solid black branch, whose group velocity is negative and which is crossed by the negative-norm dashed blue branch 
in Fig.~\ref{fig:dispersion_SiN}.  Near $D_{{\rm min},1}$ we recover the rapid transition from transmission to total reflection, as in the left plot.  We also see an enhancement of the transmission coefficient (i.e. superradiance) which is caused by the anomalous mode mixing involving the blue modes, whose $\left|\beta_{D}\right|^{2}$ is represented in dashed blue for $D \in \left[ 18 \,, \, 32 \right] \, {\rm mm}^{-1}$.
\label{fig:Scattering}}
\end{figure}

Let us now turn to the right plot of Fig.~\ref{fig:Scattering}, which shows the scattering coefficients when the incident wave is a unit norm mode living on the solid black branch of the dispersion relation in Fig.~\ref{fig:dispersion_SiN}.

Firstly, we note that in the region around $D_{\rm min, 1}$, the behavior of the scattering coefficients is essentially the same as described above for an incident mode on the dotted red branch, except that the solid black and dotted red curves have been switched. %
This is exactly as would be expected if the black and red modes are essentially decoupled from all other modes.

The situation is different, however, when $D$ is in the vicinity of $|D_{\rm cross}| \simeq 28 {\rm mm}^{-1}$
where the solid black and dashed blue curves cross, see Fig.~\ref{fig:dispersion_SiN}.
There we find 
a significant scattering between these modes. 
Since they 
have opposite norm, there is 
an {\it increase} of the transmission coefficient {\it above} $1$, i.e. it is associated with amplification of the incident wave.
This can be understood from the corresponding unitarity relation, which, since the coupling to other modes is negligible for $D$ near $D_{\rm cross}$, 
takes the form
\begin{equation}
\left|\alpha_{D}\right|^{2} - \left|\beta_{D}\right|^{2} \approx 1 \,.
\label{eq:bogoliubov}
\end{equation}
Here, $\alpha_D$ is the amplitude of the transmitted wave on the solid black branch, while $\beta_D$ is the amplitude of the outgoing wave on the dashed blue branch. 
It is clear that a non-zero $\beta_D$ requires $\alpha_D$ to have a magnitude larger than 1.

This anomalous two-mode mixing is responsible for the analogue Hawking effect. 
In the absence of enhancement, it reaches a maximum value near $D_{\rm min,1}$,
see e.g. the dashed blue curve in the upper left panel of Fig.~\ref{fig:MoreScattering}. 
This behavior was also found numerically in the wave blocking process of counter-propagating surface waves on a subcritical flow~\cite{Michel:2014zsa,Robertson:2016ocv}. 
The novelty here is that there exists a crossing point at $D_{\rm cross}$ where the relevant opposite-norm modes have exactly zero separation in Fourier space.
This allows a great enhancement of the corresponding $\beta_D$-coefficient, 
to the extent that it becomes visible even on a linear (rather than logarithmic) scale.
This is the main result of the present work. 

\subsection{Behavior of scattering coefficients on a log scale}

\begin{figure}
\includegraphics[width=0.45\columnwidth]{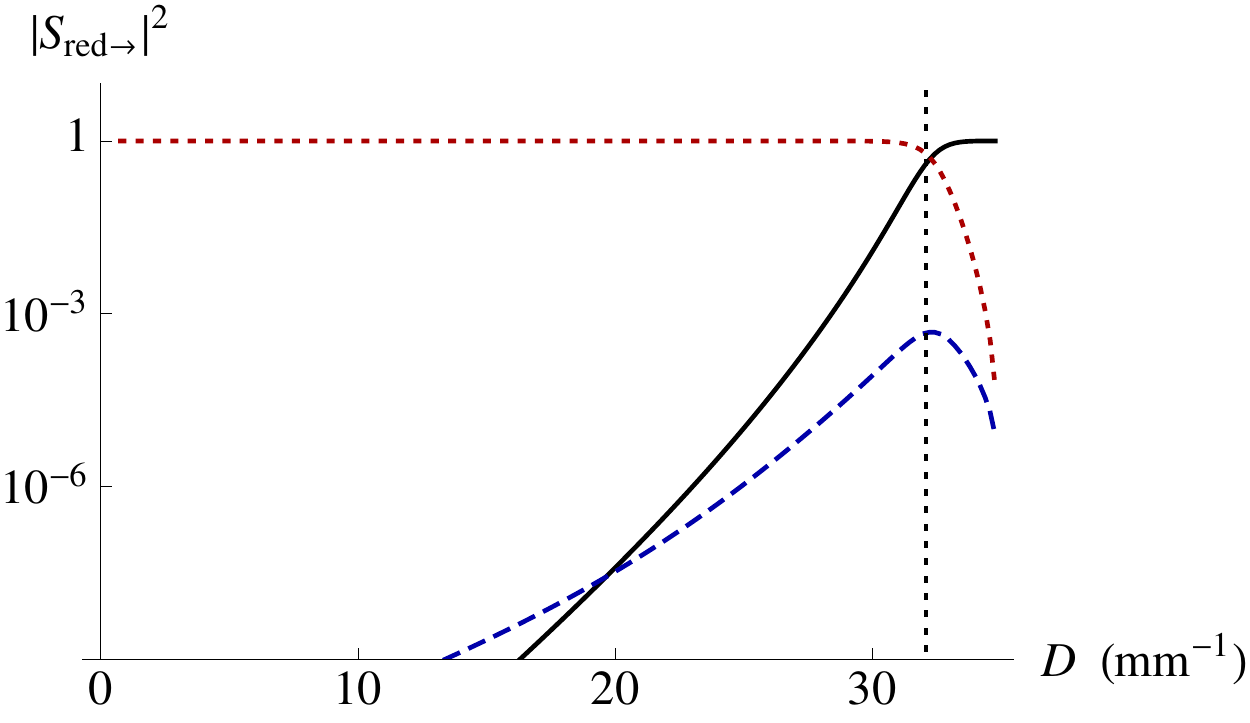} \, \includegraphics[width=0.45\columnwidth]{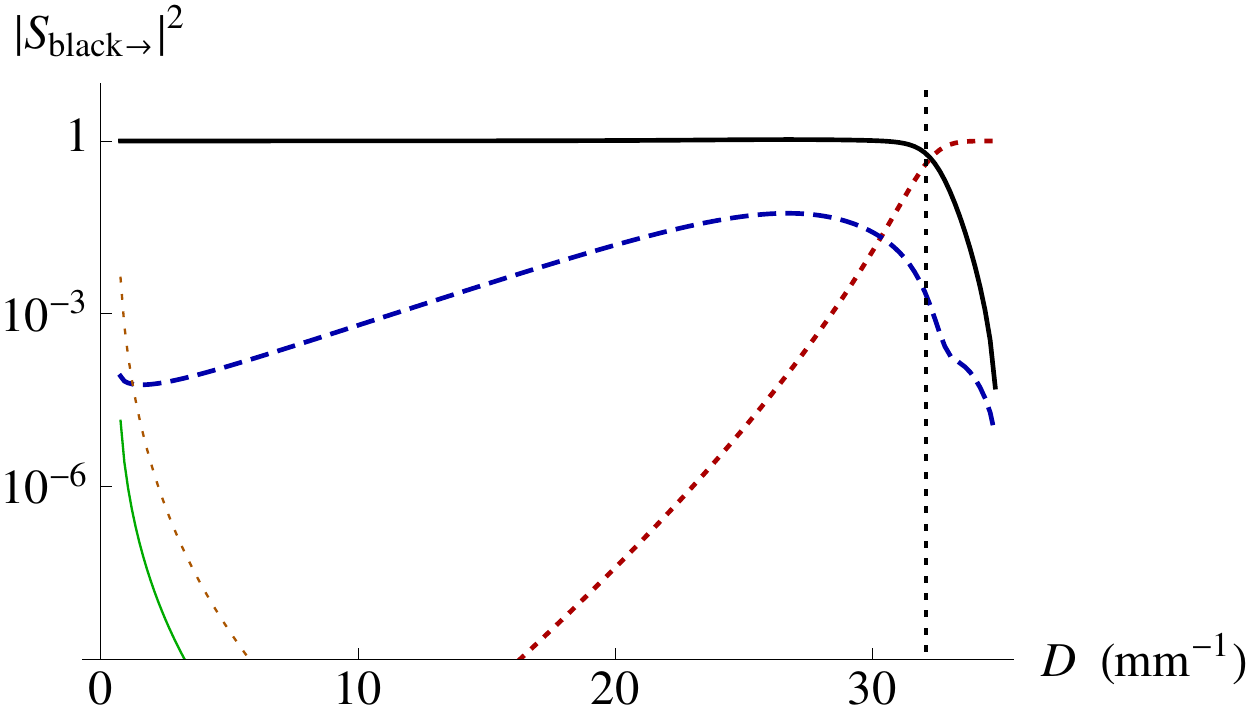}  \\
\includegraphics[width=0.45\columnwidth]{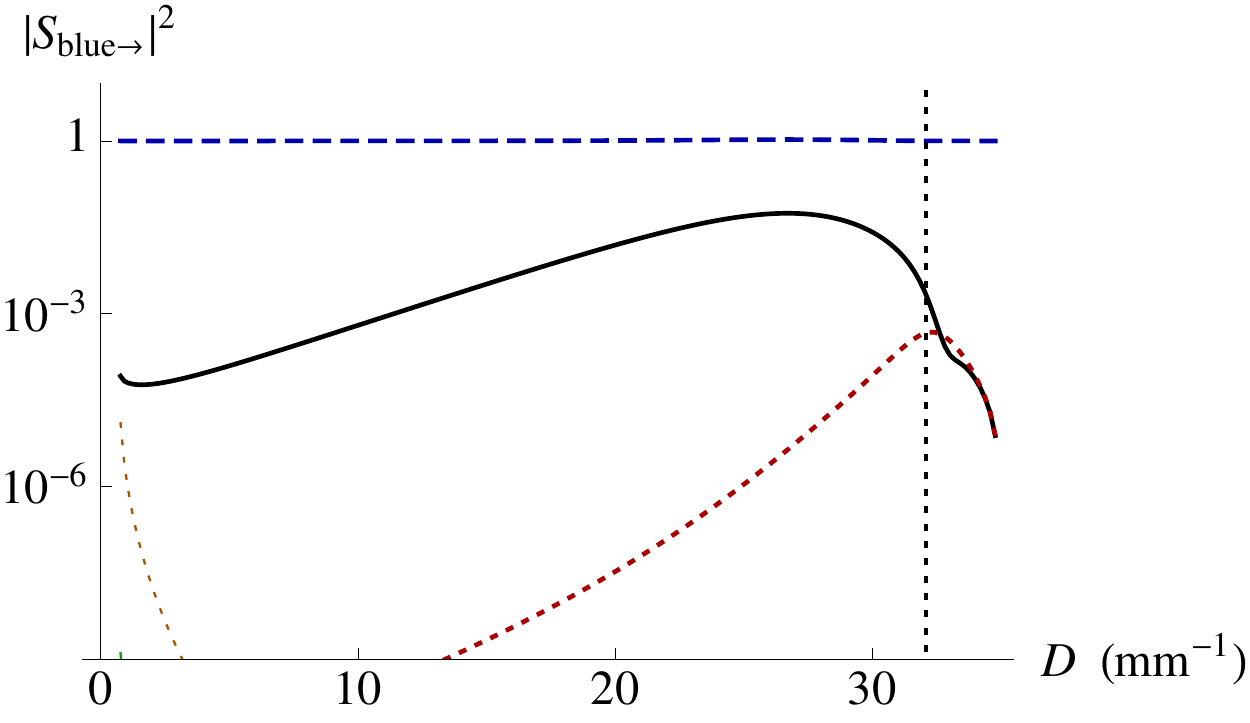} \, \includegraphics[width=0.45\columnwidth]{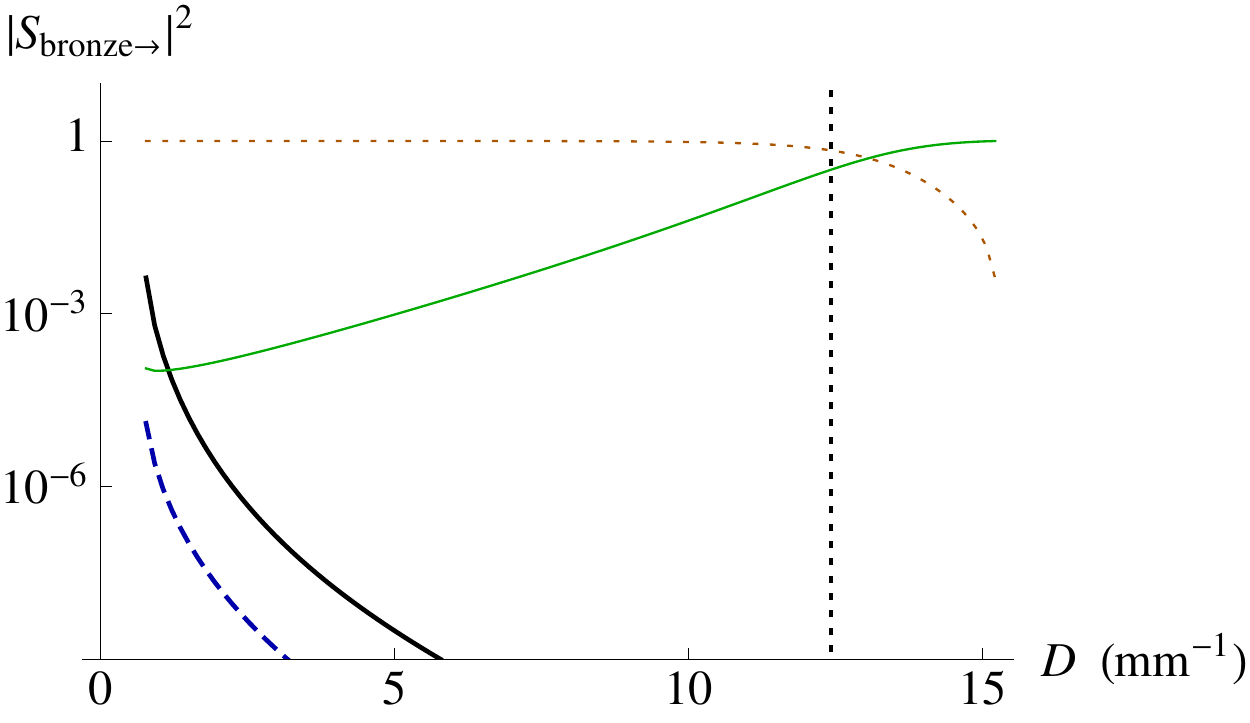}
\caption{
Scattering coefficients in the same setup as in Fig.~\ref{fig:Scattering}, but now shown on a logarithmic scale so that the small coefficients are visible.  The top row shows exactly the same scattering coefficients as Fig.~\ref{fig:Scattering}, i.e. for an incident probe wave on the dotted red branch (upper left plot) and on the solid black branch (upper right plot).  On the lower left plot, the incident probe wave lies on the dashed blue branch, while in the lower right plot it is on the light dotted bronze branch.  The dotted vertical line indicates the value of $D_{\rm min, 1}$ in all plots except the lower right, where it shows $D_{\rm min, 2}$, the extremum of the ``local'' dispersion relation on top of the pulse at which the light solid green and light dotted bronze branches merge.  Note that some scattering coefficients behave in a symmetrical way: in particular, the dashed blue curve on the upper right plot is very similar to the solid black curve on the lower left plot, while the dashed blue curve on the upper left plot is very similar to the dotted red curve on the lower left plot.
\label{fig:MoreScattering}}
\end{figure}

Because most of the scattering coefficients are much smaller than $1$, it is appropriate to plot their squared magnitudes on a logarithmic scale.
On the top row of Fig.~\ref{fig:MoreScattering}, we represent the same scattering coefficients as in Fig.~\ref{fig:Scattering}.
On the left plot, we discover that there is a small anomalous coefficient with a maximum value $\simeq 5 \times 10^{-4}$ (see the dashed blue curve).
On the right plot, we observe that the anomalous coefficients extend over the entire range of $D$, with a minimum value of $\simeq 3 \times 10^{-5}$.
We also observe the subdominant scattering coefficients 
involving modes living on the solid green and dotted bronze branches (shown in a lighter line style). %

On the bottom row, on the left plot, we represent the magnitudes of the scattering coefficients when sending a unit norm incident mode living on the dashed blue branch.
The three other curves (namely solid black, dotted red, and light dotted bronze) describe the magnitudes of anomalous scattering coefficients involving modes with opposite norms.
We notice that the solid black (dotted red) curve is similar to the dashed blue curve in the upper right (left) plot; a similar approximate symmetry was observed when studying a $U(2,1)$ $S$-matrix in atomic BEC (see Eq.~(D8) in~\cite{Macher:2009nz})~\footnote{As already mentioned (see below Eq.~(\ref{eq:bogoliubov})), the anomalous coefficient which is {\it not} enhanced by phase matching, namely the dashed blue curve in the upper left panel (or the dotted red curve in the bottom left panel), reaches its maximum value near $D_{\rm min,1}$, as was found for the anomalous coefficient governing the analogue Hawking effect in a subcritical flow over an obstacle; see the behavior of $\left|\beta_{\omega}\right|^{2}$ in the upper left plot of Fig.~5 in~\cite{Robertson:2016ocv}.  This common behavior can be understood from the presence of a ``group velocity horizon'' (i.e., a turning point)~\cite{Barcelo:2005fc} for quasiparticles, here photons with $D \in \left[D_{\rm min,1}, D_{\rm max,1}\right]$ living on the black or red branch. 
Based on this common behavior, we believe that the anomalous ``soft'' processes we are studying should be considered as a new version of the analogue Hawking effect in nonlinear optics.
\label{fn:sim2}}. 
For completeness, on the lower right plot we represent the magnitudes of the scattering coefficients when sending in a mode living on the light dotted bronze curve.
One mainly observes 
a linear mode mixing describing reflection and transmission between that branch and the corresponding modes on the light solid green one, as could have been expected from the behavior on the dispersion relation near $D_{{\rm max},2}$.

\subsection{Spontaneously emitted spectra for various soliton durations
\label{sec:emission_spectra}} 

In quantum settings, when the initial state is vacuum, the squared magnitudes  
of anomalous scattering coefficients give 
the rates of spontaneous emission of photon pairs.  More precisely, 
the number of pairs emitted per unit $D$ (the conserved wave number) per unit $z$ (the direction, conjugate to $D$, along which the wave equation is invariant) is~\cite{Corley-Jacobson-1996}
\begin{equation}
\frac{\partial^{2} N}{\partial D \, \partial z} = \frac{1}{2\pi} \left|\beta_{D}\right|^{2} \,.
\label{eq:spontaneous_emission_rate}
\end{equation}
Integrating $\left|\beta_{D}\right|^{2}/2\pi$ over all $D$ thus gives the total emission rate (per unit length) 
of photon pairs.
We now further consider the behavior of the enhanced anomalous coefficient governing the mixing of modes living on the solid black and dashed blue branches. 

\begin{figure}
\includegraphics[width=0.45\columnwidth]{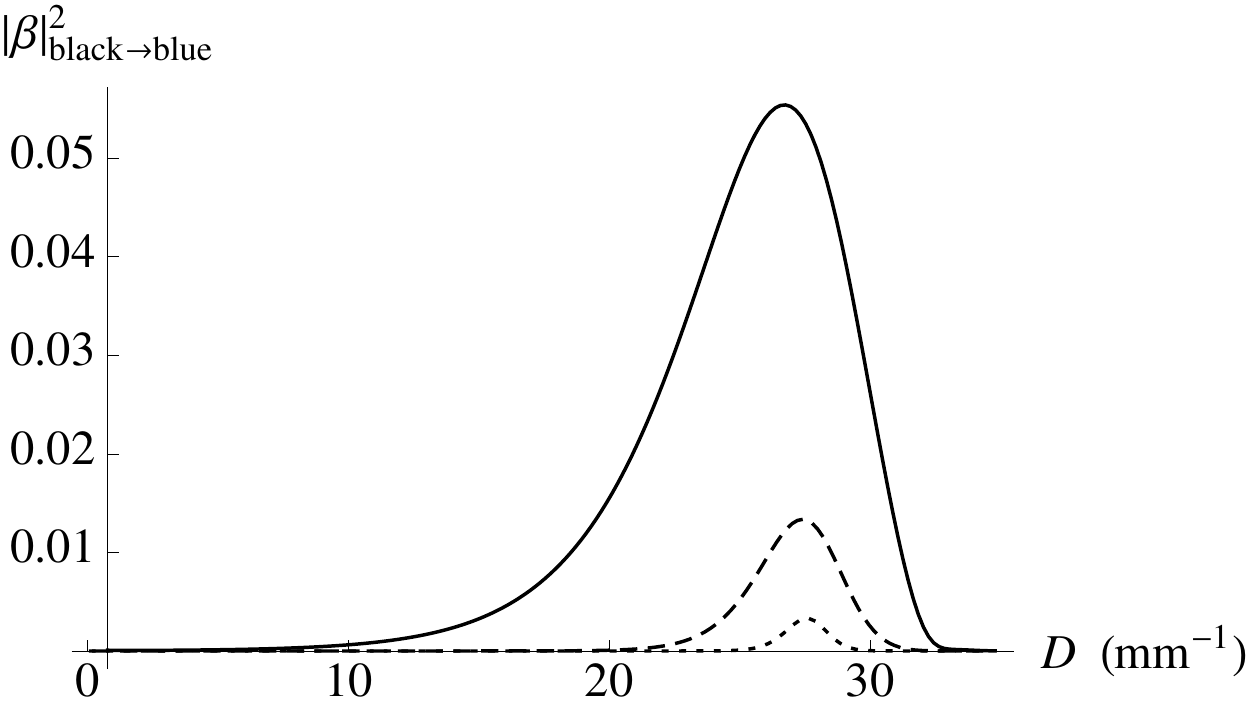}
\caption{Variation of anomalous scattering coefficient with the duration of the pulse.  The value of $\left|\beta_{D}\right|^{2}$ 
shown here describes the scattering of an incident mode on the solid black branch into an outgoing mode on the dashed blue branch.  It is given as a function of $D$ for three different pulse durations $\tau_{0}$ (defined by Eq.~(\ref{eq:soliton})): $40\,{\rm fs}$ (dotted curve), $20\,{\rm fs}$ (dashed curve) and $10\,{\rm fs}$ (solid curve).
The corresponding total emission rates are found by integrating over $D$, and for the examples here are $1.1\times 10^{-3}\,{\rm mm}^{-1}$, $8.7\times 10^{-3}\,{\rm mm}^{-1}$ and $7.5\times 10^{-2}\,{\rm mm}^{-1}$.  We thus see that the total rate scales approximately with $\tau_{0}^{-3}$, as can be rationalized using arguments presented in the text. 
\label{fig:betasq_v_tau0}}
\end{figure}

In Fig.~\ref{fig:betasq_v_tau0}, we represent on a linear scale 
$\left|\beta_{D}\right|^{2}$ 
for three different soliton durations: $\tau_{0} = 10\,{\rm fs}$, $20\,{\rm fs}$ and $40\,{\rm fs}$.  The emission spectrum is clearly 
sensitive to the value of $\tau_{0}$, and integrating over $D$ shows that the total emission rates are, respectively, $7.5\times 10^{-2}\,{\rm mm}^{-1}$, $8.7\times 10^{-3}\,{\rm mm}^{-1}$ and $1.1\times 10^{-3}\,{\rm mm}^{-1}$.  That is, the total emission rate is roughly proportional to $\tau_{0}^{-3}$.  
This can be understood as follows.  In Figure~\ref{fig:solPeak_D} in Appendix~\ref{app:homogeneous} is shown the imaginary part of the ``instantaneous'' dispersion relation at the peak of the soliton, which corresponds to unstable modes and is responsible for the strong anomalous mode mixing leading to the spontaneous emission here described.  Figure~\ref{fig:solPeak_D} indicates that both the width and height of the unstable part of the spectrum vary roughly as $\tau_{0}^{-2}$.  Its integral over $\Delta\omega$ thus varies as $\tau_{0}^{-4}$, and gives a photon production rate per unit $\tau$ per unit length.  We may thus multiply by an effective interaction time to get the production rate per unit length.  This interaction time is just the pulse duration $\tau_{0}$, leading to the total photon production rate being proportional to $\tau_{0}^{-3}$.

Since it is frequency (rather than wave number) that is measured at the output of the waveguide, 
it is appropriate to conclude this analysis 
by translating the spectra of Fig.~\ref{fig:betasq_v_tau0} into numbers of photons emitted per unit $\Delta\omega$ per unit $z$. 
Because the group velocities of the relevant modes (namely, those on the solid black and dashed blue branches) significantly differ, 
the two emission powers are quite different. Their integrals over $\Delta\omega$, however, should agree, and they will be exactly the same as the integral of $\left|\beta_{D}\right|^{2}/2\pi$ over $D$.  Indeed, it is this property -- the invariance of the integral -- that defines the emission spectrum in frequency space: 
\begin{eqnarray}
\frac{\partial^{2}N}{\partial\Delta\omega \, \partial z} & = &  \left| \frac{\partial D}{\partial\Delta\omega} \right|  \frac{\partial^{2}N}{\partial D \, \partial z} \nonumber \\
& = & \frac{1}{2\pi} \left| \frac{\partial D}{\partial\Delta\omega} \right| \, \left|\beta_{D}\right|^{2} \,.
\end{eqnarray}
The two spectra 
are shown in Fig.~\ref{fig:Spec_v_tau0}, for the same soliton durations used in Fig.~\ref{fig:betasq_v_tau0}.  In the left panel is shown the emission spectrum on the (positive-norm) {\it solid} blue branch (i.e. that with $D > 0$ and $\Delta\omega < 0$ in Fig.~\ref{fig:dispersion_SiN}), which is very narrow in frequency due to the large group velocity of this branch.  The right panel shows the emission spectrum on the (positive-norm) solid black branch (with $D<0$ and $\Delta\omega > 0$ in Fig.~\ref{fig:dispersion_SiN}), which is by contrast wide in frequency and correspondingly smaller in amplitude. 
Interestingly, 
when using the shortest soliton duration $\tau_{0} = 10\, {\rm fs}$, 
the values of the detuning 
at the maximal value of these two spectra are not exactly opposite to each other, as one might have expected from a naive use of the mode matching condition obtained when sending a CW in the WG. Indeed, for $\Delta\omega <0$ 
(along the solid blue branch) one finds that the maximum is located at $\Delta \omega = - 0.84\,{\rm PHz}$, whereas for $\Delta\omega > 0$ (along 
the solid black branch) it is localized at $\Delta \omega = 0.78\,{\rm PHz}$. 
A very similar offset is found when integrating the nonlinear equation in App.~\ref{sec:scattering-NL}, see 
the two narrow peaks on the right lower plot of Fig.~\ref{fig:stimulated}.

Finally, it is worth mentioning that the integrated power ($=7.5\times 10^{-2}\,{\rm mm}^{-1}$) obtained with a soliton duration of $\tau_{0} = 10\, {\rm fs}$ means that each soliton propagating in a WG of 1cm 
will produce in the mean $0.75$ pair of entangled pairs with one photon living on the 
(positive-norm) solid blue branch (with $\Delta\omega < 0$)
while its partner lives on the (positive-norm) solid black branch (with $\Delta\omega > 0$).  %

In Appendix~\ref{sec:scattering-NL} 
we numerically solve the nonlinear wave equation~(\ref{eq:wave_eqn_full}) forward in $z$
by sending both the soliton configuration of Eq.~(\ref{eq:soliton}) and some small perturbation. 
In doing so, we include all effects encoded in the nonlinear equation, thereby generalizing what we just did
by solving 
the linearized equation~({\ref{eq:wave_eqn_2b2}). The first outcome is that the above results are all recovered to a very good approximation. In addition, the new simulations display the 
consequences of the fact that the soliton we just used 
is not an exact solution of Eq.~(\ref{eq:wave_eqn_full}), since it exactly solves Eq.~(\ref{eq:non-lin-simplified}). 
Finally, since the new simulations deal with the full field, we are able to check the (nearly exact) conservation 
of the total number of photons, see $N_A$ of Eq.~({\ref{A1-A2}), which implies that each created pair of photons (either stimulated by sending a probe, or spontaneously produced from amplification of vacuum fluctuations) is accompanied by a corresponding decrease of the number of photons in the soliton. 

\begin{figure}
\includegraphics[width=0.45\columnwidth]{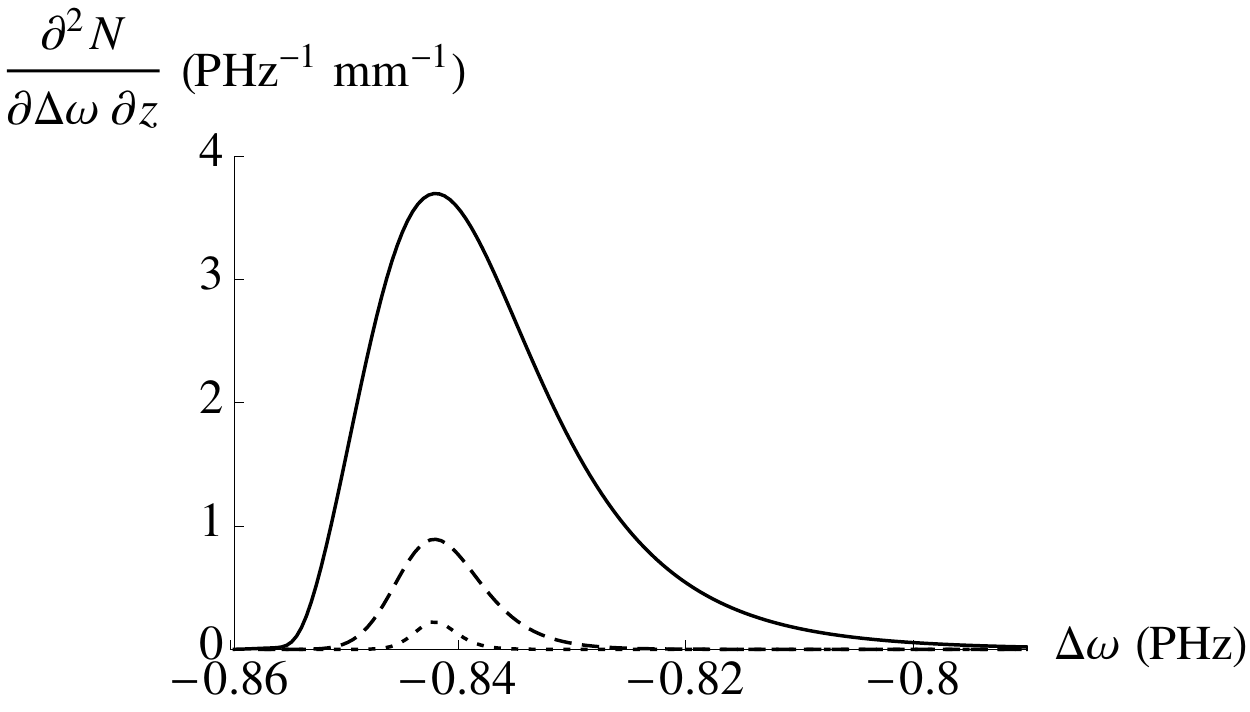} \, \includegraphics[width=0.45\columnwidth]{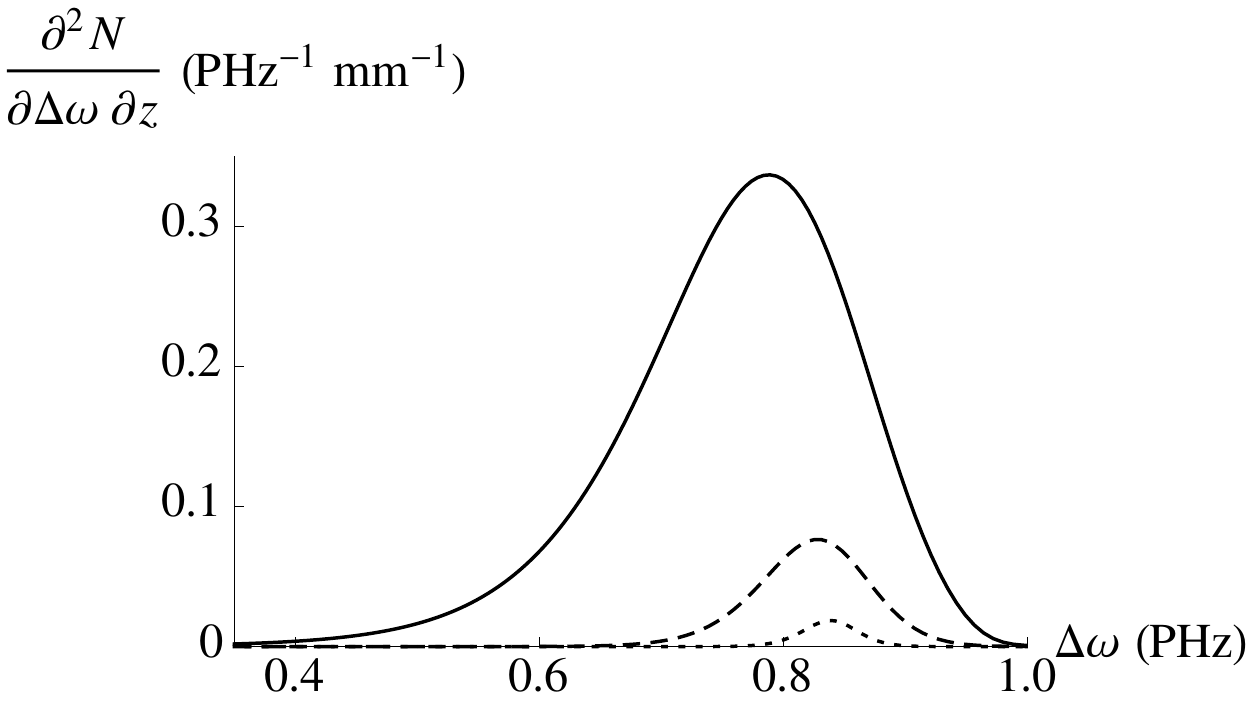}
\caption{Variation of the observed spectrum with the duration of the pulse, assuming an ingoing vacuum state so that the observed photons are generated spontaneously.  These photon number spectra correspond to the same $\left|\beta_{D}\right|^{2}$ 
shown in Fig.~\ref{fig:betasq_v_tau0}, but plotted as a function of $\Delta \omega$ rather than $D$, 
and such that their integrals over $\Delta\omega$ are equal to the integral of $\left|\beta_{D}\right|^{2}$ over $D$. 
The left plot shows the spectrum observed on the (positive-norm) solid blue branch (with $\Delta\omega < 0$), while the right plot shows that on the (positive-norm) solid black branch (with $\Delta\omega > 0$).  %
Note the significant difference in the scales used here, which is a result of the different slopes of these two branches 
in Fig.~\ref{fig:dispersion_SiN}. 
\label{fig:Spec_v_tau0}}
\end{figure}

\section{Summary and conclusions 
\label{sec:conclusion}}

In this work, we studied the scattering of linear perturbations by a soliton propagating in a nonlinear optical 
waveguide (WG).
We started 
by considering the wave equation governing the propagation of the full field (soliton + perturbations). As is usually done in these settings, the 
field is described by a slowly-varying 
envelope multiplying a 
given 
carrier wave. Hence the 
Fourier components 
of the envelope 
are characterized by
the detuned 
frequency and wave number. This equation has been simplified in two respects. First, only configurations co-propagating with the soliton are described; hence it is odd in the wave number $k>0$.  
Given the smallness of the spatial gradients (fixed by the nonlinear Kerr index and the soliton power),
this neglect is well justified. Second, we 
neglect the terms responsible for 
linear and nonlinear losses,
retardation and 
self-steepening. 
None of these 
effects 
should 
significantly modify the main results obtained with 
the simplified equation. In fact, 
we believe our results 
are robust because 
they rely 
on a crossing of two branches of the dispersion relation, often referred to as a phase 
matching condition. This crossing induces a modulation 
instability when sending a continuous carrier wave in the WG (as is reviewed in App.~\ref{app:homogeneous}), 
and an enhancement of pair creation processes when sending a soliton (as shown in Sec.~\ref{sec:scattering}). 

For concreteness, the waveguide was 
taken to be a rectangular silicon nitride waveguide on a silica substrate. This type of WG offers two advantages, namely the possibility of engineering the effective longitudinal dispersion relation (which 
we compute numerically by 
taking into account the transverse properties of the WG), and a low loss rate 
which can 
be safely ignored. 
Importantly, 
the dispersion 
is anomalous in a finite frequency window. This is relevant in two respects. 
First, 
it allows 
soliton solutions which do not significantly disperse (as 
verified in App.~\ref{sec:scattering-NL}), and which 
play the role of the background when studying scattering processes. 
Second, 
when considering the positive- and negative-norm branches of the waveguide dispersion relation defined relative to the carrier of the soliton, 
the finite extension of the anomalous 
window necessarily gives the above-mentioned crossing of 
the two branches of the dispersion relation, 
as clearly illustrated in Fig.~\ref{fig:dispersion_SiN}. 

The anomalous mode mixing coefficients encoding production of photon pairs
were first computed 
by linearizing the full equation on top of the soliton.  Keeping 
all linear terms, we pointed 
out that one of them, 
which would have been dropped if we had performed the standard rotating wave approximation (RWA), 
mixes modes of opposite detuned frequencies and opposite norm. 
It is thus necessary 
to keep this the term. 
When doing so, 
we obtained 
a wave equation which has the same structure as that governing phonon scattering in an atomic Bose condensate. We thus applied 
the same 
techniques (based on the use of mode doublets) to compute the scattering coefficients. The numerical calculation of these coefficients was 
performed for a fixed value of the detuned wave number, since it is a conserved quantity
given the stationarity 
of the background. 
We first recovered 
the expected elastic mode conversion interpolating from total transmission to total reflection which appears when the modes are blocked by the soliton. Such conversion, which occurs in a small interval of detuned wave numbers, has been described and observed in several works. Focusing on anomalous scattering coefficients which 
induce a parametric amplification (i.e. pair creation processes in quantum settings), we observed 
a significant enhancement of the coefficient in the vicinity of the crossing, see 
Fig.~\ref{fig:Spec_v_tau0}. 
As can be seen in the two plots of this figure, 
the detuned frequencies of the produced photons (which are spontaneously produced) 
are, to a good approximation, opposite to each other. 
Since their detuned wave numbers are 
also opposite, one clearly sees that the kinematics of 
the enhanced anomalous scattering on a soliton is governed by the phase 
matching condition. 
It also tells us that the pairs produced by the processes here 
considered are all extracted from pairs of ``condensed'' photons belonging to the soliton. 
When considering the pair production rate, we found it to be about 6 orders of magnitude larger than the results of previous works
on the analogue Hawking effect in nonlinear optics.  As explained in footnote~\ref{fn:soft-hard}, this discrepancy has a double origin: the crossing of the positive- and negative-norm branches of the dispersion relation, as well as the more adiabatic 
character of 
FWM with respect 
to the ``hard'' processes previously studied.
When considering the scattering on a soliton, we are thus facing two versions of the analogue Hawking effect: the ``soft'' one here presented, and the ``hard'' one studied previously.  It remains to provide a unified description where both types of anomalous scattering are simultaneously computed.

We then considered 
the 
integral over the detuned wave number so as to get 
the total pair production rate per unit propagation distance of the 
soliton. 
We found that it scales with the third power of the inverse duration of the soliton, or equivalently with $P_0^{3/2}$ where $P_0$ is the peak 
soliton power. 
For a pulse duration of $10\,{\rm fs}$ at 971\,nm (1.94\,PHz), we deduced that, in the mean, one pair will be spontaneously produced for a WG of a length equal to $1.3\,{\rm cm}$.  The two photons of the pair are respectively generated around 687\,nm ($\Delta\omega=0.8$\,PHz) and 1720\,nm ($\Delta\omega=-0.845$\,PHz) where detectors are commercially available. 
This puts the effect 
within measurable range, and would 
enable the clear demonstration of photon pair production induced by a soliton. The fact that all photons are emitted in pairs suggests using 
coincidence measurements 
to greatly increase the signal-to-noise ratio: 
one would only analyze the data where two photons are detected, and 
verify that their frequency domains lie within the expected domains shown in Fig.~\ref{fig:Spec_v_tau0}. 
If the signal-to-noise ratio is high enough and the relevant photon modes are initially in their vacuum state, 
one could 
envisage measuring 
the degree of entanglement, as briefly discussed in the last subsection of Appendix~\ref{sec:scattering-NL}.

This appendix is mainly concerned with the 
numerical integration of the nonlinear equation governing the full field (soliton + perturbations). Although this treatment completely differs from that used when dealing with the linearized wave with a fixed detuned wave number, we find 
an excellent agreement of the results.
In particular, the squared norms of the scattering coefficients extracted from nonlinear simulations closely agree in profile and 
magnitude with those obtained in the main text. This is 
non-trivial, 
because 
the evolution of the soliton is now 
dynamically determined. As a result, one observes a significant amount of Cherenkov radiation, 
which 
can 
be understood from the fact that the soliton is not 
an exact solution of the nonlinear equation. 
Morever, because we deal with the full field, we are also able to verify 
that 
pair production processes 
are always accompanied by a corresponding decrease of the number of photons in the soliton.  
Similarly, 
we also observe 
the reduction of the latter 
due to Cherenkov radiation,
as expected from the fact that the first-order equation identically conserves the total number of photons, thereby encoding only ``soft'' processes in our classification.

\section*{Acknowledgments}

We thank Florent Michel for useful discussions and for his careful reading of the manuscript. 
We are grateful to Ulf Leonhardt for useful remarks emphasizing the difference between ``soft'' and ``hard'' processes.
S.R. thanks LPT (Laboratoire de Physique Th\'{e}orique), Orsay, where most of the research work was done.
S.R. and R.P. were supported by the French National Research Agency through the Grant No. ANR-15-CE30-0017-04 associated with the project HARALAB. 
S-P.G., C.C. and S.M. acknowledge the support of the Belgian Science Policy (BELSPO) Interuniversity Attraction Pole (IAP) 7-35 Photonics@be, as well as of the Fonds de la Recherche Fondamentale Collective (Grant No. PDR.T.1084.15). 

\begin{appendices}

\section{Dispersion relation on top of a strong background 
\label{app:homogeneous}}

In this appendix we consider the case where the background 
is homogeneous and thus described by 
a plane wave. Although not original, 
see Refs.~\cite{Abdullaev-et-al-1994,Pitois-Millot-2003}, 
it is here presented in terms of the doublet formalism so as to clarify its link with the scattering described in the main text. 
Importantly, 
at the end of the appendix, we apply the same techniques to determine the effective deformation of the dispersion relation computed at the peak power of a soliton, a notion used to define the quantities $D_{\rm min, 1}$ and $D_{\rm min, 2}$ used in Figs.~\ref{fig:Scattering} and~\ref{fig:MoreScattering}. To our knowledge the results presented in 
Fig.~\ref{fig:solPeak_D} have not previously 
been presented in the literature.

\subsection{Homogeneous background} 

Starting from Eq.~(\ref{eq:wave_eqn_full}), we first consider the solution $A_{0}$ which is indepedent of $\tau$, and which thus satisfies the equation 
\begin{equation}
-i\partial_{z}A_{0} = 
\gamma \left|A_{0}\right|^{2} A_{0} \,,
\label{eq:GP}
\end{equation}
This is solved straightforwardly as follows: 
\begin{equation}
A_{0} = A_{0}^{H} \, \mathrm{exp}\left(i \delta\beta_{0}^{H} z + i \theta_{0} \right)
\label{eq:hom_soln}
\end{equation}
where $A_{0}^{H}$ and $\theta_{0}$ are constant real numbers, and where $\delta\beta_{0}^{H} = \gamma \left(A_{0}^{H}\right)^{2}$ is the nonlinear displacement of 
the wave number of the carrier wave.  
Notice that this differs by a factor of 2 from the relation between $\delta\beta_{0}$ and the peak power of a soliton; see Eq.~(\ref{eq:solitonNLshift}). 
(In the analogy between Eq. (\ref{eq:GP}) and the Gross-Pitaevskii equation, the nonlinear contribution $\delta\beta_{0}^{H}$ to the wave number of the $\tau$-independent background 
is analogous to the chemical potential of a homogeneous BEC.)  

We now proceed as in Sec.~\ref{sub:linear_wave_eqn}, writing the full solution as the sum of the background and a weak perturbation.  As in Eq.~(\ref{eq:A_decomp}), it 
is convenient to factor out the $z$-dependent 
phase of $A_{0}$: 
\begin{equation}
A = e^{i\delta\beta_{0}^{H}z+i\theta_{0}}\left(A_{0}^{H}+\delta{A}\right) \,. 
\end{equation}
Then the phase of the perturbation $\delta A$ is relative 
to that of the background, whereas its amplitude is absolute.\footnote{We could, if we wished, also use the relative amplitude in the homogeneous case considered here, as is often done in treatments of phononic perturbations in atomic BEC~\cite{Macher:2009nz}. 
In the main text, however, where a localized pulse is considered as background, this would have been 
problematic as 
the ``perturbation'' amplitude would have been 
much larger than that of the pulse asymptotically.}
To linear 
order in $\delta{A}$ and $\delta{A}^{\star}$, the wave equation~(\ref{eq:wave_eqn_full}) becomes 
\begin{subequations}\begin{eqnarray}
-i\partial_{z}\left(\delta{A}\right) & = & B(i\partial_{\tau}) \delta{A} + \delta\beta_{0}^{H} \left( \delta{A} + \delta{A}^{\star}\right) \,, \\
i\partial_{z}\left(\delta{A}^{\star}\right) & = & B(-i\partial_{\tau}) \delta{A}^{\star} + \delta\beta_{0}^{H} \left( \delta{A} + \delta{A}^{\star}\right) \,,
\label{eq:BdG}
\end{eqnarray}\end{subequations}
or, in matrix form,
\begin{equation}
-i\partial_{z} \left[ \begin{array}{c} \delta{A} \\ \delta{A}^{\star} \end{array} \right] = \left[ \begin{array}{cc} B(i\partial_{\tau}) + \delta\beta_{0}^{H} & \delta\beta_{0}^{H} \\ -\delta\beta_{0}^{H} & -B(-i\partial_{\tau}) - \delta\beta_{0}^{H} \end{array} \right] \left[ \begin{array}{cc} \delta{A} \\ \delta{A}^{\star} \end{array} \right] \,.
\label{eq:wave_eqn_matrix}
\end{equation}
This is a particular realization of Eqs.~(\ref{eq:wave_eqn_2b2}) and~(\ref{eq:diff_operator}), where here $A_{0}(\tau) = A_{0}^{H}$ is defined to be real and its magnitude is such that $\gamma \left(A_{0}^{H}\right)^{2} = \delta\beta_{0}^{H}$. 

We may 
exploit the lack of any explicit $\tau$-dependence in Eq.~(\ref{eq:wave_eqn_matrix}) and perform the Fourier transform working at fixed (conserved) $\Delta\omega$: 
\begin{subequations}\begin{eqnarray}
\delta{A}(z,\tau) & = & \int_{-\infty}^{+\infty} d\left(\Delta\omega\right) \, \widetilde{\delta{A}}_{\Delta\omega}(z) \, e^{-i\Delta\omega \,\tau} \\ 
\delta{A}^{\star}(z,\tau) & = & \int_{-\infty}^{+\infty} d\left(\Delta\omega\right) \, \widetilde{\delta{A}}^{\star}_{-\Delta\omega}(z) \, e^{-i\Delta\omega \,\tau}\,.
\label{eq:decomp}
\end{eqnarray}\end{subequations}
Then Eq. (\ref{eq:wave_eqn_matrix}) becomes:
\begin{equation}
-i\partial_{z} \left[ \begin{array}{c} \widetilde{\delta A}_{\Delta\omega} \\ \widetilde{\delta A}^{\star}_{-\Delta\omega} \end{array} \right] = \left[ \begin{array}{cc} B_{\rm even}(\Delta\omega) + B_{\rm odd}(\Delta\omega) + \delta\beta_{0}^{H} & \delta\beta_{0}^{H} \\ -\delta\beta_{0}^{H} & -B_{\rm even}(\Delta\omega) + B_{\rm odd}(\Delta\omega) - \delta\beta_{0}^{H} \end{array} \right] \left[ \begin{array}{cc} \widetilde{\delta A}_{\Delta\omega} \\ \widetilde{\delta A}^{\star}_{-\Delta\omega} \end{array} \right] \,,
\label{eq:wave_eqn_matrix_FTtau}
\end{equation}
where 
we have introduced the even and odd parts of $B(\Delta\omega)$, defined as
\begin{alignat}{2}
B_{\rm even}(\Delta\omega) = \frac{1}{2} \left( B(\Delta\omega) + B(-\Delta\omega) \right) \,, & \qquad B_{\rm odd}(\Delta\omega) = \frac{1}{2} \left( B(\Delta\omega) - B(-\Delta\omega) \right) \,.
\end{alignat}
This allows Eq. (\ref{eq:wave_eqn_matrix_FTtau}) to be written in the form
\begin{equation}
\left(-i\partial_{z} - B_{\rm odd}(\Delta\omega) \right) \left[ \begin{array}{c} \widetilde{\delta A}_{\Delta\omega} \\ \widetilde{\delta A}^{\star}_{-\Delta\omega} \end{array} \right] = \left[ \begin{array}{cc} K_{\Delta\omega} & \delta\beta_{0}^{H} \\ -\delta\beta_{0}^{H} & -K_{\Delta\omega} \end{array} \right] \left[ \begin{array}{cc} \widetilde{\delta A}_{\Delta\omega} \\ \widetilde{\delta A}^{\star}_{-\Delta\omega} \end{array} \right] \,,
\end{equation}
where $K_{\Delta\omega} \equiv B_{\rm even}(\Delta\omega) + \delta\beta_{0}^{H}$.
This equation is now very similar to that obtained for linear perturbations in a homogeneous BEC.  It can be diagonalized by introducing the variables $\varphi_{\Delta\omega}$ and $\varphi^{\star}_{-\Delta\omega}$:
\begin{equation}
\left[\begin{array}{c} \widetilde{\delta A}_{\Delta\omega} \\ \widetilde{\delta A}^{\star}_{-\Delta\omega} \end{array} \right] = \left[ \begin{array}{cc} u_{\Delta\omega} & v_{\Delta\omega} \\ v_{\Delta\omega} & u_{\Delta\omega} \end{array} \right] \left[ \begin{array}{c} \varphi_{\Delta\omega} \\ \varphi^{\star}_{-\Delta\omega} \end{array} \right] \,,
\label{eq:varphi_defn}
\end{equation}
where
\begin{equation}
\begin{array}{c} u_{\Delta\omega} \\ v_{\Delta\omega} \end{array} = \frac{\sqrt{K_{\Delta\omega} - \delta\beta_{0}^{H}} \pm \sqrt{K_{\Delta\omega} + \delta\beta_{0}^{H}}}{2\sqrt{p_{\Delta\omega}}}
\label{eq:u_v_defn}
\end{equation}
and the effective shift in the squared wave number
\begin{eqnarray}
p_{\Delta\omega}^{2} & = & K_{\Delta\omega}^{2} - \left(\delta\beta_{0}^{H}\right)^{2} \nonumber \\
& = & 2 \delta\beta_{0}^{H} B_{\rm even}(\Delta\omega) + B^{2}_{\rm even}(\Delta\omega) \,.
\label{eq:psq}
\end{eqnarray}
Finally, this yields the diagonalized matrix equation
\begin{equation}
-i\partial_{z} \left[ \begin{array}{c} \varphi_{\Delta\omega} \\ \varphi^{\star}_{-\Delta\omega} \end{array} \right] = \left[ \begin{array}{cc} D(\Delta\omega) & 0 \\ 0 & -D(-\Delta\omega) \end{array} \right] \left[ \begin{array}{c} \varphi_{\Delta\omega} \\ \varphi^{\star}_{-\Delta\omega} \end{array} \right] \,,
\end{equation}
where 
\begin{equation}
D(\Delta\omega) = p_{\Delta\omega} + B_{\rm odd}(\Delta\omega) \,. \\
\label{eq:hom_disprel}
\end{equation}
This, then, is the dispersion relation of linear perturbations on top of a constant background.  Because $p_{\Delta\omega}^{2}$ of Eq.~(\ref{eq:psq}) is not necessarily positive, $D(\Delta\omega)$ is generally complex.  Examples of its real and imaginary parts are shown in Fig.~\ref{fig:hom_D}, for realistic values of $\delta\beta_{0}^{H}$ (namely, $2.5\,\mathrm{mm}^{-1}$ and $5\,\mathrm{mm}^{-1}$).  In particular, in the imaginary part of $D(\Delta\omega)$ we recover the standard modulation instability around $\Delta\omega=0$, as well as the additional narrow instability induced by the crossing of the two branches of the (unperturbed) dispersion relation.  
It is worth reminding the reader that eigenmodes with a complex frequency necessarily have zero norm, see~\cite{Leonhardt-Kiss-Ohberg-2003-HE,Coutant-2016} and other references therein.  
This can be understood from the fact that they are formed from a linear combination of positive- and negative-norm modes with equal weights, causing the norm of the eigenmode to vanish. 

\begin{figure}
\includegraphics[width=0.45\columnwidth]{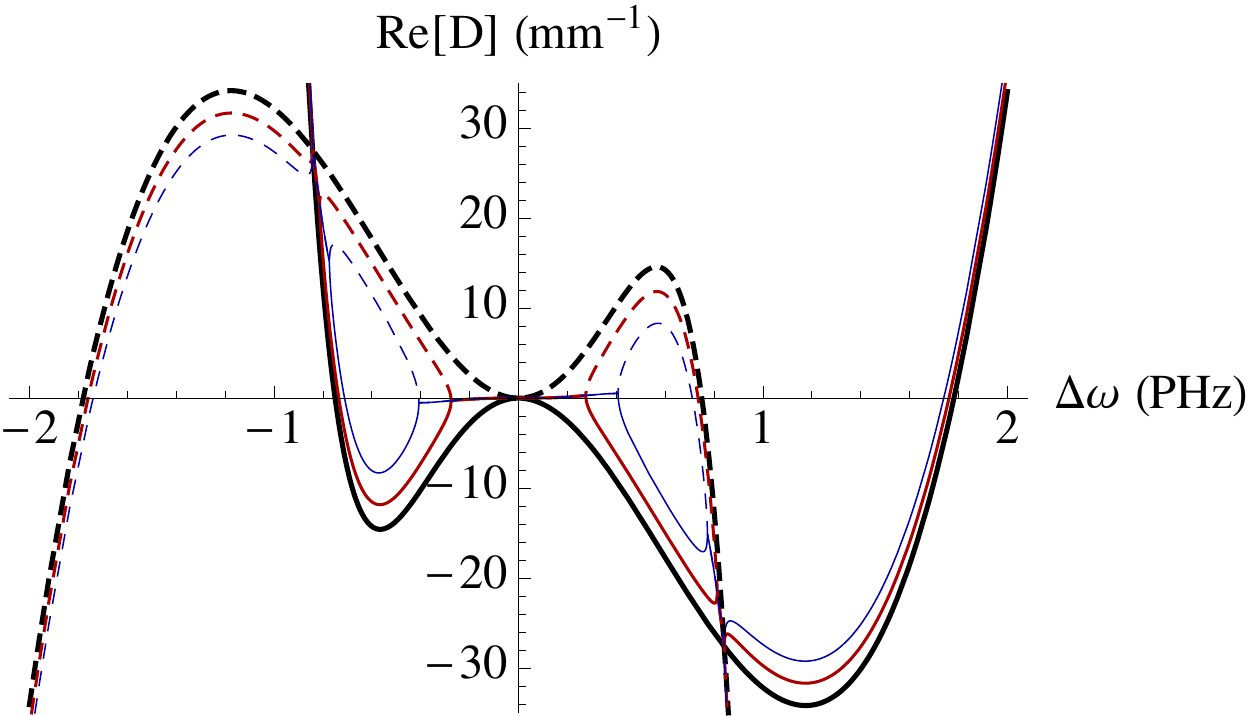} \, \includegraphics[width=0.45\columnwidth]{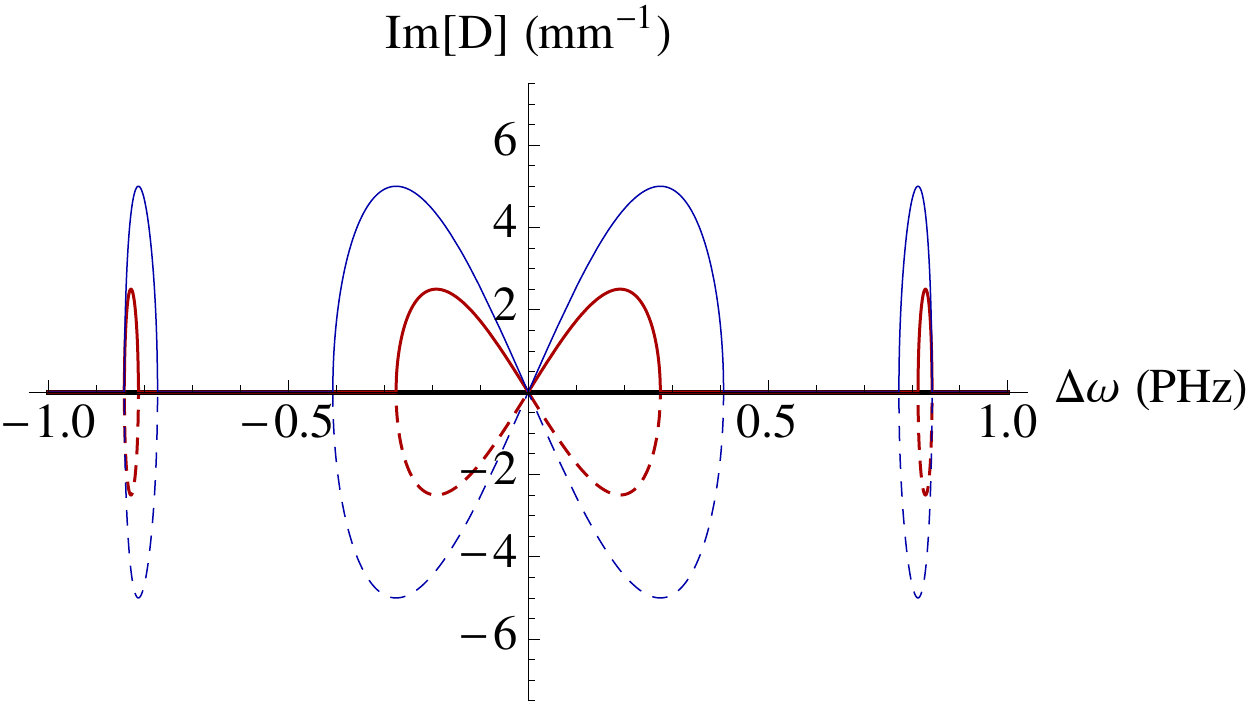}
\caption{The real (left panel) and imaginary (right panel) parts of $D$ as a function of $\Delta\omega$, for different values of $\gamma \left(A_{0}^{H}\right)^{2}$: (in order of decreasing line thickness) $0$ (black), $2.5\,\mathrm{mm}^{-1}$ (red) and $5\,\mathrm{mm}^{-1}$ (blue). These values 
correspond to the nonlinear shift $\delta\beta_{0}$ of solitons with durations $\tau_{0} = 5.3\,{\rm fs}$ and $3.7\,{\rm fs}$, respectively. 
When $D$ is real, the solid and dashed curves correspond, respectively, to the positive- and negative-norm branches of the dispersion relation.
In the right panel, we recover the standard modulation instability around $\Delta\omega = 0$, and the additional narrow instability where phase matching occurs. 
\label{fig:hom_D}}
\end{figure}

It is instructive to consider limiting cases of the dispersion relation~(\ref{eq:hom_disprel}). 
When $\gamma \left(A_{0}^{H}\right)^{2} = 0$, it 
reduces to $D(\Delta\omega) = B(\Delta\omega)$, as expected; $D(\Delta\omega)$ and $-D(-\Delta\omega)$ are just the detuned wave numbers of $\widetilde{\delta A}$ and $\widetilde{\delta A}^{\star}$, respectively.  When $\delta\beta_{0}^{H} \neq 0$, we can consider the small $\Delta\omega$ behaviour using the fact that $B(\Delta\omega) \approx \frac{1}{2} \beta_{2} (\Delta\omega)^{2}$ at small $\Delta\omega$.  To lowest order, then, $B_{\rm odd}(\Delta\omega)$ vanishes, and we have 
\begin{equation}
D(\Delta\omega) \approx \sqrt{\delta\beta_{0}^{H} \beta_{2}} \, \Delta\omega \, .  
\end{equation}
If $\beta_{2} < 0$, then modes with small $\Delta\omega$ have an imaginary $D$, and so either grow or decay exponentially with $z$.  On the other hand, if $\beta_{2} > 0$, these modes propagate with a constant velocity in the co-moving frame of the carrier wave.

\subsection{Local description on top of a soliton} 

\begin{figure}
\includegraphics[width=0.45\columnwidth]{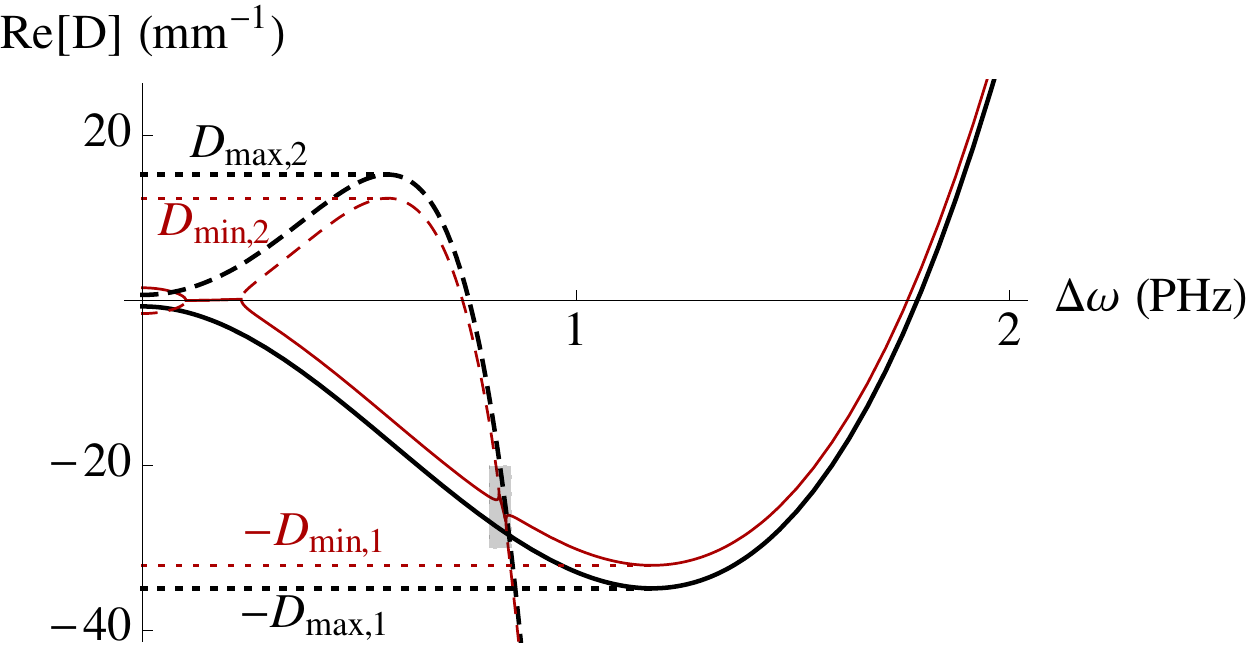} \, \includegraphics[width=0.45\columnwidth]{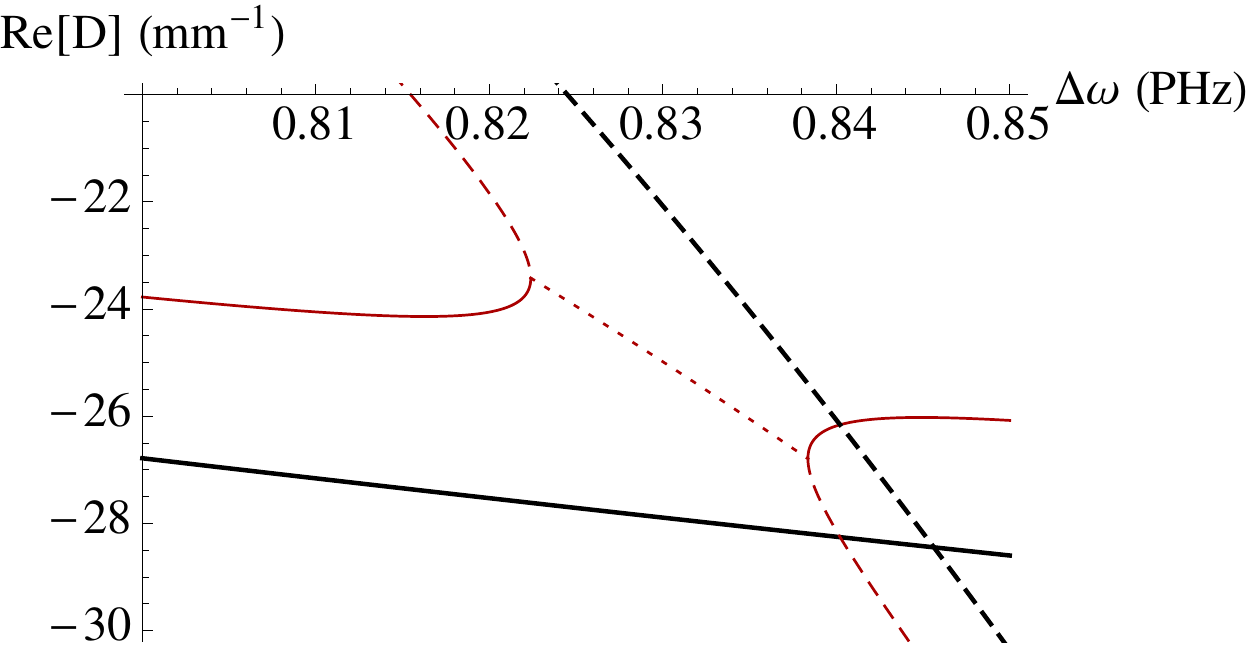} \\ 
\includegraphics[width=0.45\columnwidth]{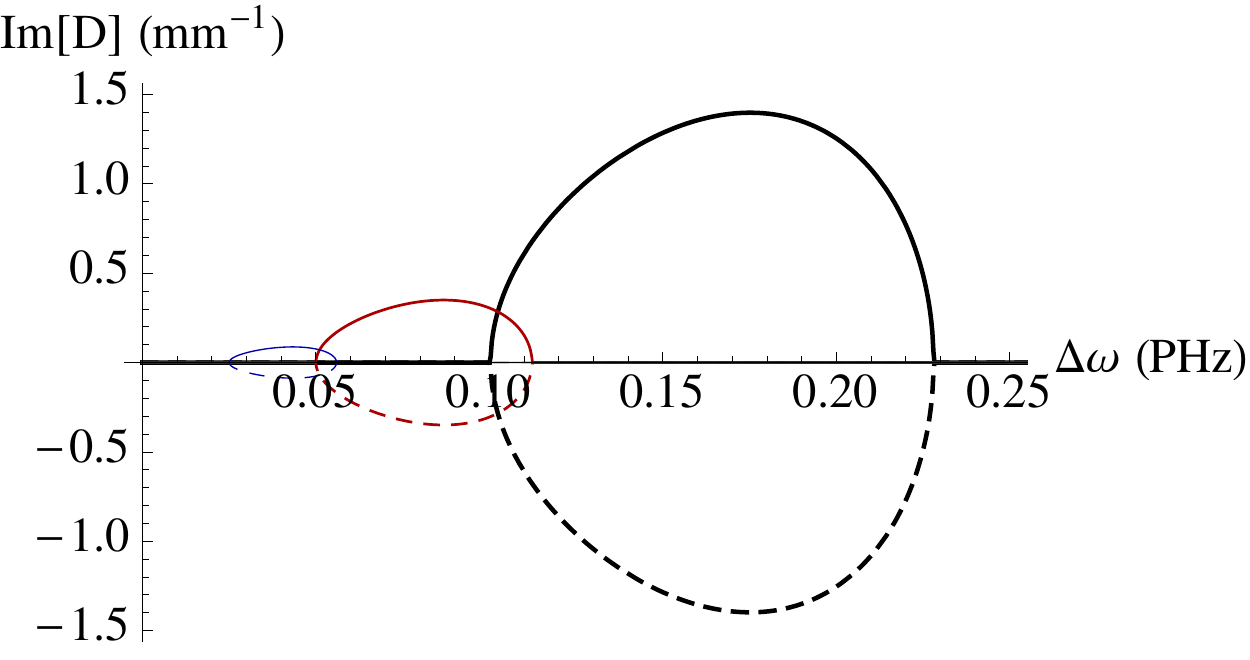} \, \includegraphics[width=0.45\columnwidth]{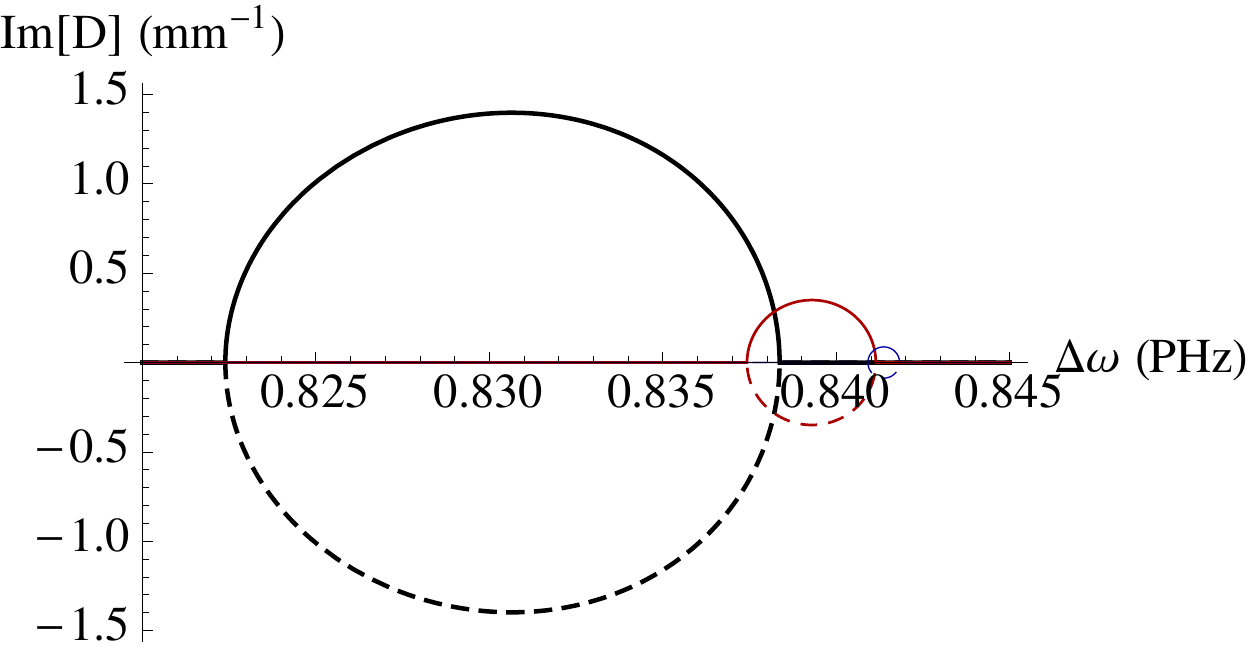}
\caption{The real (top row) 
and imaginary (bottom row) 
parts of the ``instantaneous'' dispersion relation $D\left(\Delta\omega\right)$ at the peak of the soliton, as a function of $\Delta\omega$.  In the top row, 
we show in black (thicker line) the asymptotic dispersion relation, 
and in red (thinner line) that 
at the peak of a soliton of duration $\tau_{0} = 10\,{\rm fs}$.  Solid and dashed curves correspond to positive- and negative-norm modes, respectively, while dotted curves indicate where $D(\Delta\omega)$ is complex. 
In the left panel, the extrema $D_{\rm max,1}$ and $D_{\rm min,1}$ are indicated, as are the other extrema $D_{\rm max,2}$ and $D_{\rm min,2}$.  
The right panel shows a zoom on the region near the crossing (the shaded rectangle on the left panel). 
On the bottom row are 
the imaginary parts of $D\left(\Delta\omega\right)$ in the two 
regimes where $D(\Delta\omega)$ becomes complex, 
at the peaks of solitons of three 
durations: (in order of decreasing line thickness) $\tau_{0} = 10\,{\rm fs}$ (black), $20\,{\rm fs}$ (red) and $40\,{\rm fs}$ (blue).  The left 
panel shows a descendant of the standard modulation instability on an inhomogeneous background, while the right panel shows the narrow instability that is the origin of the anomalous mode mixing studied in this paper.  The width of the former 
is seen to vary as $\tau_{0}^{-1}$ while that of the latter 
varies as $\tau_{0}^{-2}$.  Their heights instead follow the same scaling law, being both proportional to $\tau_{0}^{-2}$.
\label{fig:solPeak_D}}
\end{figure}

We now use the above analysis
to characterize the effective ``instantaneous'' (i.e. $\tau$-dependent) dispersion relation on top of the soliton.
As in the standard WKB treatment, we neglect 
the gradient of the background and keep 
only the local value of the intensity $\left|A_{0}(\tau)\right|^{2}$.
The 
difference from the above analysis lies in the fact that, while 
the wave number $\delta\beta_{0}$ of the background is a constant (now 
given by Eq.~(\ref{eq:solitonNLshift})), the intensity $\left|A_{0}(\tau)\right|^{2}$ is a function of $\tau$, and the two quantities are 
therefore disconnected. 
Moreover, as noted above, even when restricting our attention to the point at the peak of the soliton, the relationship between $\delta\beta_{0}$ and $\left|A_{0}^{\rm peak}\right|^{2}$ differs from that for a $\tau$-independent background by a factor of 2. 
As a result, the dispersion relations 
evaluated at the peak intensity of the soliton slightly differ 
from those of 
Fig.~\ref{fig:hom_D},
both in their real and imaginary parts.

The relevant matrix to be diagonalized is that of Eq.~(\ref{eq:diff_operator}) with $i \partial_{\tau} \rightarrow \Delta\omega$, and (as noted above) with $\left|A_{0}(\tau)\right|^{2}$ being treated as constant in order to extract the ``instantaneous'' dispersion relation.  Following exactly the same procedure as for the $\tau$-independent case outlined above, we find again Eq.~(\ref{eq:hom_disprel}) where now
\begin{equation}
p_{\Delta\omega}^{2} = \left(B_{\rm even}\left(\Delta\omega\right) - \delta\beta_{0} + 2 \gamma \left|A_{0}(\tau)\right|^{2}\right)^{2} - \gamma^{2} \left|A_{0}(\tau)\right|^{4} \,.
\end{equation}
At the peak of the soliton, we have $\gamma\left|A_{0}(\tau)\right|^{2} = 2\, \delta\beta_{0}$, and this becomes
\begin{eqnarray}
p_{\Delta\omega}^{2} & = & \left(B_{\rm even}\left(\Delta\omega\right) + 3\, \delta\beta_{0}\right)^{2} - 4 \left(\delta\beta_{0}\right)^{2} \nonumber \\
& = & 5 \left(\delta\beta_{0}\right)^{2} + 6 \, \delta\beta_{0} \, B_{\rm even}\left(\Delta\omega\right) + B_{\rm even}^{2}\left(\Delta\omega\right) \,.
\end{eqnarray}
Examples of the dispersion relation 
at the peak of the soliton 
are shown in Fig.~\ref{fig:solPeak_D} for the same soliton durations 
used in the main text. The corresponding extrema of the real part of the function $D(\Delta\omega)$ of Eq.~(\ref{eq:hom_disprel}) 
define the quantities called $D_{\rm min, 1}$ and $D_{\rm min, 2}$, which are 
represented by vertical dotted lines in Figs.~\ref{fig:Scattering} and~\ref{fig:MoreScattering}. 
Moreover, the variation of the width and height of the imaginary part of $D(\Delta\omega)$ 
in the vicinity of $D_{\rm cross}$ gives a good estimate for the $\tau_{0}$-dependence of the total photon emission rate, as discussed in Sec.~\ref{sec:emission_spectra}.


\section{
Nonlinear propagation of the full field configuration
\label{sec:scattering-NL}}

In this appendix, 
we numerically solve the nonlinear wave equation~(\ref{eq:wave_eqn_full}) forward in $z$
by sending both the soliton configuration of Eq.~(\ref{eq:soliton}) and some small perturbation. 
In doing so, our aim is to include all effects encoded in the nonlinear equation, thereby generalizing 
and validating 
what has been done in the main text 
by solving 
the linearized equation~({\ref{eq:wave_eqn_2b2}). 
At a deeper 
level, we also 
aim at establishing the conservations laws associated with the nonlinear evolution, which have no counterpart when dealing with Eq.~(\ref{eq:wave_eqn_2b2}). 
We shall first study 
stimulated and then 
spontaneous processes. 
The various observations 
are presented in 
separate subsections 
so that the reader can easily identify 
both the agreement with, and 
the novel elements with respect to, the linearized treatment of the main text.

\subsection{Stimulated emission by a probe wave} 
\label{sub:stimulated}}

We start by considering the stimulated case because of 
the clarity of its outcome.
As initial condition, we take the soliton $A_0(\tau)$ described in Eq.~(\ref{eq:soliton}) with duration $\tau_{0} = 10\,{\rm fs}$, plus an incoming probe wave $\delta A(\tau)$ living on the solid black branch (with $D < 0$ and $\Delta\omega >0$) in Fig.~\ref{fig:dispersion_SiN}, 
centered on the detuning 
frequency $\Delta\omega_{\rm probe} = 0.78 \,{\rm PHz}$ which is %
chosen to coincide with the maximum of the predicted emission rate 
(see the right panel of Fig.~\ref{fig:Spec_v_tau0}).
The relative amplitude of the probe wave is such that the peak power ratio $P_{\rm probe}/P_{\rm soliton} \simeq 7 \times 10^{-5}$, and it is described by a narrow 
Gaussian envelope of width $0.005\,{\rm PHz}$ in $\omega$-space.

\subsubsection{Space-time description} 

To clearly separate the evolution of the field configurations describing the soliton from those describing the probe, the simulation is performed twice using as initial 
configurations $A^{\rm in}_\pm(\tau) = A_{0}(\tau) \pm \delta A(\tau)$. 
Since the initial soliton profile is exactly the same while the 
probe wave flips sign, this allows us to extract the part of the field which is 
linear in the probe wave by taking the half difference of the two runs, 
while the part of the field which is independent of the probe wave is found 
by taking their half sum.  (Terms of quadratic order and higher in the probe field are 
negligible, 
as we have verified by using $A_{0}(\tau)$ as initial condition.) 

The difference is represented in light gray 
in Fig.~\ref{fig:stimulated}, while the sum 
is shown in thick black. 
Both have been normalized such that the peak value of the initial soliton field is $1$, both in real (left column) and in Fourier space (right column). 
On the left side, 
we clearly see the emergence of two wavepackets after the interaction with the soliton: one is short in duration with a large amplitude, 
and represents the transmitted probe wave, while the other is long in duration with a small amplitude, 
and represents the stimulated negative-norm wave on the solid blue branch (with $D>0$ and $\Delta\omega < 0$) of Fig.~\ref{fig:dispersion_SiN}. %
This can be verified by considering the right plots which represent, on a logarithmic scale,
the squared norm of the Fourier component $\tilde A(\Delta \omega)$ as a function of the detuning frequency.
It is interesting to notice that the small shift (to the left) of the negative 
detuning frequency of the stimulated emission on the blue 
branch is here recovered: namely, as was observed in Fig.~\ref{fig:betasq_v_tau0}, it 
is maximal for $\Delta\omega \simeq -0.84\,{\rm PHz}$  while 
that of the incoming probe 
is $\Delta\omega_{\rm probe} 
= 0.78\,{\rm PHz}$.

In the plots of the right column of Fig.~\ref{fig:stimulated}, 
we also clearly see the production of the ``idler'' (reflected wave), albeit with a much smaller amplitude,
in good agreement with the reflection coefficient (shown in dotted red in the upper right plot of Fig.~\ref{fig:MoreScattering})
for $D$ near 
$|D_{\rm cross}| = 28 \, {\rm mm}^{-1}$.

\begin{figure}
\includegraphics[width=0.45\columnwidth]{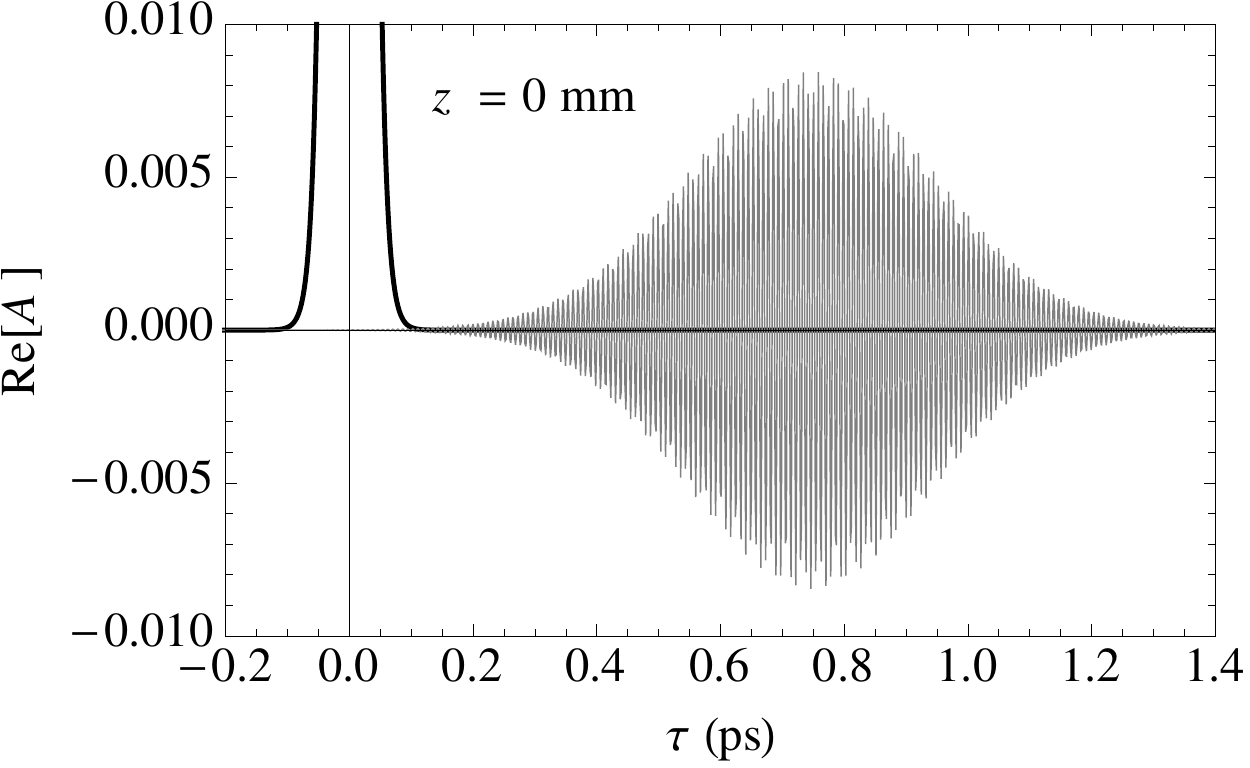} \, \includegraphics[width=0.45\columnwidth]{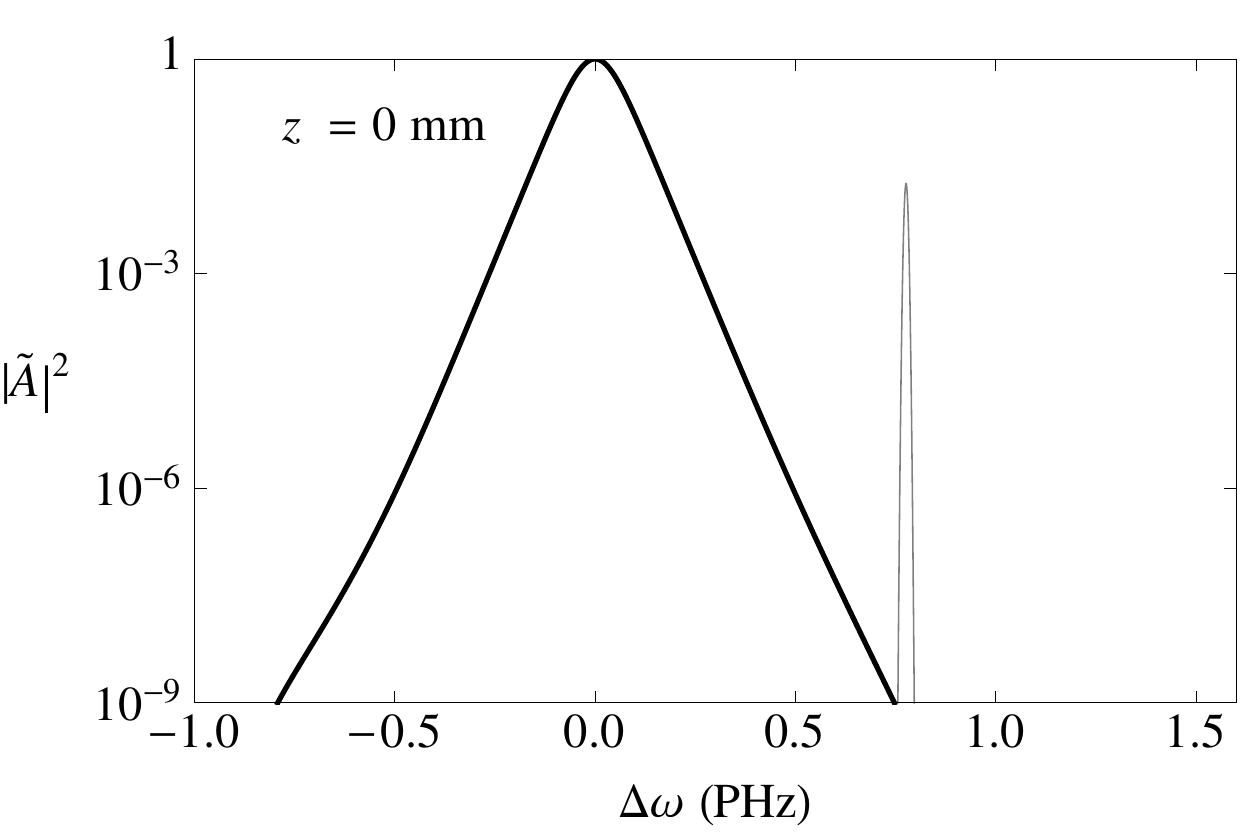} \\
\includegraphics[width=0.45\columnwidth]{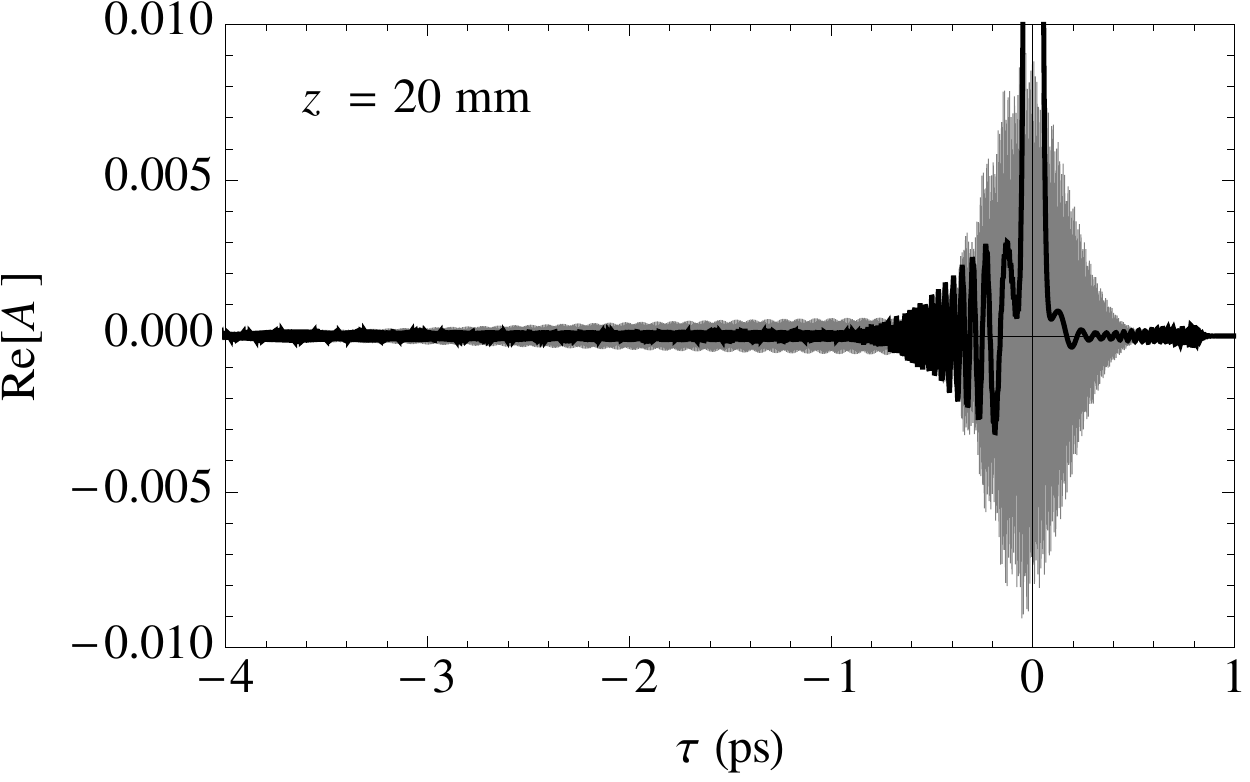} \, \includegraphics[width=0.45\columnwidth]{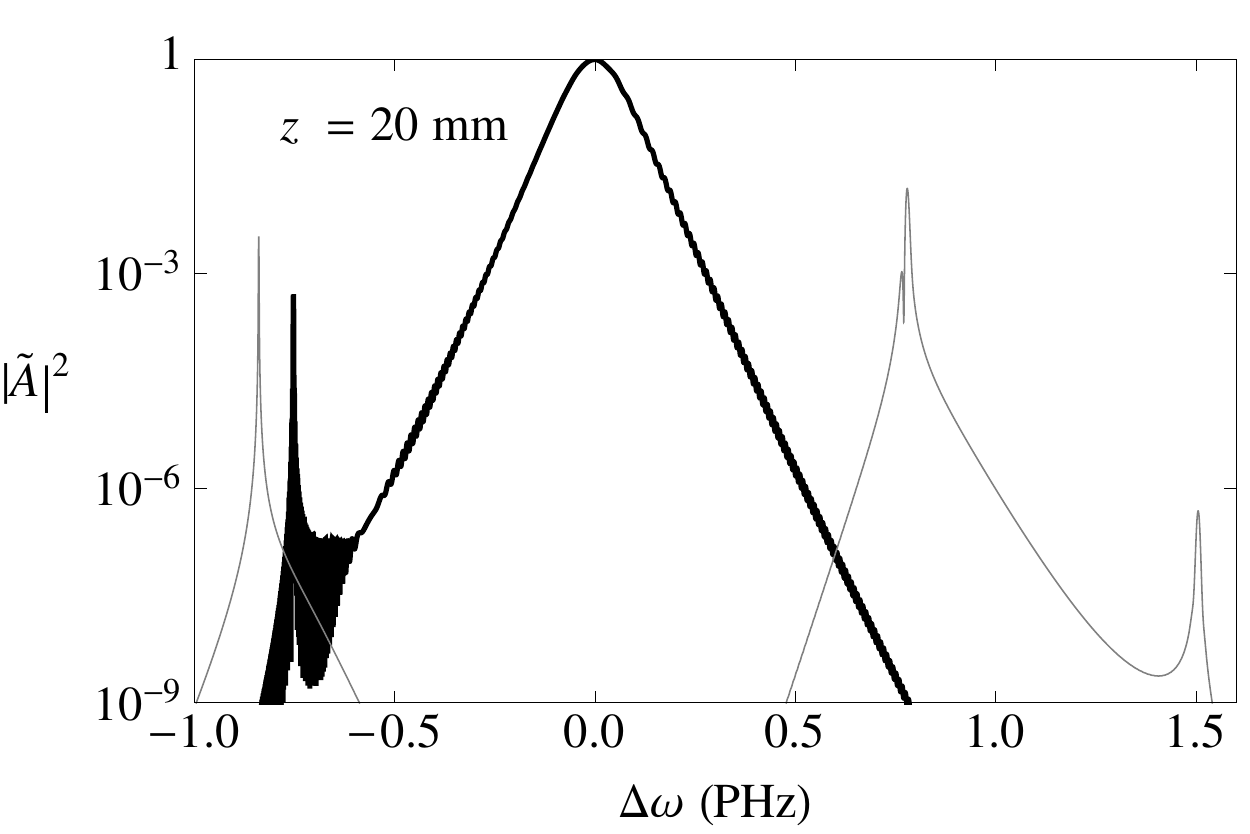} \\
\includegraphics[width=0.45\columnwidth]{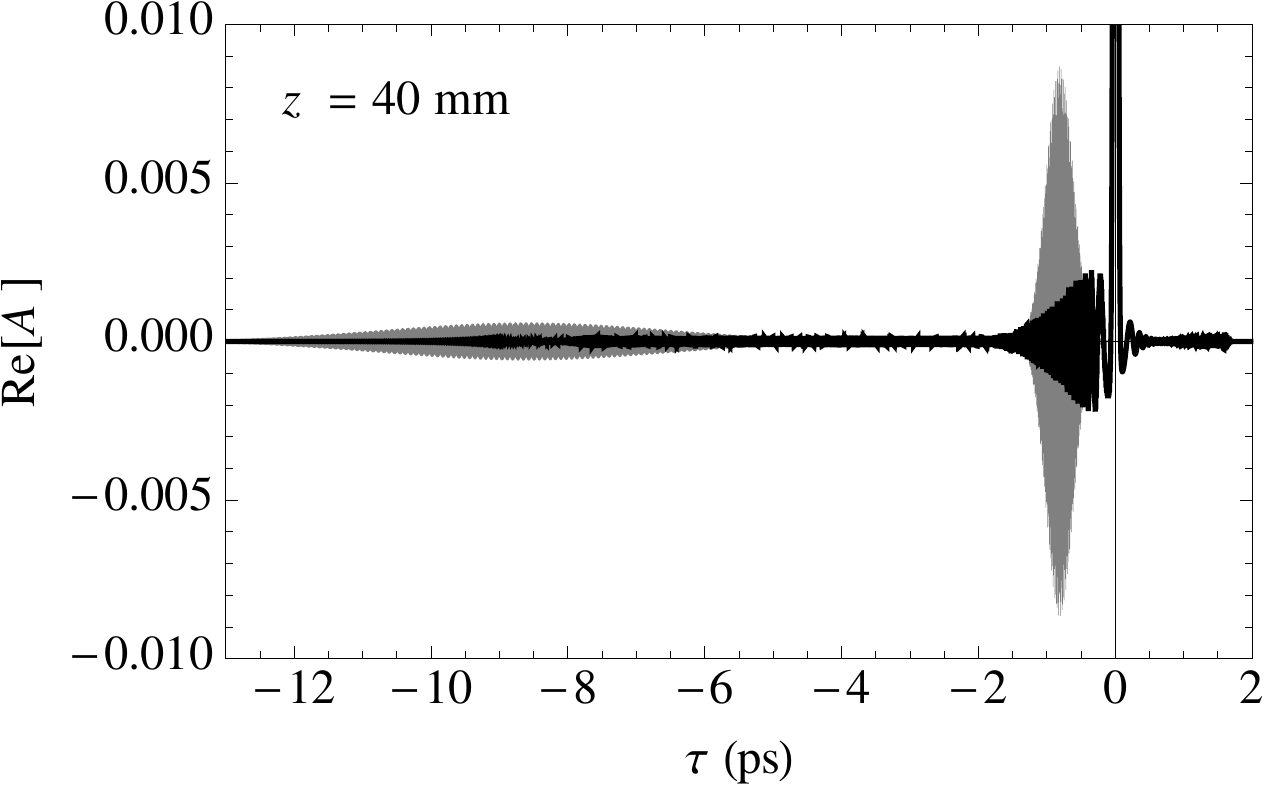} \, \includegraphics[width=0.45\columnwidth]{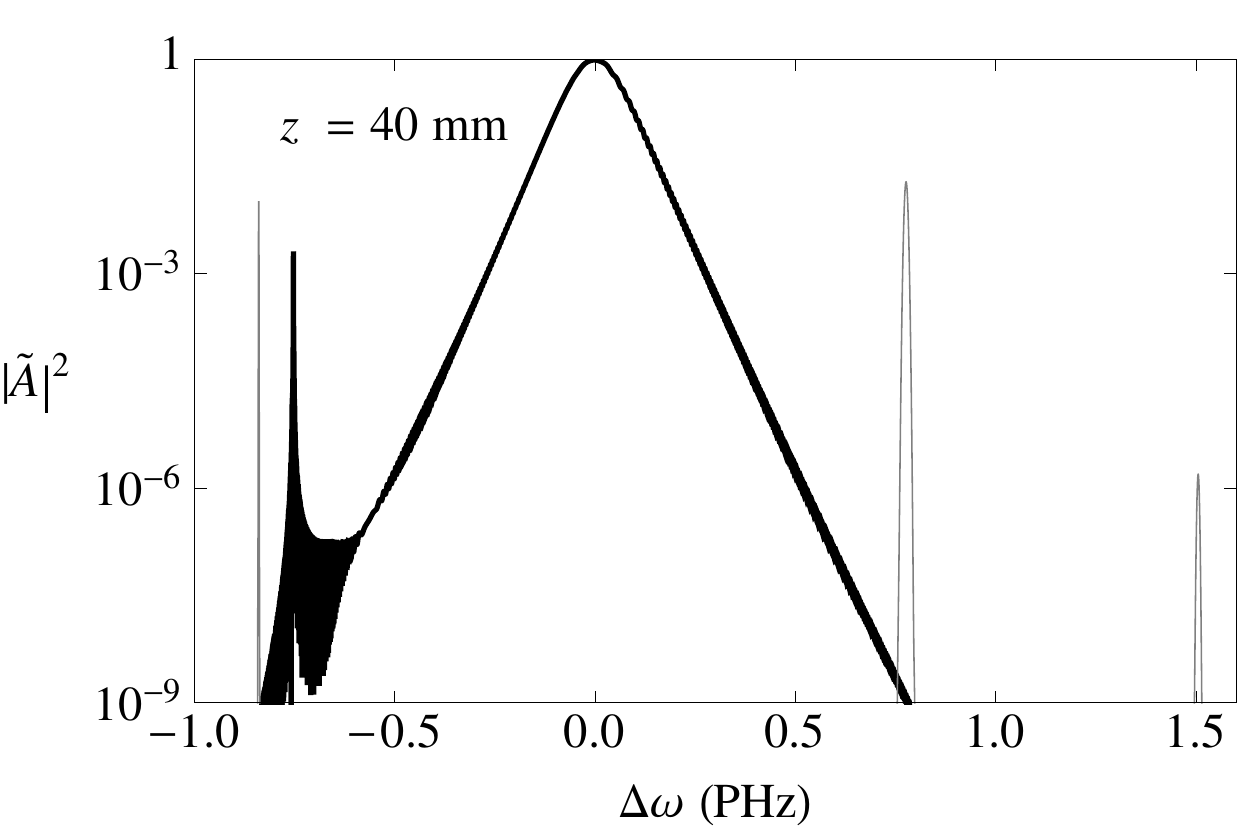}
\caption{Evolution of the full fields $A(\tau,z) = A_{0}(\tau,z) \pm \delta A(\tau,z)$, 
where $A_{0}$ initially contains the soliton field (of duration $\tau_{0} = 10\,{\rm fs}$) and $\delta A$ contains a wavepacket incident on the soliton. 
The mean field $A_{0}$ and the perturbation $\delta A$ are extracted by taking the half sum and difference, respectively, of the two runs. 
The left column shows the real part of both $\delta A$ (in light gray) 
and $A_{0}$ (in thick black), 
in units where the peak of the initial soliton field $A_{0}(\tau,z=0) = 1$. 
On the right is shown the squared magnitude of their Fourier transforms 
in $\Delta\omega$-space, again in units where the peak of $\tilde{A}_{0}(\Delta\omega, z=0) = 1$. 
The rows correspond to different values of $z$: $0\,{\rm mm}$ (top row), $20\,{\rm mm}$ (middle row) and $40\,{\rm mm}$ (bottom row). 
\label{fig:stimulated}}
\end{figure}

\subsubsection{Cherenkov radiation}

Focusing on the thick black 
curves which describe the background configurations, 
we observe, in 
both columns of Fig.~\ref{fig:stimulated}, 
the emission of Cherenkov radiation~\cite{Akhmediev-Karlsson-1995}. 
The CR is emitted in both directions, as can be understood by noting the opposite 
slopes at the two $D=0$ roots of the dispersion relation (see Fig.~\ref{fig:dispersion_SiN}).  
However, since the smaller frequency difference occurs for the root which has a negative group velocity, the CR emission rate should be greater on the left side, and this is borne out by Fig.~\ref{fig:stimulated}. 
The larger CR emission occurs at $\Delta\omega \simeq -0.75\,{\rm PHz}$, 
in good agreement with the $D=0$ intercept. 
This emission 
is due to the fact that the soliton configuration injected in the WG is not an exact solution of Eq.~(\ref{eq:wave_eqn_full}).  However, since the soliton is an exact solution of Eq.~(\ref{eq:non-lin-simplified}),
the modification $\delta A_{0}(\tau,z)$ of the background soliton solution obeys a ``forced'' equation with the following structure (see Eq.~(\ref{eq:wave_eqn_lin})):
\begin{equation}
\left\{ -i\partial_{z} - \left[ B\left(i\partial_{\tau}\right) - \delta\beta_{0} + 2 \gamma \left|A_{0}\right|^{2} \right] \right\} \, \delta A_{0} = \Delta B\left(i\partial_{\tau}\right)\,A_{0}
\end{equation}
where 
$\Delta B(\Delta\omega) \doteq B(\Delta\omega) - B_{2}(\Delta\omega)$ is the modification of the wave number with respect to the quadratic approximation of the dispersion relation 
mentioned above Eq.~(\ref{eq:non-lin-simplified}). (In obtaining this equation, we have also applied the rotating wave approximation, which amounts to neglecting the last term of Eq.~(\ref{eq:wave_eqn_lin}).) 
As a result, the CR so produced is described by a coherent (displaced) state.~\footnote{This forcing should be clearly distinguished from ``soft'' squeezing due to FWM discussed in~\cite{Tran}, as well as from its ``hard'' version 
discussed in Ref.~\cite{Rubino-et-al-2012}. 
As was noticed in~\cite{Efimov-2005}, the CR always has the same polarization as the soliton, indicating that the dominant contribution is from the forced channel. 
It is also worth mentioning that forced and squeezed 
channels exist when considering undulations downstream from an obstacle in a flume~\cite{Shen1993}, and in transonic flows in atomic BEC~\cite{Busch-2014}. 
}

\subsubsection{Local decomposition of the probe field} 

Since the nonlinearities 
governed by the last term in Eq.~(\ref{eq:wave_eqn_full}) are small, 
typically $\gamma |A_0^{\rm peak}|^2 /|D_{\rm cross}| \sim 0.05 $ 
for a soliton with  $\tau_{0} = 10\,{\rm fs}$, it is meaningful to decompose 
the probe field 
$\delta A(\tau , z)$ at each $z$ into its ``instantaneous'' Fourier components $\tilde{\delta A(\Delta \omega, z)}$. 
One can then integrate the squared magnitude of 
$\tilde{\delta A(\Delta \omega, z)}$ over the relevant detuned frequency windows (defined 
by the asymptotic dispersion relation of Fig.~\ref{fig:dispersion_SiN}) to obtain the evolution in $z$
of the transmitted, reflected and stimulated parts of the probe. 
More precisely, 
the instantaneous squared norm 
of modes living on the solid black branch (with $\Delta\omega >0$ and $D<0$) of Fig.~\ref{fig:dispersion_SiN}, called $\left|\alpha(z)\right|^{2}$, 
is found by integrating $\left|\widetilde{\delta A}\left(\Delta\omega\right)\right|^{2}$ over the interval $\Delta\omega \in \left[0\,,\,1.2\right]\,{\rm PHz}$, and dividing by the initial (i.e. $z=0$) 
value of this integral. Similarly the reflected $\left|R(z)\right|^{2}$,
and stimulated $\left|\beta(z)\right|^{2}$, contributions are obtained by integrating over the relevant domains in 
$\Delta\omega \in \left[0\,,\,1.2\right]\,{\rm PHz}$, and dividing the results by the initial 
value of this integral of $\left|\alpha(z)\right|^{2}$. By construction, when evaluated for sufficiently large $z$ so that the scattering has taken place, these three functions give the squared norm of the asymptotic scattering coefficients (integrated over the narrow width of the probe field).

These three functions  
are shown by solid lines in the left column of 
Figure~\ref{fig:stimulated_norm}.
One clearly sees that the mode mixing starts for $z \sim 10$mm and essentially ends for $z \sim 25\,$mm, as 
one might have expected from the space-time evolution of the probe field.
To quantitatively verify that  this decomposition is meaningful, we represent 
on the lower left panel of Fig.~\ref{fig:stimulated_norm} the quantity $1 - (\left|\alpha(z)\right|^{2}+ \left|R(z)\right|^{2}-  \left|\beta(z)\right|^{2})$ which contains the quadratic combination 
entering the unitarity relation, see Eq.~(\ref{eq:unitarity}). 
We note that the maximal value of this deviation $\sim 10^{-8}$ is reached for $z \simeq 19$mm, in the ``middle'' of the scattering, and is much smaller than both $\left|R(z)\right|^{2}$ and $\left|\beta(z)\right|^{2}$. Moreover, the final
value reaches a constant $\sim 2 \times 10^{-12}$ 
which is much smaller than any of the 
norms presented.~\footnote{Though the coefficient which describes scattering onto the dot-dashed yellow branch of the dispersion relation (that with $\Delta\omega < 0$ and $D<0$ in Fig.~\ref{fig:dispersion_SiN}) %
has been neglected here, this will not account for the final error because the results of the linear treatment for a soliton duration $\tau_{0} = 10\,{\rm fs}$  indicate that its squared magnitude reaches a maximum on the order of $10^{-17}$, five orders of magnitude smaller than the final error in the bottom left panel of Fig.~\ref{fig:stimulated_norm} and fifteen orders of magnitude smaller than the maximum of the squared amplitude describing stimulation of the blue branch.} 
Concomitantly, we verified that the asymptotic 
values of each contribution agree quite well with the theoretical predictions of Figs.~\ref{fig:Scattering} and~\ref{fig:MoreScattering} at the given value of $D= |D_{\rm cross}|$.

\subsubsection{Conservation of photon number and small dissipation}

Because we now 
deal with the total field $A_0 + \delta A$, we can verify that the creation of photon pairs induced by the probe field is accompanied by a corresponding decrease of the residual number of photons in the soliton.  
To this end, in the right column of Figure~\ref{fig:stimulated_norm} are shown plots related to the norm $N_{A}$ of Eq.~(\ref{A1-A2}), whose constancy in $z$ (when applied 
to the {\it total} field) implies conservation of the total number of photons.  As for the instantaneous norm content of the perturbation field, the integration is performed in $\Delta\omega$-space, which is split into a ``soliton'' region ($-0.6\,{\rm PHz} < \Delta\omega < 0.6\,{\rm PHz}$) and a ``probe'' region ($\Delta\omega > 0.6\,{\rm PHz}$); the region $\Delta\omega < -0.6\,{\rm PHz}$ is not separately presented. 
In the upper and middle panels are shown the changes in photon number in these two regions, relative to the initial number of photons in the probe; these are relevant quantities because the scattering is (to a very good approximation) linear in the probe field.  

In the upper plot, the solid curve shows the increase in the number of photons in the transmitted part of the probe field. 
One verifies that the increase (as a function of $z$) of 
the quantity $N^{\rm probe}(z)/N_{\rm in}^{\rm probe} - 1$ closely agrees with $\left|\alpha(z)\right|^{2} - 1$ which is represented in the upper left plot. 

In the middle plot, the dashed curve shows the 
{\it total} decrease of the number of photons in the soliton, which includes a steady decrease due to the CR emission.  To correct for this, we subtract $N^{\rm sol}(z)$ extracted from a simulation where the initial profile of $A_{0}$ is exactly the same while $\delta A$ is set to zero.  The result is shown by the solid curve.  One observes that the remaining decrease exactly compensates for {\it twice} 
the increase of $N^{\rm probe}$, which is as expected since each photon added to the transmitted probe field is accompanied by a partner in the negative-norm wave. 
Indeed, we have checked that the corresponding curve for the region $\Delta\omega < -0.6\,{\rm PHz}$, after having subtracted the CR contribution in the same manner as for the soliton, is exactly the same as that showing the increase of $N_{A}$ for the probe (i.e. the solid curve on the upper panel). 

The lower right panel of Fig.~\ref{fig:stimulated_norm} 
shows the relative change of $N_{A}(z)$ which gives the total number of photons (soliton $+$ probe). 
Its final value is of order $10^{-12}$, as was the final value of the deviation shown in the lower left column. 
We believe that these two residual deviations are due to accumulation of numerical errors. 
To validate this conjecture, we have increased the numerical step size $\Delta z$ by a factor of $2$ (i.e. from $0.8\times 10^{-3}\,{\rm mm}$ to $1.6\times 10^{-3}\,{\rm mm}$). We observed that the change in the total number of photons is {\it reduced} by nearly a factor of $2$,
while the deviation from unitarity 
shown in the lower left panel of Fig.~\ref{fig:stimulated_norm})
increases 
by a factor of about $20$. 

\begin{figure}
\includegraphics[width=0.45\columnwidth]{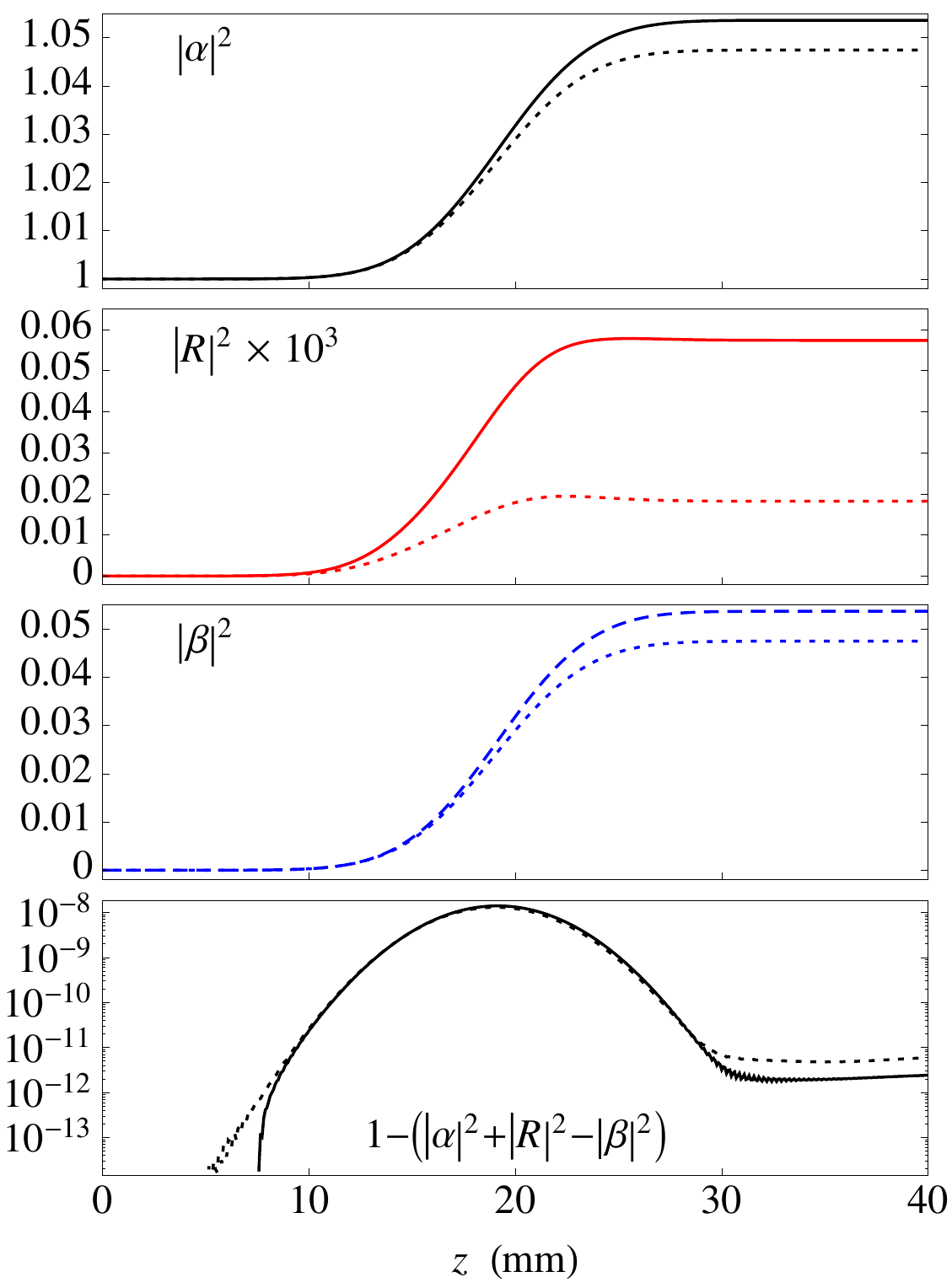} \, \includegraphics[width=0.45\columnwidth]{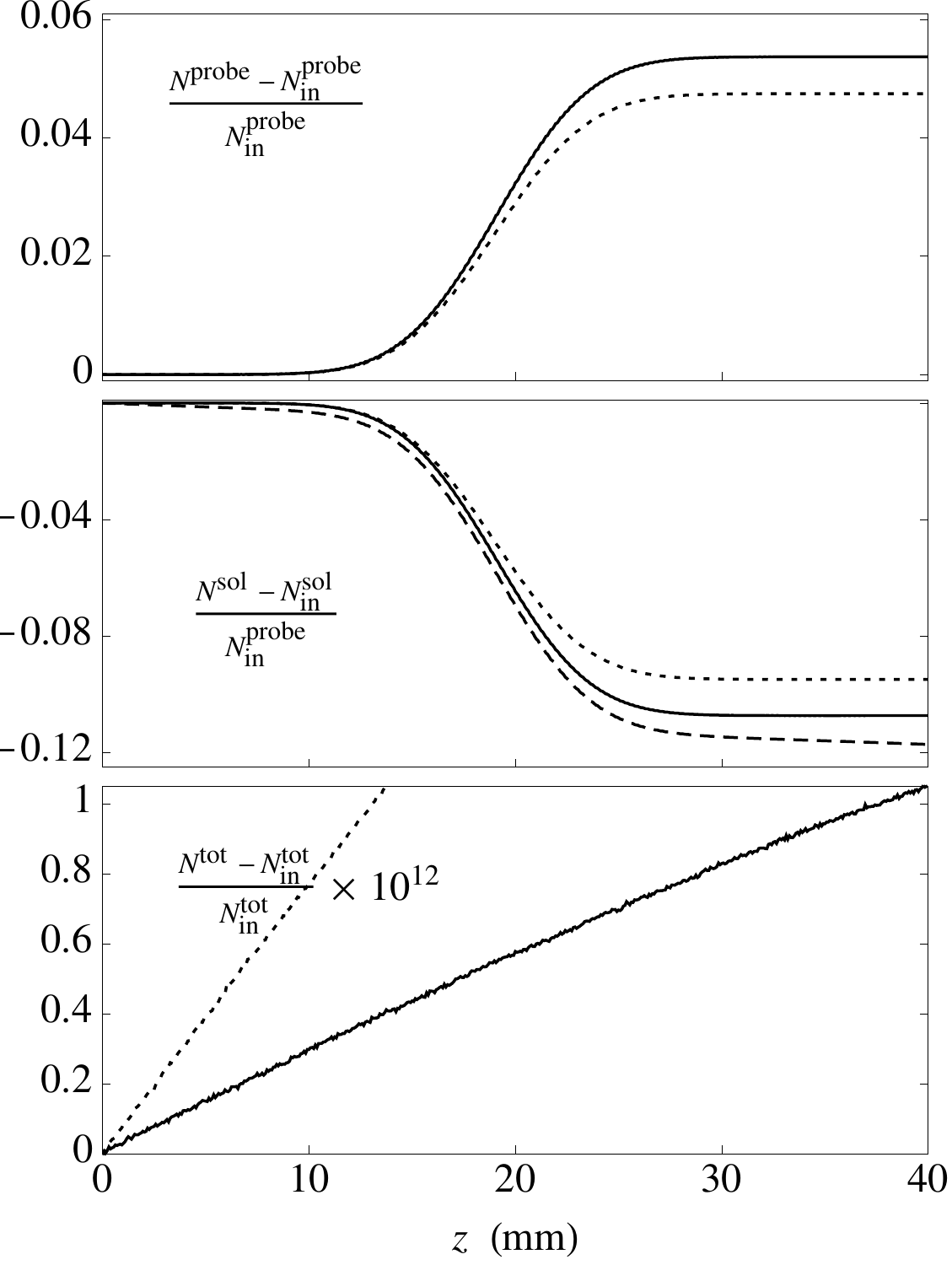}
\caption{
Conservation laws relevant to the scattering in Fig.~\ref{fig:stimulated}.
In the left column are shown the ``instantaneous'' squared magnitudes of the scattering coefficients, 
as described in the text. 
In the lower left panel is shown the ``instantaneous'' 
difference of the total norm from 1, see Eq.~(\ref{eq:unitarity}). 
The larger deviations at intermediate $z$ arise due to 
the nonlinearities.  
In the right column are shown the variations in 
$N_{A}$ (see Eq.~(\ref{A1-A2})) applied to 
the {\it total} field, 
and restricted to 
different regions of $\Delta\omega$-space.  The upper and middle panels show, respectively, the change in $N_{A}$ of the probe and of the soliton. 
The dashed curve in the middle panel takes into account both the effect of the probe on the soliton and the 
Cherenkov emission, while 
the solid curve shows the probe contribution only. 
In the lower right panel is shown the relative variation of the total photon number, integrated over all frequencies. One observes that it is 
constant up to one part in $10^{12}$, which indicates that it stems from numerical errors (given the much larger values of the scattering coefficients). 
In all panels, the dotted curves are for simulations with exactly the same initial conditions, but including a linear loss $\Gamma$ 
such that $10\%$ of the energy is lost after $40\,{\rm mm}$.  To compensate for the trivial steady decrease 
induced by the loss, we have multiplied the squared amplitudes by $e^{2 \Gamma z}$. As explained in the text, the 
residual effects visible in the plots 
are essentially accounted for 
by the reduction of the soliton power caused by $\Gamma$. 
\label{fig:stimulated_norm}}
\end{figure}

To complete this study, we have included an overall linear loss in Eq.~(\ref{eq:wave_eqn_full}). The resulting quantities
are shown by dotted lines in all panels 
of Fig.~\ref{fig:stimulated_norm}.
The loss rate $\Gamma$ was chosen such that $10\,\%$ of the total energy is lost by the time the soliton has passed through $40\,{\rm mm}$ of the waveguide~\footnote{This corresponds to a loss rate of $0.114\,{\rm dB/cm}$, a value at least 20 times higher than that reported for state-of-the-art silicon nitride waveguides~\cite{Ji-2017}.}. 
In the plots, the loss rate has been corrected for by multiplying all curves by the exponential factor $e^{\Gamma z}$.  The residual decrease seems to be due to the fact that, on account of having lost some energy by the time of the interaction between the soliton and the probe, the soliton has slightly decreased in power (increased in duration), so that the scattering coefficients are smaller in magnitude.  This interpretation is confirmed by additional simulations in which the initial soliton power is chosen so that, at the time of the interaction, the soliton width is equal to its value in the lossless simulations.  In this case, the ``instantaneous'' scattering coefficients become almost indistinguishable from their values in the lossless case. 
This establishes the robustness of our results against the introduction of a small linear loss.

\subsection{Spontaneous scattering: spectrum and entanglement
\label{non-sep}}

We now turn to the numerical simulation of spontaneous emission from the soliton, which is the only physical effect when the modes (besides those describing the soliton) are initially in their vacuum state (leaving aside the emission of CR). We aim at obtaining both the power spectrum characterizing the emitted pairs, as well as the correlations 
between the two photons in each pair. 
Since we are effectively dealing with a $U(1,1)$ mode mixing when considering the significant pair production for $D$ near $|D_{\rm cross}|$, 
the strength of these correlations should be large enough to guarantee that the bipartite state of the emitted pairs is nonseparable~\cite{Busch:2014bza}.

As a preliminary attempt~\footnote{The analysis presented here is work in progress.  We are grateful to Florent Michel for recent discussions about this method of handling vacuum fluctuations in nonlinear systems.} to characterize the strength of the correlations, 
we have found it useful to adopt the doublet formalism at the level of the nonlinear equation~(\ref{eq:wave_eqn_full}).  That is, 
guided by the 
structure of Eq.~(\ref{eq:wave_eqn_2b2}), we consider the following pair of nonlinear coupled equations:
\begin{eqnarray}
-i\partial_{z}A_{+} & = & B\left(i\partial_{\tau}\right) A_{+} + \gamma A_{-}^{\star} A_{+}^{2} \,, \nonumber \\
i\partial_{z}A_{-}^{\star} & = & B\left(-i\partial_{\tau}\right) A_{-}^{\star} + \gamma A_{+} \left(A_{-}^{\star}\right)^{2} \,.  
\label{non-lin-system}
\end{eqnarray}
This system reduces 
to Eq.~(\ref{eq:wave_eqn_full}) whenever $A_{+} = A_{-}= A_0$. 
In addition, when writing the doublet $A = [A_{+}, A_{-}^*]$ as 
a common background term plus perturbations on top of that background, 
i.e., $A= [A_0 + \delta A _{+}, A_0^* + \delta A _{-}^*]$, 
then to linear order in 
the perturbations 
the system reduces to Eq.~(\ref{eq:wave_eqn_2b2}). Hence the system here considered correctly describes both the nonlinear evolution (in $z$) of the background configurations $A_0(z,\tau)$ and the linear evolution of the perturbations described by the doublet
$w = [\delta A _{+},\delta A _{-}^*] $. 

The expressions on the right-hand side of 
Eq.~(\ref{non-lin-system}) are the Euler equations of the following Lagrange density: 
\begin{equation}
L(z) =  {\rm Re} \left\{ \int {\rm d}\tau \, \left(-i A_{-}^* \partial_{z}A_{+} - K\right) \right\} \,,
\end{equation}
where the new generator of $z$-translations is the real part of (cf. Eq.~(\ref{K_A}))
\begin{equation}
K  = \int {\rm d}\tau \left\{ A_-^* B\left(i\partial_{\tau}\right)A_+ + \frac{\gamma}{2} (A_-^* A_+ )^{2}\right\} \,. 
\end{equation}
Therefore, 
for any pair of coupled solutions $(A_{+}, A_{-}^{\star})$, 
the generalized version of the 
scalar product of Eq.~(\ref{A1-A2}), i.e., ${\rm Re} \left\{ \int\,{\rm d}\tau\, A_{-}^{*} A_+ \right\}$,
is 
identically conserved, 
as follows from the $U(1)$ invariance of $L$ under $A_\pm \to A_\pm e^{i \phi}$.

To introduce the notion of vacuum fluctuations, 
we use the following asymmetric initial conditions:
\begin{eqnarray}
A_{+}(\tau,z=0) & = & A_{0}(\tau) \,, \nonumber \\
A_{-}^{\star}(\tau,z=0) & = & A_{0}^{\star}(\tau) + \delta A_{-}^{\star}(\tau) \,.
\label{eq:spontaneous_init_conds}
\end{eqnarray}
The field $A_{0}(\tau)$ represents 
the initial configuration of the coherent background state, which is taken to be the soliton solution of Eq.~(\ref{eq:soliton}) in the forthcoming simulations
(exactly as in the top row of Fig.~\ref{fig:stimulated}). 
The initial value of the perturbation 
is described by the doublet $\bar w_{\rm in} 
= [0,\delta A_{-}^{*}(\tau)]$, 
which 
represents a configuration describing vacuum fluctuations. 
By this we mean the following. The upper component of the doublet identically vanishes, as 
is the case in the second line of Eq.~(\ref{eq:unit-norm-mode}). 
The lower component is given by~\footnote{In the simulations, we apply effective limits to the integral to avoid exciting unphysical branches of the dispersion relation, which appear when large values of $D$ are folded back into the first Brillouin zone defined by the step size $\Delta z$.  We set to zero all modes with $\left|D\right| \gtrsim 40\,{\rm mm}$, corresponding to the integral range $\Delta\omega \sim \left[ -1\,{\rm PHz}\,,\, 2\,{\rm PHz} \right]$.  On top of this, after taking the integral of Eq.~(\ref{eq:vac_fluc}) to get the initial fluctuations in $\tau$-space, we also apply a smooth (double tanh-shaped) filter so as to remove $\delta A_{-}^{\star}$ in the range $\tau \sim \left[-0.2\,{\rm ps} \,,\, 0.2\,{\rm ps}\right]$. This allows us to study only the forward evolution of vacuum fluctuations that are not initially on the soliton; without the filter, we observe a significant transient expulsion of fluctuations from the soliton.} 
\begin{equation}
\delta A_{-}^*(\tau) = \int\! d\Delta \omega\, a^*_{\Delta \omega} { e^{i \Delta \omega \tau} \over 2\pi}\,  . 
\label{eq:vac_fluc}
\end{equation}
In this expression, each amplitude $a^*_{\Delta \omega}$ is a random complex number governed by a Gaussian ensemble
with a vanishing mean value and a variance which is independent of $\Delta\omega$. 
These therefore encode 
the creation of 
one 
photon in each Fourier mode.~\footnote{Planck's constant $\hbar$ provides an absolute normalization for the field 
fluctuations, 
and an absolute meaning as to the number of photons in any Fourier mode.  However, if effects which are nonlinear in the fluctuation $\delta A_{-}$ are negligible, we may scale it up in magnitude for computational convenience, so long as the photon number is calculated relative to its initial value, this being always set to $1$.} 
The contents of this field are defined only probabilistically, and in principle the extraction of mean values requires averaging over many realizations of $\delta A_{-}^{*}$, as is done when using the truncated Wigner approximation~\cite{Castin-2002,Robertson:2018gwi}, 
(TWA).

The advantage of this approach is that, as for the linearized wave equation, anomalous scattering occurs between the two components of the doublet, while normal scattering does not.  Therefore, only the spontaneously produced photons will appear in the upper component $A_+$, making their emission rates easy to extract. In fact, 
comparing this treatment to the TWA, 
the extraction of small numbers of photons 
is here much simpler, 
because in the TWA 
one deals with the expectation value of a symmetrized product of the field operator (that is, an anti-commutator). As a result, to get the emission rate
from vacuum, one should remove the contribution (of the commutator) which is present even in the absence of pair creation. When the emission rate is very small, this method requires many realizations, as it 
involves extracting 
a small difference between two terms dominated by a contribution which is independent of the number of pairs produced.

\subsubsection{Emission spectra}

In practice we proceed by first evaluating the Fourier components of $A_{+}(\tau)$ 
at different values of late $z$. From these, we 
extract an average creation rate per unit $z$ per unit $\Delta\omega$ that we 
compare with the predictions of the linearized wave equation already shown in Fig.~\ref{fig:Spec_v_tau0}.  It should be noted that, for any single discrete value of $\Delta\omega$, the Fourier components oscillate rapidly when varying $\Delta \omega$ since we deal with a single realization of a Gaussian ensemble. 
These oscillations can be suppressed either by 
considering an ensemble of initial realizations of $\delta A_{-}$, as one does when using the truncated Wigner approximation~\cite{Castin-2002,Robertson:2018gwi}, 
or, since we use a high resolution in $\Delta\omega$, by binning over many Fourier components.  For the sake of efficiency, we adopt the latter approach.  The results are shown in Figure~\ref{fig:spontaneous_spectra}.  Even after binning, the numerical results show quite large oscillations when using a linear scale, but the convergence towards the predictions 
of the linearized equation becomes evident when plotted on a logarithmic scale. Hence the above method works efficiently 
in the sense that 
all Fourier components (and hence all values of the detuning parameter $D$) are 
dealt with at once. 

\begin{figure}
\includegraphics[width=0.45\columnwidth]{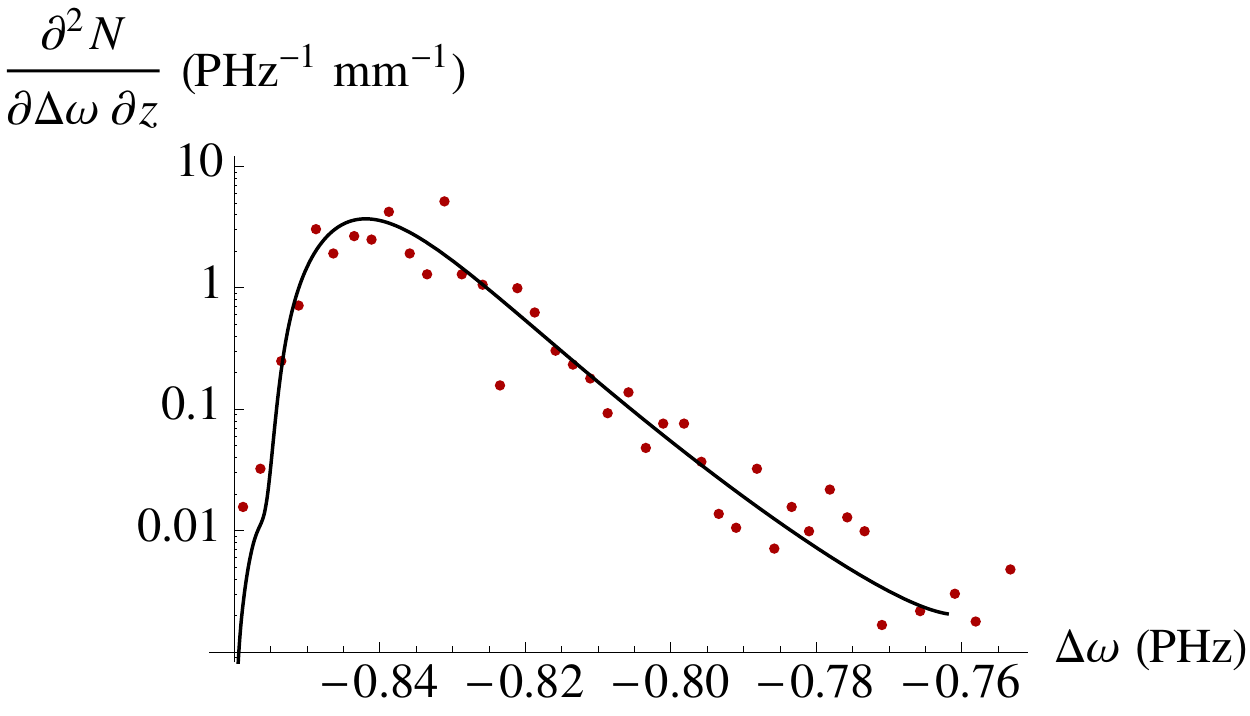} \, \includegraphics[width=0.45\columnwidth]{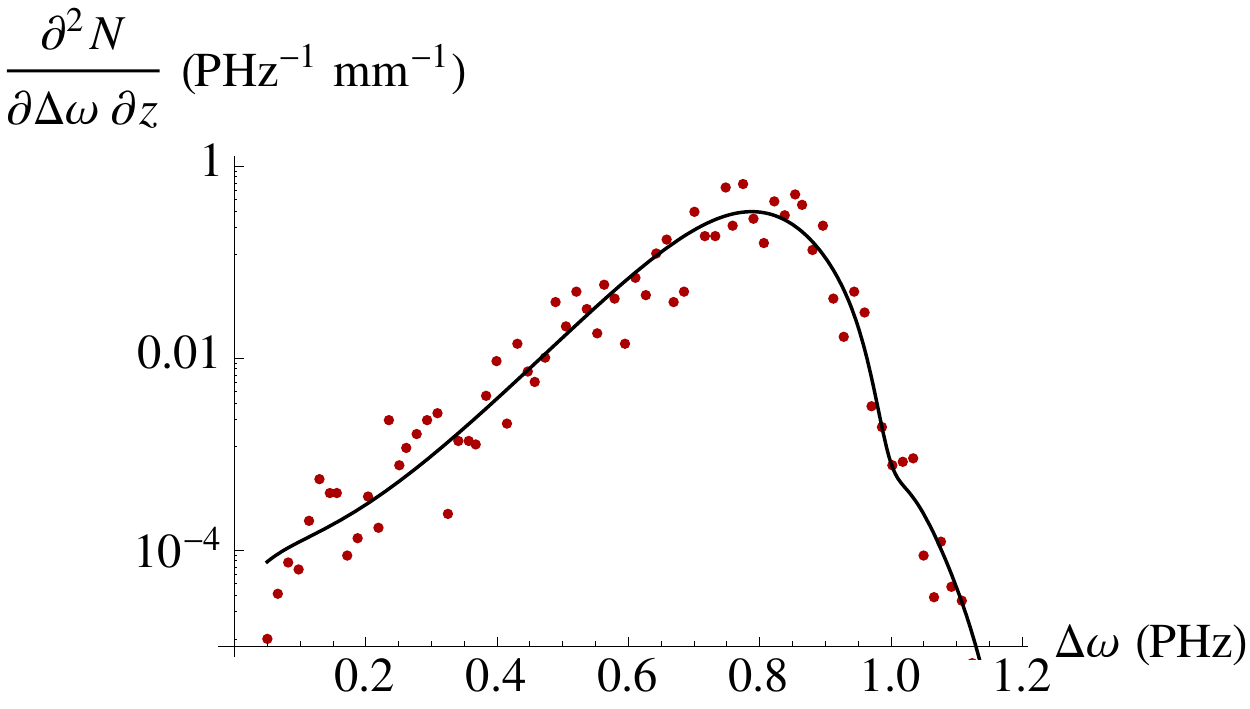}
\caption{Emission rates of spontaneously created photon pairs.  We use the Fourier transform in $\tau$ of the linear perturbation $\delta A_{+}(z,\tau)$ generated in the upper component of the doublet when the lower component is fed by initial vacuum fluctuations (see the initial conditions given in Eqs.~(\ref{eq:spontaneous_init_conds})). 
The soliton is removed by applying a window filter which vanishes for $\tau$ close to zero and is equal to $1$ far from the soliton. 
We take an average growth rate of the spectrum with respect to $z$, and bin many discrete frequencies together to smooth out the profile (which is highly oscillatory for a single realization).  Since the two peaks have very different spreads in frequency, we use two different binning rates: the discrete frequencies used in the simulation have a spacing of about $1.26 \times 10^{-4} \, {\rm PHz}$, and we bin these into groups of $40$ for the left plot, and groups of $600$ for the right plot.  The black curves show the predictions from the linearized treatment at fixed $D$; they are exactly the solid curves already presented in Fig.~\ref{fig:Spec_v_tau0}.
\label{fig:spontaneous_spectra}}
\end{figure}

In Figure~\ref{fig:spontaneous_DFT}, we represent the squared magnitude of the positive-norm component $A_{+}(\Delta\omega,D)$, the soliton having been removed using a filter which vanishes for $\tau$ near zero. 
We have applied a 
Fourier transform in both 
$\tau$ and $z$ 
so as to get the power spectrum of the spontaneously created pairs in the dispersion relation plane, see Fig.~\ref{fig:dispersion_SiN}.
To avoid spurious signals which are very wide in $D$, we have applied a regularizing function which vanishes at $z=0$ and $z=40\,{\rm mm}$, thereby removing a strong discontinuity when the data is 
treated (as we do) 
as periodic in $z$. 
One observes that the field configurations here considered only 
live along the positive-norm branch. This is precisely 
the 
sought-for result from having put to zero the initial value of $\delta A_{+}$ 
when using a doublet $(A_+,\, A_-)$ to describe and propagate configurations describing the soliton plus vacuum fluctuations. 
It is clear that the strongest signals appear on the solid black ($\Delta\omega>0$, $D<0$) and solid blue ($\Delta\omega<0$, $D>0$) branches of the dispersion relation, with a faint signal on the dotted red (idler) branch that decreases with increasing $\Delta\omega$. 

\begin{figure}
\includegraphics[width=0.6\columnwidth]{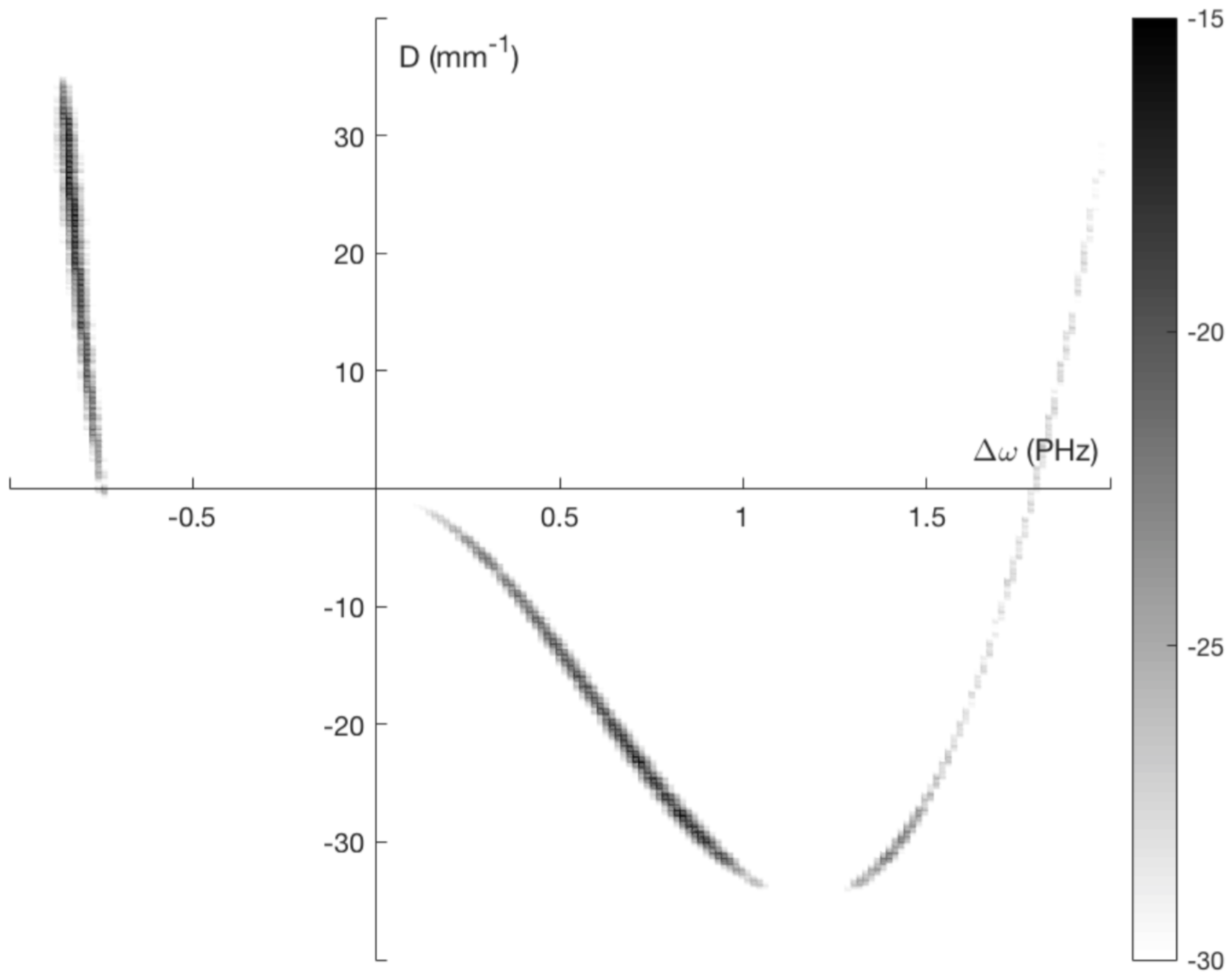}
\caption{Squared magnitude of the Fourier transform in both $z$ and $\tau$ of the linear perturbation $\delta A_{+}(z,\tau)$, for the same simulation used in Fig.~\ref{fig:spontaneous_spectra} 
(i.e. where only the lower component of the doublet is fed by initial vacuum fluctuations). 
The soliton is removed by applying a 
filter which vanishes for $\tau$ close to zero and is equal to $1$ far from the soliton. 
We use a logarithmic scale to make most of the relevant data visible. As expected the signal is found along the positive norm dispersion relation, and is maximal near the mode crossing taking place for $|D_{\rm cross} |= 28.4 {\rm mm}^{-1}$.
\label{fig:spontaneous_DFT}}
\end{figure}

\subsubsection{Correlations in emitted pairs}

Finally, we study the correlations between the photons emitted in pairs. To this end, we
consider 
the connected part of the first-order correlation function 
\begin{equation}
g_{1}\left(\Delta\omega\,,\,\Delta\omega^{\prime}\right) \doteq \left\langle \widetilde{A}_{+}\left(\Delta\omega\right) \, \widetilde{A}_{-}\left(\Delta\omega^{\prime}\right) \right\rangle - \left\langle \widetilde{A}_{+}\left(\Delta\omega\right) \right\rangle \, \left\langle \widetilde{A}_{-}\left(\Delta\omega^{\prime}\right) \right\rangle \,.
\end{equation}
The use of the two-component formalism means $g_{1}$ is not necessarily symmetric when dealing with a single random configuration to describe vacuum fluctuations. Hence, 
we define its symmetrized version using a geometric mean:
\begin{equation}
g_{1}^{\rm symm.}\left(\Delta\omega\,,\,\Delta\omega^{\prime}\right) \doteq \sqrt{g_{1}\left(\Delta\omega\,,\,\Delta\omega^{\prime}\right) \, g_{1}\left(\Delta\omega^{\prime}\,,\,\Delta\omega\right)} \,.
\label{g1sym}
\end{equation}

To obtain the correlation map shown in  Figure~\ref{fig:correlation_map}, we have proceeded 
in three steps. \begin{itemize} 
\item
First, we take an average over late values of $z$ (over the final $100$ steps of the simulation), which sufficiently suppresses products of modes which do not have equal and opposite $D$.  
\item
We then bin neighboring Fourier modes together, suppressing some of the randomness due to the particular realization for the initial value of $\delta A_{-}$. 
\item
 Lastly, although only one set of random amplitudes for $\delta A_{-}$ is generated, eight different runs are performed: the sign of all fluctuations are flipped; the sign of fluctuations living only on the solid black branch ($\Delta\omega>0$, $D<0$) are flipped; and the sign of fluctuations living only on the solid blue branch ($\Delta\omega<0$, $D>0$) are flipped. 
\end{itemize}
This combination of initial conditions efficiently suppresses spurious correlations which appear for a single realization, but which should not 
survive in a proper ensemble average taken over many realizations.

The norm 
of 
$ g_{1}^{\rm symm.}(\Delta\omega\,,\,\Delta\omega^{\prime})$ so calculated is shown using a linear scale on the left panel in Figure~\ref{fig:correlation_map}. On the right panel, we show the complete set of correlations between positive-norm modes with detuning wave numbers $\Delta \omega_j^+$ which have equal and opposite values of $D$. By direct comparison one finds, as expected, that 
the strongest correlations occur between the 
black and blue branches of the dispersion relation.
As expected as well, 
we also see a smaller 
correlation between the dotted red (idler) and solid blue branches (which also have opposite values of $D$).

\begin{figure}
\includegraphics[width=0.45\columnwidth]{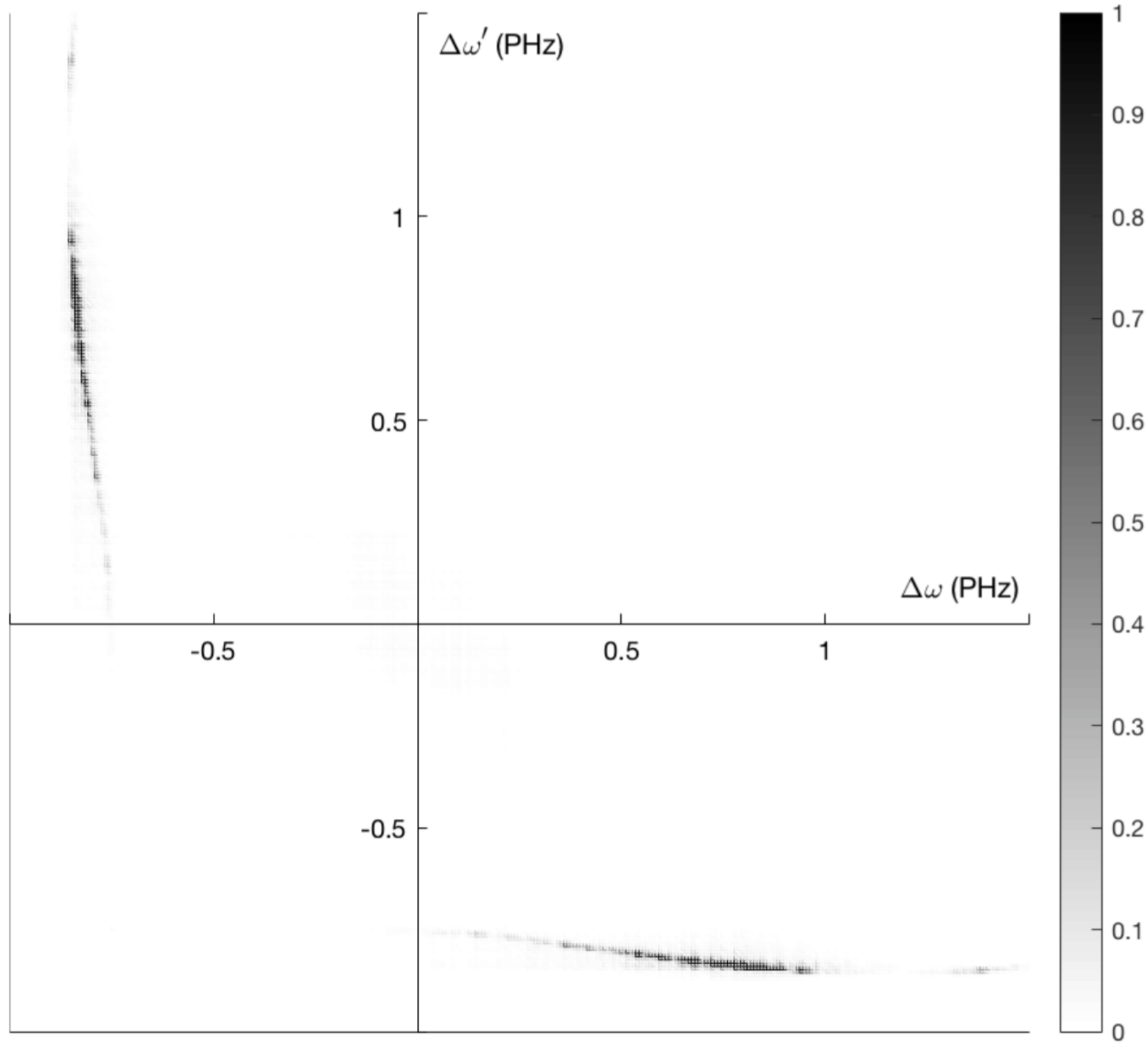} \, \includegraphics[width=0.45\columnwidth]{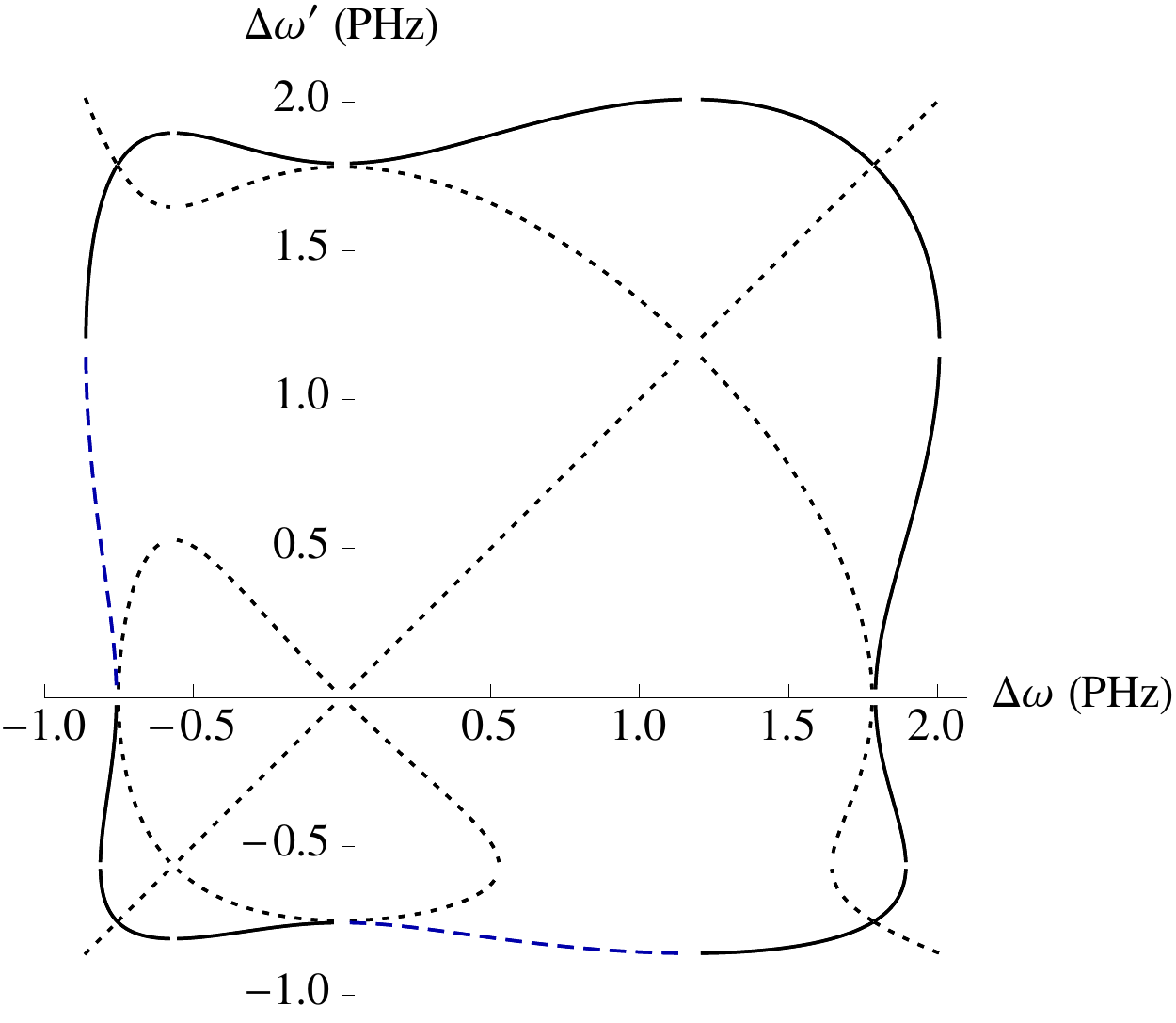}
\caption{Correlations between spontaneously produced photons when the initial state is vacuum.  On the left 
is shown the magnitude of the first order correlation of Eq.~(\ref{g1sym}) numerically obtained by integrating 
Eqs.~(\ref{non-lin-system}) and by extracting an average by the procedure described in the text. The normalization
of the intensity is such that the maximal value is 1. 
The right panel shows the expected loci of the correlations in the $\left(\Delta\omega,\Delta\omega^{\prime}\right)$-plane.  The dotted lines show normal correlations involving products of the form $\tilde{A}^{\star}(\Delta\omega) \tilde{A}(\Delta\omega)$, while the solid (and dashed) lines show anomalous correlations having the form $\tilde{A}(\Delta\omega) \tilde{A}(\Delta\omega)$.  The dashed blue lines show where the correlations between the black and blue branches lie, and correspond to those seen in the left panel. 
\label{fig:correlation_map}}
\end{figure}

\end{appendices}

\bibliography{biblio}

\end{document}